\documentclass{jfm}

\usepackage{graphicx}
\usepackage{natbib}
\usepackage{color}
\usepackage{latexsym}
\usepackage{subfig}
\usepackage{natbib}
\usepackage{upgreek}
\usepackage{mathrsfs}
\usepackage{float}
\usepackage{caption}
\usepackage{soul}
\usepackage{textcomp}
\usepackage{stmaryrd}
\usepackage{amsmath}
\usepackage{mathtools}
\usepackage{epstopdf, epsfig}
\usepackage{amsmath,etoolbox}
\usepackage{bigints}
\usepackage{lipsum}

\usepackage{xargs}                      
\usepackage[pdftex,dvipsnames]{xcolor}  
\usepackage[colorinlistoftodos,prependcaption,textsize=small]{todonotes}

\AtBeginEnvironment{pmatrix}{\setlength{\arraycolsep}{2pt}}

\newcommandx{\raunak}[2][1=]{\todo[linecolor=red,backgroundcolor=red!25,bordercolor=red,#1]{#2}}

\newcommandx{\anirban}[2][1=]{\todo[linecolor=blue,backgroundcolor=blue!25,bordercolor=blue,#1]{#2}}

\def\eee{{\rm e}}
\def\ii{{\rm i}}

\def\be{\begin{equation}}
\def\ee{\end{equation}}
\def\bea{\begin{eqnarray}}
\def\eea{\end{eqnarray}}
\def\bse{\begin{subequations}}
\def\ese{\end{subequations}}
\def\bsea{\begin{subeqnarray}}
\def\esea{\end{subeqnarray}}
\def\dunderline#1{\underline{\underline{#1}}}

\def\({\left (}
\def\){\right )}
\def\[{\left [}
\def\]{\right ]}
\def\<{\left <}
\def\>{\right >}

\affiliation{
$^1$ Environmental and Geophysical Fluids Group, Department of Mechanical Engineering, Indian Institute of Technology, Kanpur, U.P. 208016, India.}

\pubyear{2012}
\volume{xxx}
\pagerange{xxx--yyy}
\title[Bragg resonances and wave triad interactions in two-layered shear flows]{On Bragg resonances and wave triad interactions in two-layered shear flows}

\shortauthor{R. Raj and A. Guha}

\author{Raunak Raj\aff{1} \and Anirban Guha\aff{1}\corresp{\email{anirbanguha.ubc@gmail.com}}}

\affiliation{\aff{1}Environmental and Geophysical Fluids Group, Department of Mechanical Engineering,
Indian Institute of Technology, Kanpur, U.P. 208016, India.}

\begin{document}

\maketitle

\begin{abstract}
The standard resonance conditions for  Bragg scattering as well as weakly nonlinear wave triads have been traditionally derived in the absence of any background velocity. In this paper, we have studied how these resonance conditions get modified when uniform, as well as various piecewise linear velocity profiles, are considered for two-layered shear flows.  Background velocity can influence the resonance conditions in two ways (i) by causing Doppler shifts, and (ii) by changing the intrinsic frequencies of the waves. For Bragg resonance, even a uniform velocity field changes the resonance condition. Velocity shear strongly influences the resonance conditions since, in addition to changing the intrinsic frequencies, it can cause unequal Doppler shifts between the surface,  pycnocline, and the bottom. Using multiple scale analysis and Fredholm alternative, we analytically obtain the equations governing both the Bragg resonance and the wave triads. We have also extended the Higher Order Spectral method, a highly efficient computational tool usually used to study triad and Bragg resonance problems, to incorporate the effect of piecewise linear velocity profile. A significant aspect, both in theoretical and numerical fronts, has been extending the potential flow approximation, which is the basis of studying these kinds of problems, to incorporate piecewise constant background shear.

\end{abstract}


\section{Introduction}
`Wave triad interaction' -- the nonlinear  interaction  between three waves (or modes) satisfying a certain resonant condition, is a fundamental mechanism of energy transfer in fluid flows due to the nonlinear nature of the governing Navier-Stokes equations. 
In a two-layered density stratified flow in the absence of  background velocity, \cite{Ball} showed that two counter-propagating surface gravity waves can give rise to an interfacial gravity wave by forming a wave triad. 
Although Ball had ruled out the possibility of the existence of any other triads involving two surface modes, such interactions were later observed between three co-propagating modes -- two surface waves and one interfacial wave \citep{Baker}. In fact, two counter-propagating interfacial gravity waves can also give rise to a surface gravity wave \citep{Wen,hill1996subharmonic}. Remarkably enough, a rippled bottom topography can act like a neutral, stationary wave and mediate nonlinear energy transfer between two  waves -- a phenomenon known as the `Bragg resonance' \citep{davies1982reflection,mei1985resonant,kirby1986general}.
Bragg resonance strongly affects the wave spectrum in  continental shelves  and coastal regions \citep{Ball}, and also modifies the shore-parallel sandbars \citep{heathershaw1985resonant,elgar2003bragg}.
The study on Bragg resonance was performed in a two-layered density stratified flow by \cite{Alam1}. They showed that second order nonlinearity causes a surface wave propagating over a rippled bottom to transfer energy to (i) an interfacial wave propagating in the same direction (of the surface wave), (ii) an interfacial wave propagating in the opposite direction, or (iii) a surface wave propagating in the same direction, depending on the wavenumber of the bottom ripple. Similar results were also obtained for an interfacial wave. \cite{Alam1} also studied  interactions up to the third order of nonlinearity, thereby giving rise to various classes of Bragg resonance. The numerical simulations for the same were performed using a Higher Order Spectral (HOS) code \citep{Alam2}, which was initially developed for a single layered flow over bottom topography by \cite{Dommermuth}.  Although the equations governing a single triad can also be  analytically obtained without much difficulty up to the second order of nonlinearity,  numerical simulation allows one to incorporate multiple triads up to several orders of nonlinearity. 
In most of the above-mentioned analytical and numerical (e.g.\  HOS) studies on wave triads or Bragg resonances, the base velocity was assumed to be absent. 
This is because these analytical and numerical treatments were based on the potential flow theory. The primary advantage of using the potential flow assumption is that it leads to an outstanding simplification --  one can  solve for the interfaces only. This  allows a deeper insight into the  complex nonlinear problem of resonant triad interactions and subsequent energy transfer. A general base flow falls beyond the purview of the potential flow theory, neither in such flows the dynamics remain confined at the interfaces.

Since atmospheric and oceanic flows usually have base velocities \citep{vallis2017atmospheric}, application of the `standard' potential flow theory in such flows may be  an over-simplification. {Furthermore, The velocity present in the ocean, exspecially in the littoral region and estuaries can be substantial \citep{geyer2017hydraulics}.  Further, it is also well known that the shear can affect the dynamics of the problem \citep{peregrine1976interaction}.} In order to accomodate shear in the study of wave triad interaction, in this work,  we have considered a two-layered density stratified flow in the presence of a piecewise linear base velocity profile. {While piecewise profiles similar to the ones we have considered here have been widely studied in the context of linear instabilities \citep{Drazin,vallis2017atmospheric}, studies involving nonlinear waves and instabilities in presence of piecewise velocity shear are very limited}.  We have shown that such kind of velocity profiles can be included under the umbrella of the \emph{extended} potential flow theory. Therefore, the dynamics is still localized at the interfaces, even though there is a base velocity present.
Piecewise linear base velocity implies that the base vorticity is layerwise constant. Here,  no vorticity is generated in the perturbed flow {except at the interfaces}. In other words, if the initial disturbances are irrotational, the perturbed flow {in the bulk} remains irrotational forever, despite the fact that the base flow is vortical. This fundamental concept has also allowed us to use and extend the general framework of the HOS method by incorporating a piecewise linear velocity profile. In the case of wave triad interaction, adding a constant base velocity doesn't change the dynamics of the problem because all the frequencies are merely Doppler shifted. It can also be intuitively seen that adding a uniform flow `$U$' is similar to moving in a reference frame with a velocity `$U$', and change of the reference frame should not change the dynamics of a problem. Any non-trivial base velocity profile, however,  will break the otherwise symmetric nature of the dispersion relation of  surface/interfacial gravity waves. Addition of a constant base velocity  leads to a significant alteration in the resonance conditions  for Bragg resonance \citep{kirby1988current}; here the Doppler shift is not simply equivalent to changing of the reference frame because of the involvement of the bottom topography. The fact that the bottom topography is at rest while the surface and the interface have some base flow results in unequal Doppler shift between the surface/interface and the bottom topography. 

Significant changes occur when a uniform shear is present in each layer. When  there is a jump in the base vorticity (i.e.\ shear)  across an interface, it  leads to vorticity waves. In addition, if there is a buoyancy jump at the same interface, we get vorticity-gravity waves \citep{Harnik}. Interaction between an interfacial vorticity wave (with no buoyancy jump) and a surface gravity wave was the focus of a recent study by \cite{Drivas}. Due to the presence of shear, the surface and the interface  move with different base velocities, which   significantly alters the conditions for the formation of resonant triads. Therefore, we expect that the problems involving triad interactions and Bragg resonances are remarkably enriched when piecewise linear base velocity field is present.

The paper is organized as follows. In \S \ref{sec:2}, we have shown the applicability of potential flow theory to a piecewise linear velocity profiles. Furthermore, we have derived the modified evolution equations, which has been subsequently applied to the HOS Code in order to incorporate the velocity field. We also use the evolution equations to obtain the dispersion relation of a general two-layered flow with a velocity field. This is followed by a perturbation expansion of the variables till $\mathcal{O}(\epsilon^2)$, and using the Fredholm's alternative, we obtain the analytical solution for amplitude variation both for the case of Bragg resonance and wave triad interaction. {Here, the expansion parameter $\epsilon$ measures the steepness of the wave and following \cite{Alam2}, we assume the steepness of the wave and the bottom to be of the same order.} In \S \ref{sec:3},  we have explored the effect of different types of velocity fields on different types of Bragg resonance triads using  dispersion relations. In \S \ref{sec:4}, we have briefly explained the effect of velocity field on wave triad interactions. We have devoted the \S \ref{sec:5} to the numerical code and simulation. In this section, we  have described the HOS code, which we have extended to incorporate piecewise linear velocity profiles. After validating the code, we have shown some numerical simulations to corroborate our analytical derivations. Finally, we summarize and conclude the paper in \S \ref{sec:6}.


\section{Theory}\label{sec:2}

The kinematic boundary conditions and the dynamic boundary conditions for the water wave problem are nonlinear, suggesting that waves can interchange energy between them through a nonlinear interaction. This nonlinear exchange of energy between the waves, known as the wave triad interaction, is maximum when the waves involved satisfy a specific resonance condition. Although the energy exchange is a weakly nonlinear phenomenon, the condition for triad interactions can simply be obtained from the linear dispersion relations. The condition for the resonance between waves of wavenumbers $(k_1, k_2 \textrm{ and } k_3)$ and frequencies $(\omega_1, \omega_2 \textrm{ and } \omega_3)$ is
\begin{subequations}
\begin{align}
k_3&=k_1\pm k_2,\\
\omega_3&=\omega_1\pm\omega_2.
\end{align}
\end{subequations}
The above condition basically means that if on the $k$--$\omega$ plane, waves are denoted by the vectors $(k_1,\omega_1)$, $(k_2,\omega_2)$ and ($k_3,\omega_3$), then these vectors are linearly dependent \citep{Ball}. Further, when two waves  exchange energy with each other via mediation of the bottom ripples, which acts as a stationary wave with zero frequency, it is known as the Bragg resonance. Here, the resonance condition becomes
\begin{subequations}
\begin{align}
k_2&=k_1\pm k_b,\\
\omega_1&=\omega_2.
\end{align}
\end{subequations}

For the case of no velocity, the dispersion relation is a biquadratic polynomial in $\omega$, and is given as \citep{Ball,Alam1}
\begin{equation}
\omega^4(R+\coth{kh_u}\coth{kh_l})-\omega^2gk(\coth{kh_u}+\coth{kh_l})+g^2k^2(1-R)=0.
\label{eq:biq}
\end{equation}
\begin{figure}
\centering\includegraphics[width=120mm]{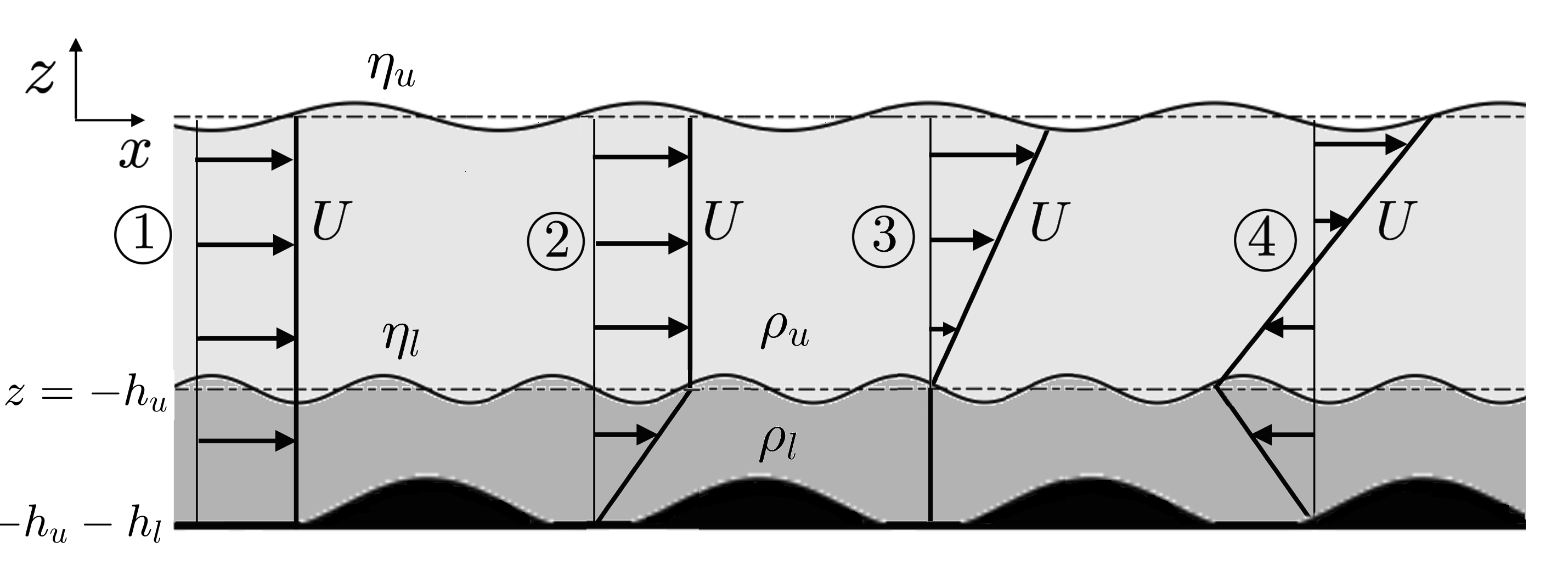}
\caption{Schematic of a two-layered density stratified flow in presence of bottom topography and various kinds of simple velocity profiles, labelled by \textcircled{1}: uniform flow, \textcircled{2}: constant shear in the bottom layer, \textcircled{3}: constant shear in the top layer, and \textcircled{4}: constant shear in both layers.
 }
\label{fig:1}
\end{figure}

\noindent Here $R\equiv \rho_u/\rho_l$ is the density ratio, and $h_u$ and $h_l$ are   respectively the depths of the upper and  lower layers. Throughout the paper, the subscripts $u$ and $l$ respectively denote  `upper' and `lower'. The implication of (\ref{eq:biq}) being a biquadratic in $\omega$ is that the leftward travelling waves and the rightward travelling waves are symmetric, i.e. the difference between the two is simply a matter of a change in the sign of $\omega$. However, in the presence of a uniform velocity $U$, the intrinsic frequencies of the waves are Doppler shifted by an amount `$Uk$'. Further, if the velocity field is a function of the vertical coordinate `$z$', then there can be a presence of a vorticity wave also, which will alter the intrinsic frequency of the waves as well, and the biquadratic and symmetric nature of the dispersion relation will be lost. We have classified the velocity profiles into 4 categories: (i) a uniform flow (ii) shear only in the lower layer (iii) shear only in the upper layer (iv) shear in both the layers. These cases have been shown in  figure \ref{fig:1}. In the first case, the surface and the interface are not Doppler shifted with respect to each other but they are Doppler shifted with respect to the bottom. This should mean that the condition for wave triad interaction will not change but the condition for Bragg resonance should get altered. Further, there won't be any change in the intrinsic frequencies of any of the waves present in the system. In the second case, shear is only present in the lower layer. This case is similar to the first one with reference to the Doppler shifts, i.e. both the surface, and the interface between $\rho_u$ and $\rho_l$ (hereafter, simply referred to as  `interface' or `pycnocline'), are Doppler shifted equally with respect to the bottom, but additionally, the intrinsic frequencies of waves will change due to the presence of a shear jump at the interface. In the third case,  shear is present only in the upper layer and hence the surface and the interface   are Doppler shifted; moreover, there is a presence of a shear jump at the interface too. Hence, the intrinsic frequencies of the waves will also change.  In the last case, shear is present in both  layers. Hence, there is a shear jump both at the interface and at the surface and the surface and the interface are Doppler shifted unequally with respect to the bottom. It is also to be noted here that the velocity difference between the surface and the interface, i.e. the second and the fourth cases might lead to linear instabilities as well due to the formation of counter--propagating system \citep{Guha,Shete}. 
However, such linear instabilities, for moderate values of shear are restricted to high wavenumbers and don't have appreciable growth rates. In any case, we would be focussing on the nonlinear interactions only.

It was shown in \cite{guha2017waves} that in the presence of a piecewise linear velocity profile, there is no perturbation vorticity generation {in the fluid bulk and vorticity is  generated exclusively at the interfaces}. This means that if the {bulk} flow is initially irrotational,  then it will remain so forever, similar to the scenario of no background velocity. Further, if there is a density difference $(\rho_1, \rho_2)$ as well a shear difference $(\Omega_1, \Omega_2)$ across any \emph{general} (hence subscripts `$1$' and `$2$' are used, instead of `$u$' and `$l$') interface $z=h_0+\eta(x,t)$ moving in a velocity field $U=U(z)$, then the dynamic boundary condition at any interface $z=h_0+\eta(x,t)$ is given by (See appendix \ref{app:A} for derivation)
\begin{multline}
\rho_1 \left[ \phi_{1,t}+\frac{1}{2} \left( \phi_{1,x}^2+\phi_{1,z}^2\right)+U\phi_{1,x}-\Omega_1\psi_1+g\eta\right]
         =\\ \rho_2\left[\phi_{2,t}+\frac{1}{2} \left( \phi_{2,x}^2+\phi_{2,z}^2\right)+U\phi_{2,x}-\Omega_2\psi_2+g\eta\right].
\end{multline}
Here, $\phi_1$ and $\phi_2$ are respectively the perturbation velocity potentials of fluids `$1$' and `$2$', while $\psi_1$ and $\psi_2$ are the same for the streamfunctions, which can be obtained using the respective velocity potentials. The comma in the subscript denotes partial derivative; for example, $\eta_{1,x}\equiv \partial \eta_1/\partial x$. In the above equation, the terms $U \phi_{1,x}$  and $U\phi_{2,x}$  are the `Doppler shift' terms  indicating that the interface is moving in a velocity field $U$. The terms $\Omega_1\psi_1$ and $\Omega_2\psi_2$ appear due to the presence of the constant shears $\Omega_1$ and $\Omega_2$ on either side of the interface. Rest all other terms are usual and appear in the absence of velocity as well. Similarly, the kinematic boundary condition for the same interface will be given by
\begin{equation}
\eta_{,t}+(U+\phi_{,x})\eta_{,x}=\phi_{,z}.
\end{equation}
Here, the term $U\eta_{,x}$ is the Doppler shift term.  We will apply both kinematic and dynamic boundary conditions to the surface and the interface in figure \ref{fig:1}.  The above equations are applicable at the interface i.e. at $z=h_0+\eta(x,t)$. More accurately, the LHS of the dynamic boundary condition is evaluated just above the interface $z=h_0+\eta(x,t)$ whereas, the RHS is evaluated just below the interface. On the other hand, for the kinematic boundary condition, there are two separate equations -- one above the interface and one below it. However, quite often in this paper, we would use the Taylor expansion to evaluate the variables at the mean level i.e. $z=h_0$. In particular, the velocity $U(z)$ just above the interface would be given as
\begin{equation}
U=U_0+\Omega_1\eta,     \label{eq:Tay1}
\end{equation}
and just below the interface it will be 
\begin{equation}
U=U_0+\Omega_2\eta,     \label{eq:Tay2}
\end{equation}
where $U_0=U(h_0)$.
\subsection{Framework}
\label{sec:Triad_Framework}

Here we give a general framework that consists of a system of equations at $\mathcal{O}(\epsilon)$ and $\mathcal{O}(\epsilon^2)$, which is obtained using  perturbation analysis and the method of multiple scales for a periodic wave train. We have kept the system quite general so as to use the system of equations for the purpose of wave triad interaction (see \S \ref{sec:Triad_Analytical}) and  Bragg resonance (see \S \ref{sec:Triad_Bragg}). 

We consider a two-interface system with piecewise constant density and vorticity in each layer, see figure \ref{fig:1}. The velocity profile is continuous, but the derivative of velocity may have a discontinuity at the density interface. Total depth of the system is $H$ and $H=h_u+h_l$. The fluid above the surface is assumed to be a zero density fluid and $R$ is the density ratio at the interface $(R\equiv\rho_u/\rho_l)$. The base velocity profile is piecewise linear, and has the values $U=\{U_u, U_l, U_b\}$ at $z=\{0, -h_u, -h_u-h_l\}$ respectively. The vertical ($z$) axis points upwards, hence the gravity (g) is along the negative $z$-direction. The elevations of the surface and the pycnocline from their respective mean level are $\eta_u(x,t)$ and $\eta_l(x,t)$. Similarly, the elevation of the bottom topography is $\eta_b(x)$ from its mean level at $z=-h_u-h_l$. As mentioned already,  for piecewise linear base velocity profile perturbed by  irrotational initial disturbances, the vorticity generation is limited to the interfaces and the bulk flow remains irrotational. This allows us to introduce the velocity potentials $\phi_u$ and $\phi_l$ respectively in the upper  and the lower layers. Hence, the continuity equation reduces  to the Laplace equation
\begin{subequations}
\begin{align}
    \nabla^2 \phi_u &=0\qquad\textrm{$-h_u+\eta_l<z<\eta_u$},\label{HOS_Lap1}\\
    \nabla^2 \phi_l&=0\qquad  \textrm{$-h_u-h_l+\eta_b<z<-h_u+\eta_l$}.\label{HOS_Lap2}    
\end{align}
\end{subequations}
The kinematic boundary conditions  are
\begin{subequations}
\begin{align}
	\eta_{u,t}+(U+\phi_{u,x})\eta_{u,x}&=\phi_{u,z} \qquad \textrm{at $z=\eta_u$},\label{KB1}\\
    \eta_{l,t}+(U+\phi_{u,x})\eta_{l,x}&=\phi_{u,z} \qquad  \textrm{at $z=-h_u+\eta_l$},\label{KB2}\\
    \eta_{l,t}+(U+\phi_{l,x})\eta_{l,x}&=\phi_{l,z} \qquad\;  \textrm{at $z=-h_u+\eta_l$},\label{KB3}\\
    (U+\phi_{l,x})\eta_{b,x}&=\phi_{l,z} \qquad\; \textrm{at $z=-h_u-h_l+\eta_b$}.\label{KB4}
\end{align}
Likewise, the dynamic boundary conditions are as follows:
\begin{align}
	\phi_{u,t}+\frac{1}{2} \left( \phi_{u,x}^2+\phi_{u,z}^2\right)+U\phi_{u,x}-\Omega_u\psi_u+g\eta_u&=0 \qquad \textrm{at $z=\eta_u$}, \label{DB1}\\
    \begin{split}
    	 \rho_u \left[ \phi_{u,t}+\frac{1}{2} \left( \phi_{u,x}^2+\phi_{u,z}^2\right)+U\phi_{u,x}-\Omega_u\psi_u+g\eta_l\right]\\
         -\rho_l\left[ \phi_{l,t}+\frac{1}{2} \left( \phi_{l,x}^2+\phi_{l,z}^2\right)+U\phi_{l,x}-\Omega_l\psi_l+g\eta_l\right]&=0 \qquad \textrm{at $z=-h_u+\eta_l$}. \label{DB2}
    \end{split}
\end{align}
\end{subequations}
{We are interested in obtaining the solutions up to a first order of \emph{nonlinearity}. Hence, we perform a perturbation expansion till $\mathcal{O}(\epsilon^2)$, where the expansion parameter $\epsilon$ measures the wave steepness. It is also assumed that the steepness of the bottom topography is $\mathcal{O}(\epsilon)$.}
\begin{subequations}
\begin{align}
\phi_u(x,z,t)&=\epsilon\phi_u^{(1)}(x,z,t,\tau)+\epsilon^2\phi_u^{(2)}(x,z,t,\tau),\\
\phi_l(x,z,t)&=\epsilon\phi_l^{(1)}(x,z,t,\tau)+\epsilon^2\phi_l^{(2)}(x,z,t,\tau),\\
\eta_u(x,t)&=\epsilon\eta_u^{(1)}(x,t,\tau)+\epsilon^2\eta_u^{(2)}(x,t,\tau),\\
\eta_l(x,t)&=\epsilon\eta_l^{(1)}(x,t,\tau)+\epsilon^2\eta_l^{(2)}(x,t,\tau).
\end{align}
\end{subequations}
Here we have assumed that the potentials and elevations have a slow time scale `$\tau$' associated with them such that $\tau=\epsilon t$. The superscripts $(1)$ and $(2)$ respectively denote the  $\mathcal{O}(\epsilon)$ and $\mathcal{O}(\epsilon^2)$ terms. Further, we expand the velocity potential $\phi$ and the streamfunction $\psi$ in a Taylor series about the respective mean surface/interface, which at $\mathcal{O}(\epsilon)$ gives the following set of equations:
\begin{subequations}\label{eq:Order1}
\begin{align}
	\phi^{(1)}_{u,z}-\eta^{(1)}_{u,t}-U_u\eta^{(1)}_{u,x}&=0\qquad \textrm{at $z=0$},\label{eq:Order1_1}\\
    \phi^{(1)}_{u,z}-\eta^{(1)}_{l,t}-U_l\eta^{(1)}_{l,x}&=0 \qquad  \textrm{at $z=-h_u$},\\
    \phi^{(1)}_{l,z}-\eta^{(1)}_{l,t}-U_l\eta^{(1)}_{l,x}&=0 \qquad  \textrm{at $z=-h_u$},\\
    \phi^{(1)}_{u,t}+U_u\phi^{(1)}_{u,x}-\Omega_u\psi^{(1)}_u+g\eta^{(1)}_u&=0 \qquad \textrm{at $z=0$},\\
 R\left[ \phi^{(1)}_{u,t}+U_l\phi^{(1)}_{u,x}-\Omega_u\psi^{(1)}_u+g\eta^{(1)}_l\right]\\
         -\left[ \phi^{(1)}_{l,t}+U_l\phi^{(1)}_{l,x}-\Omega_l\psi^{(1)}_l+g\eta^{(1)}_l\right]&=0 \qquad \textrm{at $z=-h_u$}, \\
\phi^{(1)}_{l,z}&=0 \qquad \textrm{at $z=-h_u-h_l$}\label{eq:Order1_2}
\end{align}
\end{subequations}
Additionally, we use the eigenfunction expansions with slowly varying amplitudes satisfying the respective Laplace equations. Thus for $j=\{1,2,..\}$ and $m=\{1,2\}$, where the subscript $j$ denotes the $j$-th wavenumber and the superscript $(m)$ denotes the order of nonlinearity, we get
\begin{subequations}\label{eq:Eigen}
\begin{align}
    \phi_{uj}^{(m)}&= \[A_j^{(m)}(\tau )\frac{\cosh{k_j(z+h_u)}}{\cosh{(k_jh_u)}} + B_j^{(m)}(\tau )\frac{\sinh{k_jz}}{\cosh{(k_jh_u)}} \] \eee^{\ii(k_jx-\omega_j t)}+\mathrm{c.c.},\label{eq:Eigen1}\\
    \phi_{lj}^{(m)}&=\[C_j^{(m)}(\tau )\frac{\cosh{k_j(z+h_u+h_l)}}{\cosh{(k_jh_l)}}+D_j^{(m)}(\tau )\frac{\sinh{k_j(z+h_u+h_l)}}{\cosh{(k_jh_l)}}  \] \eee^{\ii(k_jx-\omega_j t)} +\mathrm{c.c.},  \\
    \psi_{uj}^{(m)}&= \ii\[A_j^{(m)}(\tau )\frac{\sinh{k_j(z+h_u)}}{\cosh{(k_jh_u)}} + B_j^{(m)}(\tau )\frac{\cosh{k_jz}}{\cosh{(k_jh_u)}} \] \eee^{\ii(k_jx-\omega_j t)}+\mathrm{c.c.},\\
            \psi_{lj}^{(m)}&=\ii\[C_j^{(m)}(\tau )\frac{\sinh{k_j(z+h_u+h_l)}}{\cosh{(k_jh_l)}}+D_j^{(m)}(\tau )\frac{\cosh{k_j(z+h_u+h_l)}}{\cosh{(k_jh_l)}} \] \eee^{\ii(k_jx-\omega_j t)} +\mathrm{c.c.}, \\
\eta_{uj}^{(m)}&= a_j^{(m)}(\tau )\eee^{\ii(k_jx-\omega_j t)}+\mathrm{c.c.},\\
    \eta_{lj}^{(m)}&=b_j^{(m)}(\tau )\eee^{\ii(k_jx-\omega_j t)} +\mathrm{c.c.},\label{eq:Eigen2}
\end{align}
\end{subequations}
where  c.c.\ denotes complex conjugate. Substituting the above equations \eqref{eq:Eigen1}--\eqref{eq:Eigen2} in the equations \eqref{eq:Order1_1}--\eqref{eq:Order1_2} at $\mathcal{O}(\epsilon)$, we obtain a set of linear equations corresponding to any given wavenumber $k_j$ at $\mathcal{O}(\epsilon)$, {the homogenenous part of which is}
\begin{equation}
\dunderline{\mathfrak{D}}(\omega_j,k_j)\mathbf{x}^{(1)}_j=0.
\end{equation}
Here, the vector $\mathbf{x}^{(1)}_j\equiv \left[A_j^{(1)}, B_j^{(1)}, C_j^{(1)}, D_j^{(1)}, a_j^{(1)},b_j^{(1)}\right]^\dagger$, and the matrix $\dunderline{\mathfrak{D}}(\omega_j,k_j)$ is given by

 \footnotesize{$${\hskip -15pt}\left[ {\begin{array}{cccccc}
   \dfrac{k_j}{\coth(k_jh_u)}  &\dfrac{k_j}{\cosh(k_jh_u)} & 0 &0& \ii\omega_j^{\{1\}} & 0\\[10pt]
   0& k_j &0&0 &  0&   \ii\omega_j^{\{2\}}   \\[10pt] 
    0& 0 &   \dfrac{k_j}{\coth(k_jh_l)} &k&0 &\ii\omega_j^{(2)}  \\[10pt]
    -\ii\omega_j^{(1)}-\dfrac{\ii\Omega_u}{\coth{(k_jh_u)}} &-\dfrac{\ii\Omega_u}{\cosh{(k_jh_u)}}&0&0&g&0 \\[10pt]
   -\dfrac{\ii R\omega_j^{\{2\}}}{\cosh(k_jh_u)} & \dfrac{\ii R\omega_j^{\{2\}}}{\coth(k_jh_u)}-\ii R\Omega_u &\ii\omega_j^{\{2\}}+\dfrac{\ii\Omega_l}{\coth{k_jh_l}}&\dfrac{\ii\omega_j^{(2)}}{\coth{k_jh_l}}+\ii\Omega_l &0&g(R-1) \\[10pt]0&0&0&\dfrac{k_j}{\cosh{k_jh_l}}&0&0\end{array}} \right]{\hskip -7pt}$$}
 \normalsize
 
\noindent where $\omega_j^{\{1\}}=\omega_j-U_uk_j\quad;\quad\omega_j^{\{2\}}=\omega_j-U_lk_j$. The dispersion relation of the above system is obtained by setting the determinant of the above matrix to zero and is given by the equation 
 \begin{equation}
\mathfrak{D}(\omega_j,k_j)=0,
\end{equation}
where $\mathfrak{D}(\omega_j,k_j)$ is the determinant of the matrix $\dunderline{\mathfrak{D}}(\omega_j,k_j)$. 

{
In addition to the homogenous  solution, we also have the particular solutions at $\mathcal{O}(\epsilon)$  due to the velocity difference between the bottom and the fluid above it. In such a case, the time independent surface elevation, capturing the non-homogeneity introduced by the mean flow's interaction with the bottom, is  given by 
\begin{equation}
     \hat{\eta}_u=-\frac{U_uU_bU_l^2k_b^6}{\cosh{(k_bh_u)}\cosh^2{(k_bh_l)}\mathfrak{D}(0,k_b)}\hat{\eta}_b,
\end{equation}
and other coefficients, i.e.\ $\hat{\eta}_l$, $A, B, C, D$ are given in the appendix \ref{app:B}. 
}

At $\mathcal{O}(\epsilon^2)$, we  obtain the following equations:
\begin{subequations}
\begin{align}
	\phi^{(2)}_{u,z}-\eta^{(2)}_{u,t}-U_u\eta^{(2)}_{u,x}&=p_1+q_1\qquad  \textrm{at $z=0$},\label{eq:Order2_1}\\
    \phi^{(2)}_{u,z}-\eta^{(2)}_{l,t}-U_l\eta^{(2)}_{l,x}&=p_2+q_2 \qquad  \textrm{at $z=-h_u$},\\
    \phi^{(2)}_{l,z}-\eta^{(2)}_{l,t}-U_l\eta^{(2)}_{l,x}&=p_3 +q_3\qquad  \textrm{at $z=-h_u$},\\
    \phi^{(2)}_{u,t}+U_u\phi^{(2)}_{u,x}-\Omega_u\psi^{(2)}_u+g\eta^{(2)}_u&=p_4+q_4 \qquad \textrm{at $z=0$},\\
 \begin{split}
 \left[ \phi^{(2)}_{u,t}+U_l\phi^{(2)}_{u,x}-\Omega_u\psi^{(2)}_u+g\eta^{(2)}_l\right]\\
         -\left[ \phi^{(2)}_{l,t}+U_l\phi^{(2)}_{l,x}-\Omega_l\psi^{(2)}_l+g\eta^{(2)}_l\right]&=p_5+q_5  \qquad \textrm{at $z=-h_u$}, 
 \end{split}\\
\phi^{(2)}_{l,z}&=p_6+q_6 \qquad \textrm{at $z=-h_u-h_l$}.\label{eq:Order2_2}
\end{align}
\end{subequations}
The RHS terms $p_1,p_2,p_3,p_4,p_5$ and $p_6$ are the products of two $\mathcal{O}(\epsilon)$ terms and for neatness, we have removed the superscript `$(1)$' from the variables. They are given as
\begin{subequations}
\begin{align}
p_1&=\eta_{u,x}(\phi_{u,x}+\Omega_u\eta_{u})-\eta_u\phi_{u,zz}\qquad\qquad\qquad\qquad \qquad\qquad\quad\textrm{$z=0$},\\
p_2&=\eta_{l,x}(\phi_{u,x}+\Omega_u\eta_{l})-\eta_l\phi_{u,zz}\qquad\qquad\qquad\qquad \qquad\qquad\qquad \textrm{$z=-h_u$},\\
p_3&=\eta_{l,x}(\phi_{l,x}+\Omega_l\eta_{l})-\eta_l\phi_{l,zz}\qquad\qquad\qquad\qquad \qquad \qquad\qquad\textrm{$z=-h_u$},\\
p_4&=-\eta_u(\phi_{u,tz}+U_u\phi_{u,xz}+\Omega_u\phi_{u,x})-\frac{1}{2}\[(\phi_{u,x})^2+(\phi_{u,z})^2\]+\Omega_u\eta_u\psi_{u,z} \quad \textrm{$z=0$},\\
\begin{split}
p_5&=R\[\eta_l(\phi_{u,tz}+U_l\phi_{u,xz}+\Omega_u\phi_{u,x})-\frac{1}{2}\[(\phi_{u,x})^2+(\phi_{u,z})^2\]+\Omega_u\eta_l\psi_{u,z}\] \\
&\quad-\[\eta_l(\phi_{l,tz}+U_l\phi_{l,xz}+\Omega_l\phi_{l,x})-\frac{1}{2}\[(\phi_{l,x})^2+(\phi_{l,z})^2\]+\Omega_l\eta_l\psi_{l,z}\]\qquad \;\textrm{$z=-h_u$},
\end{split}\\
p_6&=\eta_{b,x}(\phi_{l,x}+\Omega_l\eta_{b})-\eta_b\phi_{l,zz}\qquad\qquad\qquad\qquad\qquad \qquad\qquad\textrm{$z=-h_u-h_l$}.
    \end{align}
\end{subequations}
Here, the RHS terms comprise of terms due to non-linearity of the boundary condition as well as due to taylor expansion about the mean level. The  RHS terms $q_1,q_2,q_3,q_4,q_5$ and $q_6$ are the time derivatives of the $\mathcal{O}(\epsilon)$ terms:
\begin{subequations}
\begin{align}
q_1&=\eta_{u,\tau},\\
q_2&=\eta_{l,\tau},\\
q_3&=\eta_{l,\tau},\\
q_4&=-\phi_{u,\tau},\\
q_5&=-R\phi_{u,\tau}+\phi_{l,\tau}, \\
q_6&=0.
\end{align}
\end{subequations}
The set of equations obtained till here are very general and  works both for the case of wave triad interaction and  Bragg resonance. This is because till here, we haven't made any assumption on the wavenumbers present in the system or if those wavenumbers satisfy any particular resonance condition. Hence, we will be using the above framework to obtain the analytical solutions for wave triad interaction in \S \ref{sec:Triad_Analytical} as well as Bragg resonance in \S \ref{sec:Triad_Bragg}.

\subsection{Analytical solution for wave triad interaction}
\label{sec:Triad_Analytical}

We assume that initially at $\mathcal{O}(\epsilon)$, the system has only 3 wavenumbers  $\{k_1,k_2, k_3\}$ and corresponding  frequencies $\{\omega_1,\omega_2,\omega_3\}$, satisfying the resonance condition. Without any loss of generality, the resonance condition is given by $k_1=k_2+k_3$ and $\omega_1=\omega_2+\omega_3$. The surface elevation expressed as a sum of these three modes read
\begin{equation}
\eta_u^{(1)}(x,t,\tau)=a_1^{(1)}(\tau )\eee^{\ii(k_1x-\omega_1 t)}+a_2^{(1)}(\tau )\eee^{\ii(k_2x-\omega_2 t)}+a_3^{(1)}(\tau )\eee^{\ii(k_3x-\omega_3 t)}+\mathrm{c.c}.
\end{equation}
The other functions  $\phi_u^{(1)},\phi_l^{(1)},\psi_u^{(1)},\psi_l^{(1)}$ and $\eta_l^{(1)}$ can  also be written in a similar fashion. Substituting this in the equations at $\mathcal{O}(\epsilon)$, we would obtain the following set of linear equations:
\begin{equation}\label{eq:Fred1}
\dunderline{\mathfrak{D}}(\omega_1,k_1)\mathbf{x}^{(1)}_1=0\quad;\quad\dunderline{\mathfrak{D}}(\omega_2,k_2)\mathbf{x}^{(1)}_2=0\quad;\quad\dunderline{\mathfrak{D}}(\omega_3,k_3)\mathbf{x}^{(1)}_3=0.
\end{equation}
The vector $\mathbf{x}^{(1)}_j\equiv \left[A_j^{(1)}, B_j^{(1)}, C_j^{(1)}, D_j^{(1)}, a_j^{(1)},b_j^{(1)}\right]^\dagger$ and the matrix $\dunderline{\mathfrak{D}}(\omega,k)$ are given in \S \ref{sec:Triad_Framework}. We further proceed to substitute   \eqref{eq:Eigen1}--\eqref{eq:Eigen2} in  \eqref{eq:Order2_1}--\eqref{eq:Order2_2} at $\mathcal{O}(\epsilon^2)$.
Here, the LHS of the equations obtained at $\mathcal{O}(\epsilon^2)$ is similar to those obtained at $\mathcal{O}(\epsilon)$. On substitution, we collect the terms corresponding to each wavenumber $k_1,k_2$ and $k_3$ after using the resonance condition $k_1=k_2+k_3$ and $\omega_1=\omega_2+\omega_3$. We  obtain  equations of the form
\begin{subequations}\label{eq:Fred2}
\begin{align}
\dunderline{\mathfrak{D}}(\omega_1,k_1)\mathbf{x}^{(2)}_1=\mathbf{v}_1a^{(1)}_2a^{(1)}_3+\mathbf{r}_1a^{(1)}_{1,\tau}\;,\\
\dunderline{\mathfrak{D}}(\omega_2,k_2)\mathbf{x}^{(2)}_2=\mathbf{v}_2\bar{a}^{(1)}_3a^{(1)}_1+\mathbf{r}_2a^{(1)}_{2,\tau}\;,\\
\dunderline{\mathfrak{D}}(\omega_3,k_3)\mathbf{x}^{(2)}_3=\mathbf{v}_3a^{(1)}_1\bar{a}^{(1)}_2+\mathbf{r}_3a^{(1)}_{3,\tau}\;,
\end{align}
\end{subequations}
where overbar denotes complex conjugate. The vector $\mathbf{x}^{(2)}_j\equiv \left[A_j^{(2)}, B_j^{(2)}, C_j^{(2)}, D_j^{(2)}, a_j^{(2)}, b_j^{(2)}\right]^\dagger$ and the terms of the vector $\mathbf{v}_j$ and $\mathbf{r}_j$ are given in the appendix \ref{app:B}. The vector $\mathbf{v}_j$ comes from the coefficient of $\exp{[\ii(k_jx-\omega_jt)]}$ present in the product of two $\mathcal{O}(\epsilon)$ terms.  Similarly, the vector $\mathbf{r}_j$ comes from the time derivatives of $\mathcal{O}(\epsilon)$ terms. It might be noted here that the product terms contain combinations of various terms such as $A^{(1)}_ia^{(1)}_j, B^{(1)}_ia^{(1)}_j,C^{(1)}_iC^{(1)}_j$ etc; however, we have converted each of these products into the product $a^{(1)}_ia^{(1)}_j$, i.e. in terms of products of amplitude of surface elevation by using the null space of the respective matrix $\dunderline{\mathfrak{D}}(\omega_j,k_j)$. Similarly, the slow time derivatives are also converted in terms of the slow time derivative of the surface elevations, i.e. $a^{(1)}_{j,\tau}$.

Using the Fredholm alternative in the context of the sets of equations \eqref{eq:Fred1} and \eqref{eq:Fred2}, we deduce that the solutions for $\mathbf{x}^{(2)}_i$ exist if and only if the vectors $\mathbf{v}_i$ are orthogonal to the null space of the transpose of the respective matrices $\dunderline{\mathfrak{D}}(\omega_j,k_j)$. Denoting the null space of the transpose of the matrix $\dunderline{\mathfrak{D}}(\omega_j,k_j)$ by $\mathbf{n}_j$, we finally get a set of three equations:
\begin{subequations}
\begin{align}
\mathbf{n}_1\cdot \(\mathbf{v}_1a_2^{(1)}a_3^{(1)}+\mathbf{r}_1 a_{1,\tau}^{(1)}\)&=0,\\
\mathbf{n}_2\cdot \(\mathbf{v}_2\bar{a}_3^{(1)}a_1^{(1)}+\mathbf{r}_2a_{2,\tau}^{(1)}\)&=0,\\
\mathbf{n}_3\cdot \(\mathbf{v}_3a_1^{(1)}\bar{a}_2^{(1)}+\mathbf{r}_3a_{3,\tau}^{(1)}\)&=0,
\end{align}
\end{subequations}
which finally gets reduced to
\begin{align}
a_{1,\tau}^{(1)}= \beta_1a_2^{(1)}a_3^{(1)}\qquad;\qquad a_{2,\tau}^{(1)}=\beta_2\bar{a}^{(1)}_3a_1^{(1)}\qquad;\qquad a_{3,\tau}^{(1)}= \beta_3a^{(1)}_1\bar{a}^{(1)}_2,
\end{align}
where
\begin{equation}
\beta_j=-\frac{\mathbf{n}_j\cdot \mathbf{v}_j}{\mathbf{n}_j\cdot\mathbf{r}_j}.
\end{equation}

\subsection{Analytical Solution for Bragg resonance}
\label{sec:Triad_Bragg}

The equations for the case of Bragg resonance can also be obtained using the same framework as in \S \ref{sec:Triad_Analytical}. However, in the case of Bragg resonance, only two propagating waves are involved, the third one is the bottom ripple. We assume that the participating waves to have the wavenumbers $\{k_1,k_2\}$ with frequencies $\{\omega_1,\omega_2\}$ and the bottom with the wavenumber $k_b$. Substituting the normal modes, we would get a set of linear equations at $\mathcal{O}(\epsilon)$:

\begin{equation}
\dunderline{\mathfrak{D}}(\omega_1,k_1)\mathbf{x}^{(1)}_1=0\quad;\quad\dunderline{\mathfrak{D}}(\omega_2,k_2)\mathbf{x}^{(1)}_2=0.
\end{equation}
Here the vector $\mathbf{x}^{(1)}_j\equiv [A_j^{(1)}, B_j^{(1)}, C_j^{(1)}, D_j^{(1)}, a_j^{(1)},b_j^{(1)}]^\dagger$ and the matrix $\dunderline{\mathfrak{D}}(\omega,k)$ are the same as that in \S \ref{sec:Triad_Analytical}. We assume that at $\mathcal{O}(\epsilon)$, the surface consists of only two modes, $k_1$ and $k_2$. Hence we write $\eta^{(1)}_u(x,t)$ as
\begin{align}
\eta_u^{(1)}(x,t,\tau)&=a_1^{(1)}(\tau )\eee^{\ii(k_1x-\omega_1 t)}+a_2^{(1)}(\tau )\eee^{\ii(k_2x-\omega_2 t)}+\mathrm{c.c.},\\
\eta_b(x)&=a_b\eee^{\ii k_1x}+\mathrm{c.c.}
\end{align}
The other functions  $\phi_u^{(1)},\phi_l^{(1)},\psi_u^{(1)},\psi_l^{(1)}$and $\eta_l^{(1)}$ containing the wavenumbers `$k_1$' and `$k_2$' can also be written similarly. Substituting this in the equations at $\mathcal{O}(\epsilon)$, we would obtain a set of linear equations
\begin{equation}
\dunderline{\mathfrak{D}}(\omega_1,k_1)\mathbf{x}^{(1)}_1=0\quad;\quad\dunderline{\mathfrak{D}}(\omega_2,k_2)\mathbf{x}^{(1)}_2=0.
\end{equation}
The vector $\mathbf{x}^{(1)}_j\equiv [A_j^{(1)}, B_j^{(1)}, C_j^{(1)}, D_j^{(1)}, a_j^{(1)},b_j^{(1)}]^\dagger$ and the matrix $\dunderline{\mathfrak{D}}(\omega,k)$ are the same as that in \S \ref{sec:Triad_Analytical}. We further proceed to substitute the equations \eqref{eq:Eigen1}--\eqref{eq:Eigen2} in the equations \eqref{eq:Order2_1}--\eqref{eq:Order2_2} at $\mathcal{O}(\epsilon^2)$. Assuming $k_1+k_2=k_b$ and $\omega_1+\omega_2=0$, we  obtain 
\begin{align}
\dunderline{\mathfrak{D}}(\omega_1,k_1)\mathbf{x}^{(2)}_1=\mathbf{v}_1a_b\bar{a}^{(1)}_2+\mathbf{r}_1a^{(1)}_{1,\tau},\\
\dunderline{\mathfrak{D}}(\omega_2,k_2)\mathbf{x}^{(2)}_2=\mathbf{v}_2a_b\bar{a}^{(1)}_1+\mathbf{r}_2a^{(1)}_{2,\tau}.
\end{align}
Denoting the null space of transpose of $\dunderline{\mathfrak{D}}(k_j,\omega_j)$ by $\mathbf{n}_j$ and using the Fredholm alternative, we get the following set of equations:
\begin{align}
a_{1,\tau}^{(1)}= \beta_1a_b\bar{a}_2^{(1)}\qquad;\qquad a_{2,\tau}^{(1)}=\beta_2a_b\bar{a}^{(1)}_1,
\end{align}
where
\begin{equation}
\beta_j=-\frac{\mathbf{n}_j\cdot \mathbf{v}_j}{\mathbf{n}_j\cdot\mathbf{r}_j}.
\end{equation}
Furthermore, when $k_1-k_2=k_b$ and $\omega_1-\omega_2=0$, we get
\begin{align}
a_{1,\tau}^{(1)}= \beta_1a_ba_2^{(1)}\qquad;\qquad a_{2,\tau}^{(1)}=\beta_2\bar{a}_ba^{(1)}_1,
\end{align}
in which $\beta_j$ remains the same as before.

\section{Bragg resonance in the presence of a velocity field}
\label{sec:3}

In a single layered flow in the absence of a velocity field, there can be only one condition for  Bragg resonance -- when the wavenumber of the bottom is twice the wavenumber of the surface wave, i.e. $k_b=2k_s$. In such a case, an oppositely travelling surface mode having the same frequency as that of the incident wave is generated by the resonant forcing of the bottom. However, in a two-layered flow, several other resonant pairs are possible \citep{Alam1}. As mentioned previously, in the presence of a pycnocline, there exists four different modes of propagation -- two oppositely travelling surface (or external) modes and two oppositely travelling interfacial (or internal) modes. Any of these modes, depending on the wavenumber of the bottom ripples, may resonate with any other mode in the system, subject to the fulfilment of the resonance conditions.  In the absence of  a velocity field, there is an inherent symmetry in the weakly nonlinear wave interaction owing to the symmetric (or  biquadratic) nature of the dispersion relation. This means that if a rightward travelling surface mode of wavenumber $k_i$ interacts with the bottom of wavenumber $k_b$ to resonantly generate a leftward travelling interfacial mode of wavenumber $k_r$, then a leftward travelling surface mode of wavenumber $k_i$ will also interact with the same bottom of wavenumber $k_b$ to resonantly generate  a rightward travelling interfacial mode of wavenumber $k_r$. In the presence of a velocity field, however, this `right-left symmetry' of the interaction is destroyed. 

\begin{figure}
\centering\includegraphics[width=120mm]{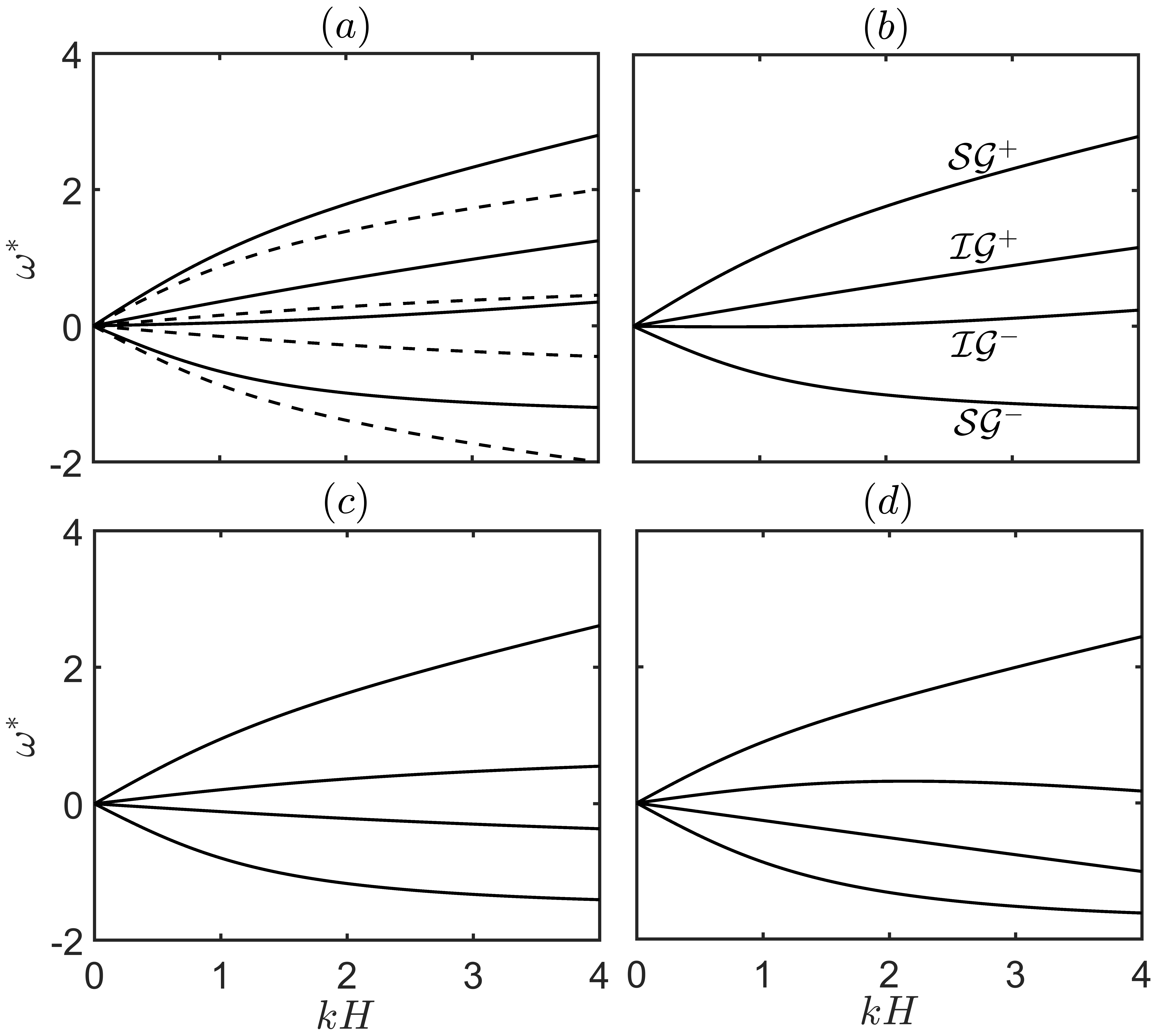}
\caption{Dispersion relation for various velocity profiles with $h_u/h_l=1$ and $R=0.90$. (a) $U_u^*=U_l^*=U_b^*=0.2$, (b) $U_u^*=U_l^*=0.2,U_b^*=0$, (c) $U_u^*=0.2,U_l^*=U_b^*=0$, and  (d) $U_u^*=0.2,U_l^*=-0.2,U_b^*=0.2$. $\mathcal{SG^{\pm}}$ denotes surface or external mode, $\mathcal{IG^{\pm}}$ denotes interfacial or internal mode, and the $+$ and $-$ signs respectively imply the direction of wave propagation. }
\label{fig:Dispersion11}
\end{figure}

The presence of velocity field may also change the intrinsic frequency of the waves. It may also cause a relative Doppler shift between the interfaces. When there is a uniform flow (case 1 of figure \ref{fig:1}), there is neither a change in the intrinsic frequency of the waves nor is there a relative Doppler shift between the surface and the interface. However, the bottom ripples are Doppler shifted with respect to the surface and the interface. The dispersion curves for this case has been plotted in figure \ref{fig:Dispersion11}(a)  in solid lines. In the dotted lines, we have plotted the dispersion curves without any velocity field. On the vertical axis is the non-dimensionalised frequency ($\omega^*\equiv\omega/\sqrt{g/H} $)  and on the horizontal axis is the non-dimensional wavenumber $kH$. The non-dimensionalised velocity is $U^*\equiv U/\sqrt{gH}$, where $H=h_u+h_l$. All the branches of Doppler shifted dispersion curves are simply $U^*kH$ away from the respective branches without the velocity field. 

Figure \ref{fig:Dispersion11}(b) shows the dispersion curve for the case when the shear is in the lower layer only (case 2 of figure \ref{fig:1}). Thus the Doppler shift component is the same for both external and internal modes, but the only way this differs from case 1 is  the presence of shear in lower layer, which has changed the intrinsic frequencies of both external and internal modes. 

For the case of shear only in the upper layer (case 3 of figure \ref{fig:1}), instead of the pycnocline, the surface undergoes a Doppler shift.  Because of shear jump,  intrinsic frequencies of both the branches change. It can be seen from the dispersion curve (figure \ref{fig:Dispersion11}(c)) that the branches $\mathcal{SG^+}$ and $\mathcal{SG^-}$ are highly non-symmetrical due to presence of the velocity $U_u$ at the surface. There is a small change in the intrinsic frequency as well, however, it is not evident from the dispersion curves.

In figure \ref{fig:Dispersion11}(d) we have plotted the dispersion curve for the case when both the layers have  shear (case 4 of figure \ref{fig:1}). For this case, we have assumed the shear to be positive in the upper layer and negative in the lower layer. Thus, the external mode is Doppler shifted positively whereas the internal mode is negatively Doppler shifted.

\subsection{Shear in the lower layer}
\label{sec:shear_lower}
\begin{figure}
        \centering
        \subfloat[]{\includegraphics[width=6cm]{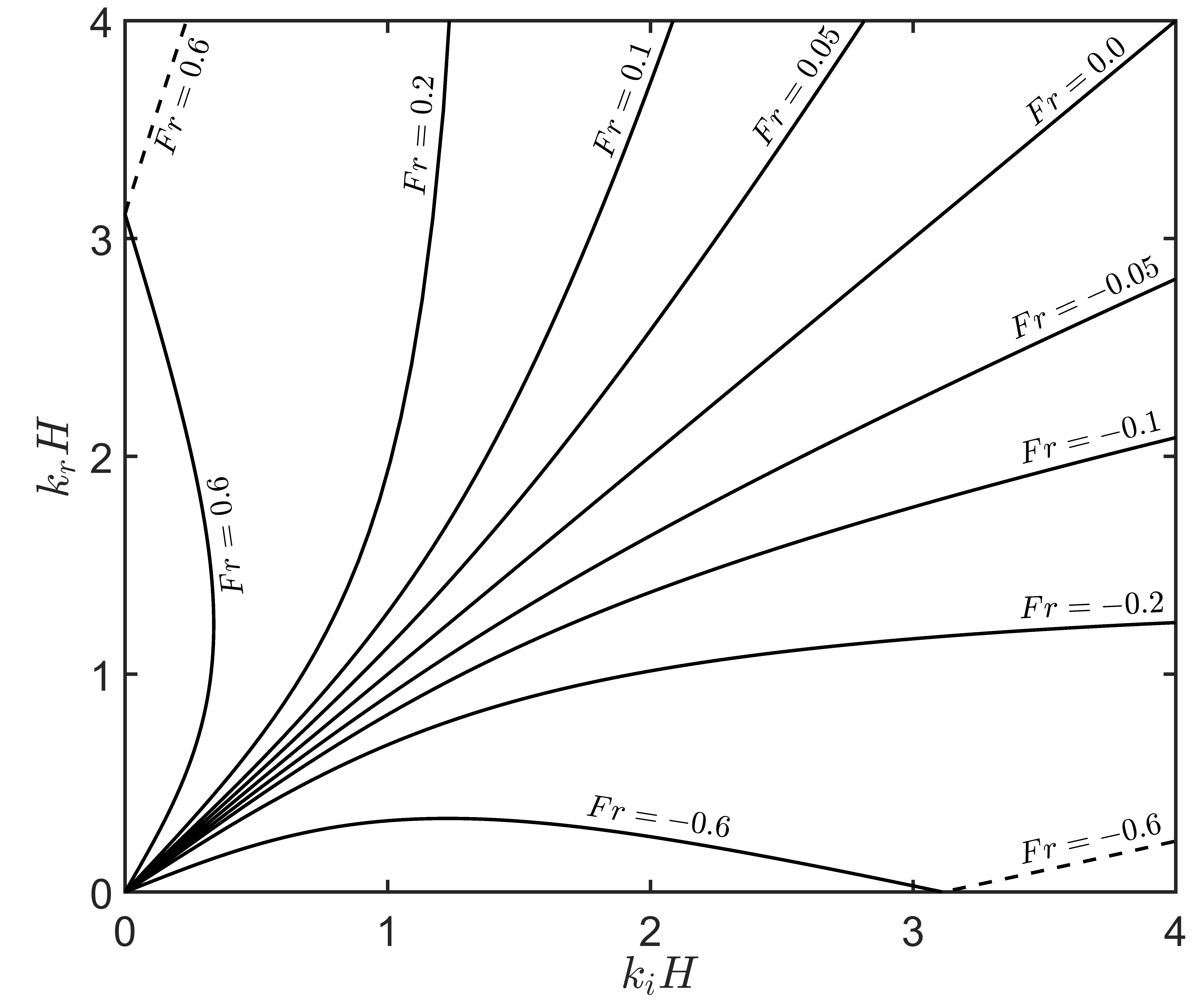}
}
\subfloat[]{\includegraphics[width=6cm]{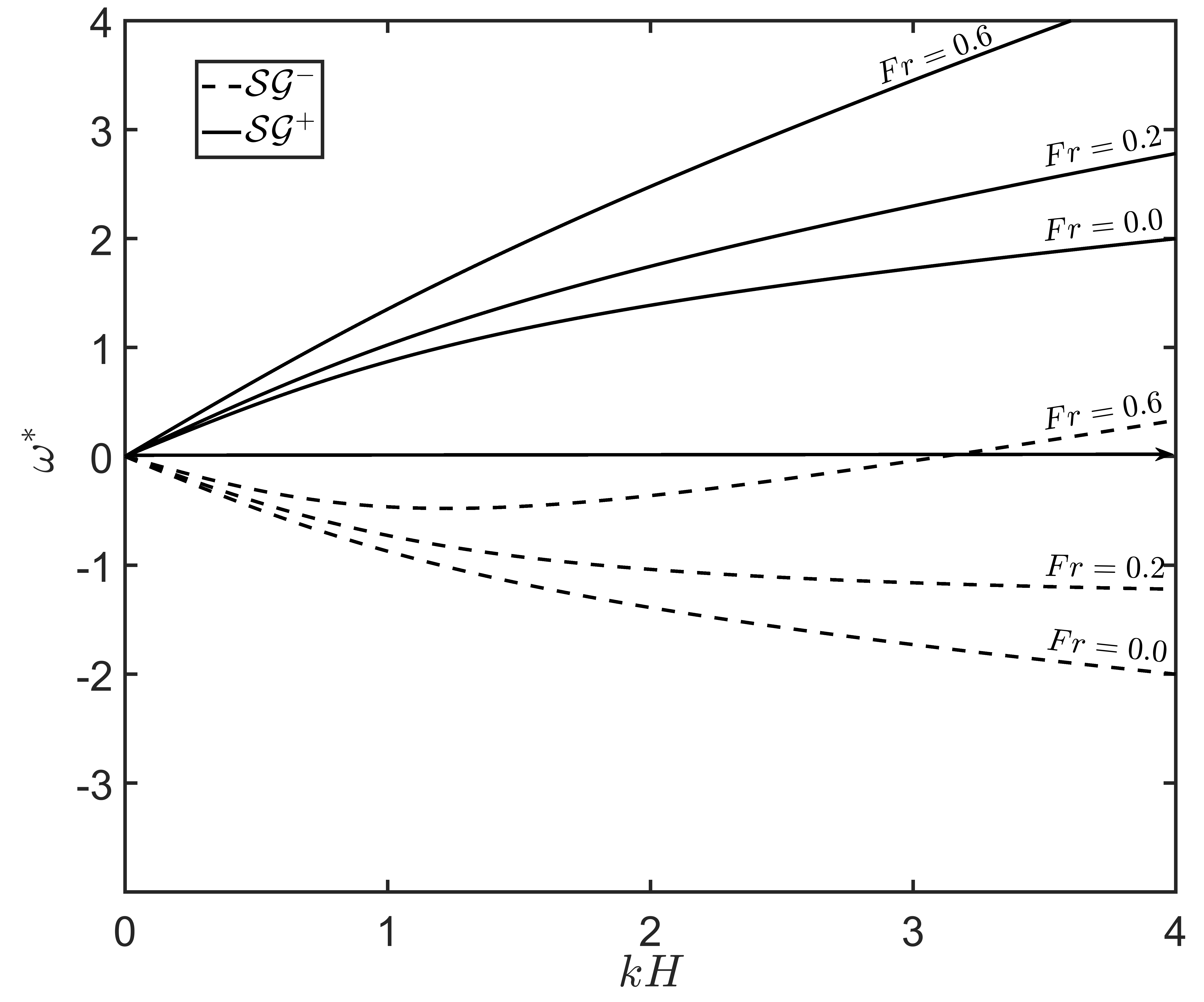}
}
        \caption{(a) Different combinations of $k_r$ on $\mathcal{SG^-}$ such that $k_i$ is on $\mathcal{SG^+}$, performed for various values of $Fr\equiv U_u/\sqrt{gH}$ for the case of shear in the lower layer. Here $R=0.95$ and $h_u/h_l=1/3$. For solid lines, $k_b=k_i+k_r$ but for dashed lines, $k_b=|k_i-k_r|$. (b) Dispersion relations for the same case for three values of $Fr$. Here solid lines represent $\mathcal{SG^+}$ modes and dashed lines represent $\mathcal{SG^-}$.  }
        \label{fig:Bragg1}
\end{figure}

Here, we analyse the case when shear is present only in the lower layer and the local velocity at the bottom is zero (case 2 of figure \ref{fig:1}). Therefore, the surface modes and the interfacial modes are Doppler shifted by an equal amount with respect to the bottom ripple. Presence of shear will also result in a change in the intrinsic frequencies of the waves. For the case of shear in the lower layer, firstly we investigate the triads formed by  two surface modes, i.e. $\mathcal{SG^+}$ and $\mathcal{SG^-}$. We have taken the incident wave $k_i$ on $\mathcal{SG^+}$ and the resonant wave $k_r$ on $\mathcal{SG^-}$. Changing the Froude number changes the resonance condition, as is evident from figure \ref{fig:Bragg1}(a). As   mentioned earlier, in the absence of shear, all the Bragg resonance triads having $k_i$ on the $\mathcal{SG^+}$ branch will resonate the waves on the $\mathcal{SG^-}$ branch having $k_r=k_i$. This corresponds to the straight line labelled $Fr=0.0$ in figure \ref{fig:Bragg1}(a). Increasing the $Fr\;(\equiv U_u/\sqrt{gH})$ will mean that the surface  will be positively Doppler shifted with respect to the bottom ripples. For any given positive velocity, at some value of $k$, the dispersion curve $\mathcal{SG^-}$ is bound to cross the $k$-axis; see figure \ref{fig:Bragg1}(b). However, while plotting, we have kept the values of $k$ restricted because for higher values of $k$, even though the resonance condition is satisfied,  the rate of energy exchange falls off because the waves are unable to `feel' the bottom. We see that for $Fr=0.2$, the $\mathcal{SG^-}$ branch shifts upwards. This is naturally reflected in the change in the resonance condition in  figure \ref{fig:Bragg1}(b), in which we have plotted the two branches of the dispersion relation\footnote{The dispersion relation is a fourth order polynomial in $\omega$ but we have plotted only two branches on which the resonance is being studied, i.e. $\mathcal{SG^+}$ and $\mathcal{SG^-}$ in this case.}. If $Fr$ is further increased, then for a given $k_i$ on $\mathcal{SG^+}$, there can be up to 3 values of $k_r$ on $\mathcal{SG^+}$ which would form the triad. This is the reason that for $Fr=0.6$ curve in figure \ref{fig:Bragg1}, for a single $k_iH$, there exists 3 values of $k_rH$ for which resonance condition is met. Two of these triads will be formed if the bottom's wavenumber is $k_b=k_i+k_r$ (shown by the solid line). However, the third $k_r$ would lie on the part of $\mathcal{SG^-}$ for which $\omega>0$ and for such a triad (shown in broken lines in figure \ref{fig:Bragg1}(a) for $Fr=0.6$), the bottom's wavenumber would be $k_r-k_i$. We note here in passing that these triads represented by the broken lines (in figure \ref{fig:Bragg1}(a), not in \ref{fig:Bragg1}(b)) are not `usual' triads but are `explosive' triads. In such triads, both the incident wave and the resonant wave grow simultaneously, while the total energy of the system still remains conserved. This is due to the existence of negative energy waves \citep{Cairns}. These `explosive' triads have been explored for capillary--gravity waves by \citet{mchugh1992stability} as well as the authors of this paper \citep{raj2018explosive}.

Further, in  figure \ref{fig:Bragg1}(b) we have also plotted the change in the dispersion curves of $\mathcal{SG^-}$ and $\mathcal{SG}^+$ for $Fr=(0,0.2,0.6)$ for $kH<4$. It can be seen that within this window of $kH$, for a given $\omega_i$ on $\mathcal{SG^+}$, there can be only one $k_i$ (lines parallel to k-axis i.e. $\omega=\pm \omega_0$ intersects any given $\mathcal{SG^+}$ at exactly one point). But for a given $|\omega_r|$ on $\mathcal{SG^-}$, for $Fr=0.6$, there can be three values of $k_r$ satisfying the dispersion relation, two values are negative and one positive (lines parallel to k-axis i.e. $\omega=\pm \omega_0$ may intersect any given $\mathcal{SG^-}$ at either one point or at three points).


Although we have discussed the modification in the resonance condition for a positive $Fr$, a very similar thing happens for a negative $Fr$. In figure \ref{fig:Bragg1}, whereas for a positive $Fr$, there may exist up to three $k_r$ on $\mathcal{SG^-}$ for a given $k_i$ on $\mathcal{SG^+}$, for a negative $Fr$ (see $Fr=0.6$), three different $k_i$ on $\mathcal{SG^+}$ may resonate the same wavenumber $k_r$ on $\mathcal{SG^-}$ (see, $Fr=-0.6$). Because the dispersion curves in question, i.e. $\mathcal{SG^+}$ and $\mathcal{SG^-}$ are symmetric for $Fr=0$, the symmetry is also maintained for a positive and a negative $Fr$.

It might be noticed that the value of $Fr$ needed for any appreciable change in the resonance condition varies from moderate to large. The reason for this is that for surface gravity waves, the intrinsic frequency is quite large and to Doppler shift the intrinsic frequency, a local velocity of similar magnitude is needed. For  example, to get three possible resonant waves having $k_rH<4$ for a given $k_iH$, a Froude number of approximately 0.5 is needed. However, to Doppler shift the interfacial gravity waves on the pycnocline, a significantly smaller Froude number is sufficient because the intrinsic phase speeds of the interfacial waves are significantly low.



\begin{figure}
        \centering
        \subfloat[]{\includegraphics[width=6cm]{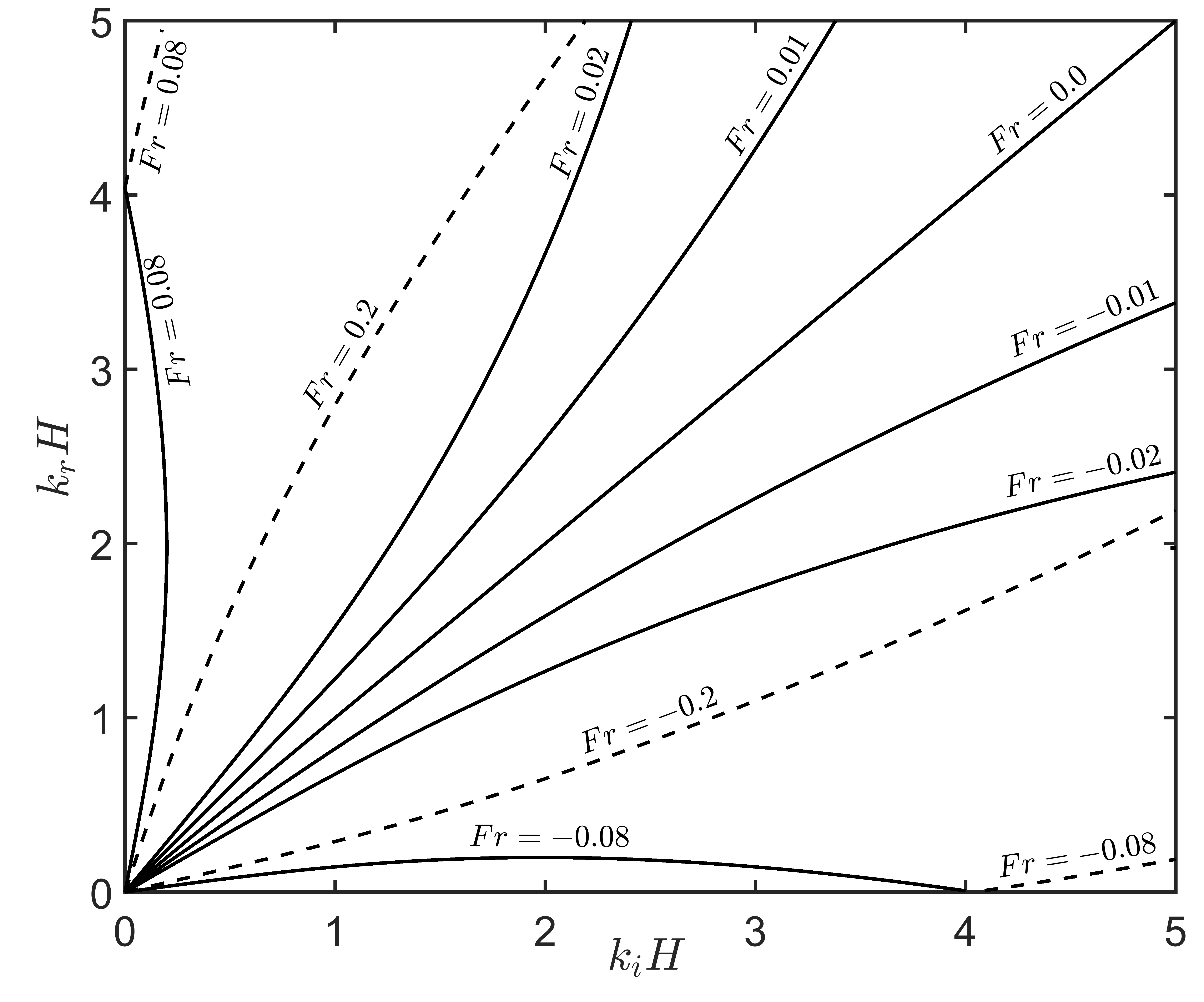}
}
\subfloat[]{\includegraphics[width=6cm]{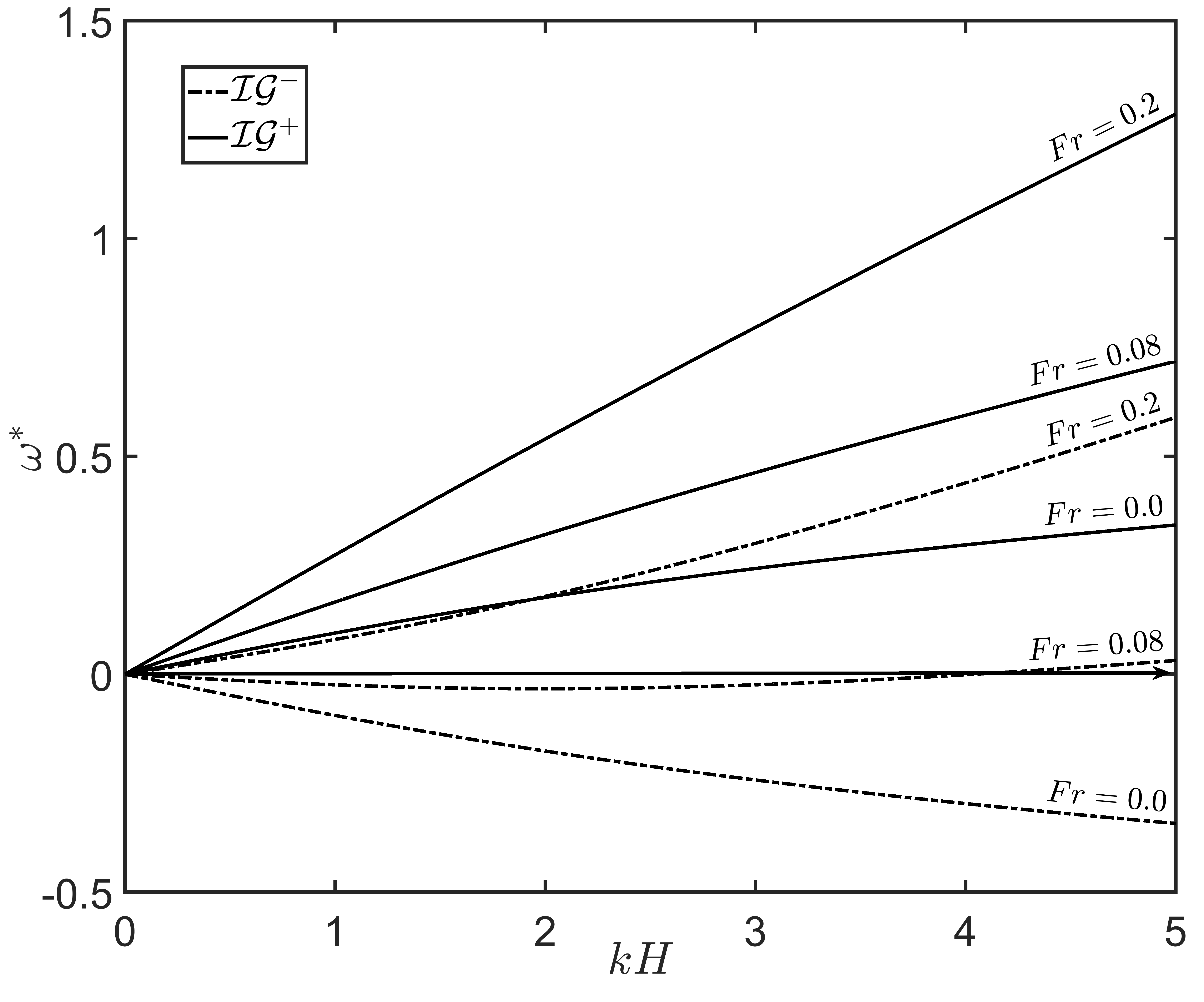}
}
        \caption{(a)  Different combinations of $k_r$ on $\mathcal{IG^-}$ such that $k_i$ is on $\mathcal{IG^+}$, for various values of $Fr$ when shear is in the lower layer. Here $R=0.95$ and $h_u/h_l=1/3$. For solid lines, $k_b=k_i+k_r$ but for dashed lines, $k_b=|k_i-k_r|$. (b) Dispersion relations for the same case for three values of $Fr$. Here solid lines represent $\mathcal{IG^+}$ modes and dashed lines represent $\mathcal{IG^-}$.   }
        \label{fig:Bragg2}
\end{figure}


We move on to the incident/resonant wave pairs formed by two interfacial modes, i.e. by the waves on $\mathcal{IG^+}$ and $\mathcal{IG^-}$ for the case of shear in the upper layer (case 3 of figure \ref{fig:1}). The pycnocline is not only Doppler shifted with respect to the bottom, but it also has a discontinuity in shear across it. This signifies the presence of vorticity-gravity waves at the pycnocline and a significant change in the intrinsic frequency as well. A figure similar to the previous case showing combinations of $k_i$ (on $\mathcal{IG^+}$) and $k_r$ (on $\mathcal{IG^-}$) has been plotted in the figure \ref{fig:Bragg2}(a) restricting the non-dimensionalised wavenumber to 5. Naturally, at $Fr=0$, the resonance condition is symmetric but the resonance condition changes greatly even for a small amount of mean flow. As we increase the $Fr$, for a small $k_i$ on $\mathcal{IG^+}$, the resonance condition is met by a larger $k_r$ on $\mathcal{IG^-}$ (see curves labelled $Fr=0.01,0.02$ of the figure \ref{fig:Bragg2}(a)). On increasing $Fr$ further, again we see the existence of three $k_r$ values for a given $k_i$, similar to the resonance between $\mathcal{SG^+}$ and $\mathcal{SG^-}$ ($Fr=0.08$, figure \ref{fig:Bragg2}(a,b)). However, if we further keep on increasing the $Fr$, then the complete $\mathcal{IG^-}$ curve will become positive (shown in the figure \ref{fig:Bragg2}(b), $Fr=0.2$) and in such a case, only one resonant wave for any given $k_i$ on $\mathcal{SG^-}$ will exist (dashed line labelled $Fr=0.2$ in  figure \ref{fig:Bragg2}(a)). The dashed line implies that the wavenumber of the bottom ripple for such a triad is $k_b=k_i-k_r$ unlike the usual case $k_b=k_i+k_r$ for the solid lines in figure \ref{fig:Bragg2}. Again, similar to the previous case, the triads marked by the dashed lines are the explosive triads. The positive and a negative $Fr$ result in symmetric cases as shown in the figure \ref{fig:Bragg2} 

{We put this in the context of a real ocean of depth $H=100$ m having pycnocline at $h_u=25$ m from surface. These data are similar to those used by \citet{Alam2}. In the `no-flow' situation, an interfacial wave having a wavelength $\lambda_i\sim200$ m will resonate an oppositely travelling wave of wavelength $\lambda_r\sim200$ m. However, in the presence of a small velocity of $U_u=U_l=0.31$ m/s opposite to the direction of the incident wave, the resonant wave would have a wavelength $\lambda_r\sim140$ m.}  

\begin{figure}
        \centering
        \subfloat[]{\includegraphics[width=6cm]{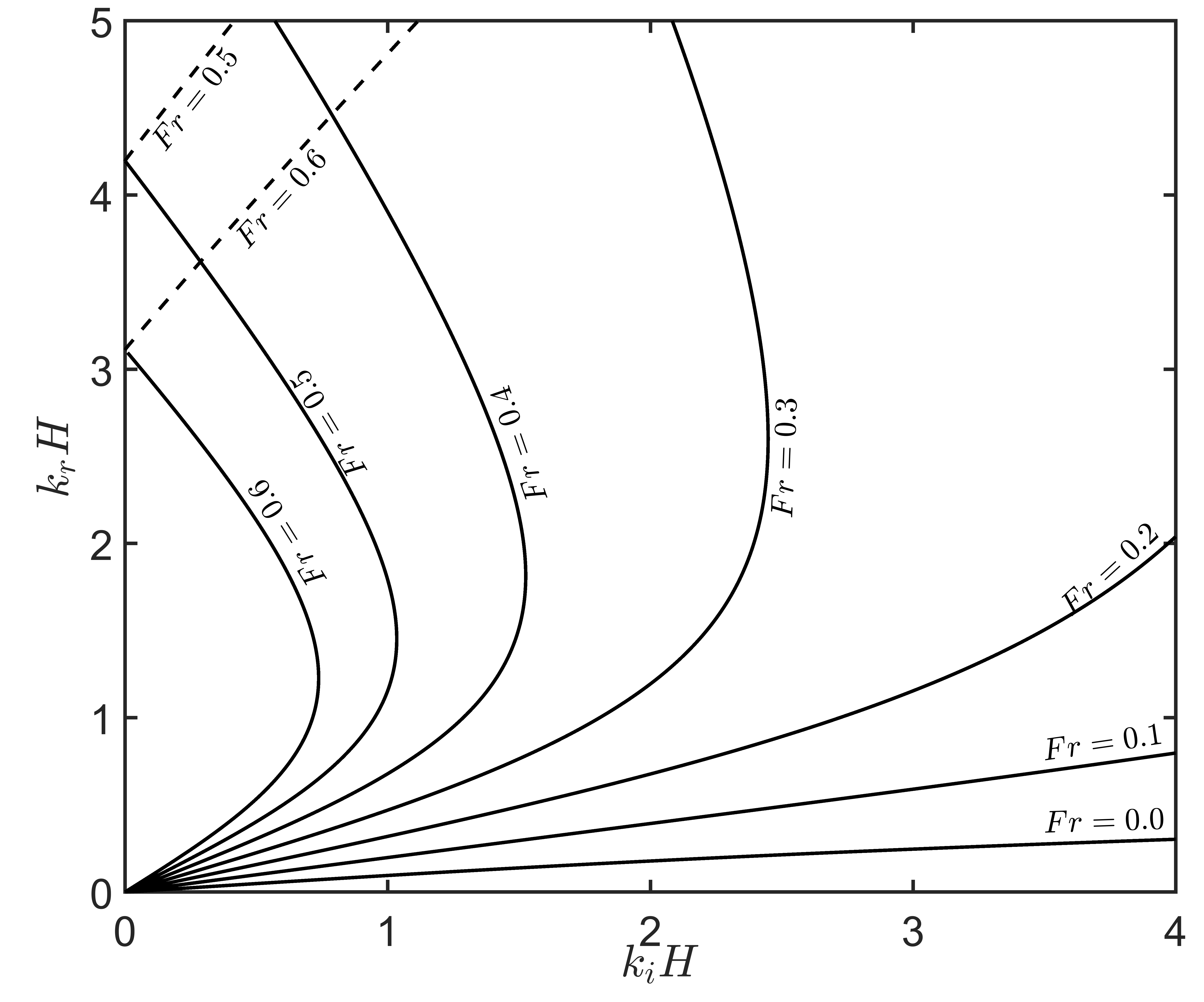}
}
\subfloat[]{\includegraphics[width=6cm]{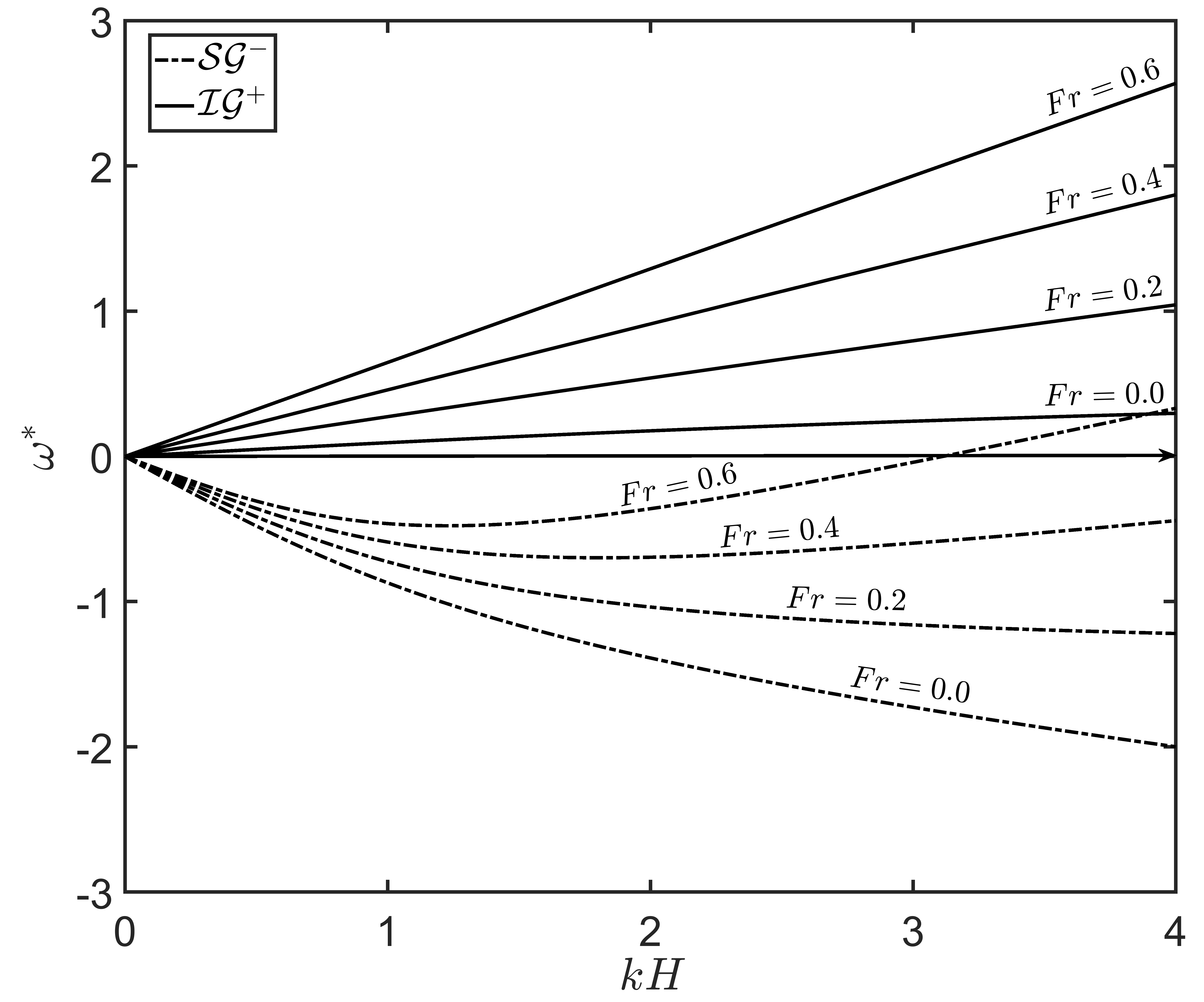}
}
        \caption{(a) Different combinations of $k_r$ on $\mathcal{SG^-}$ such that $k_i$ is on $\mathcal{IG^+}$, for various positive $Fr$ values when shear is in the lower layer. Here $R=0.95$ and $h_u/h_l=1/3$. For solid lines, $k_b=k_i+k_r$ but for dashed lines, $k_b=|k_i-k_r|$. (b) Dispersion relations for the same case for different increasingly positive $Fr$. Here solid lines represent $\mathcal{IG^+}$ modes and dashed lines represent $\mathcal{SG^-}$.  }
        \label{fig:Bragg3}
\end{figure}

The third sub-case for the case of shear in the upper layer is the resonant interaction between surface and interfacial mode having opposite intrinsic frequency. This means that the incident/resonant pair is either $\mathcal{IG^+}/\mathcal{SG^-}$ or $\mathcal{IG^-}/\mathcal{SG^+}$. Without a loss of generality, we will discuss only the $\mathcal{IG^+}/\mathcal{SG^-}$ pair; see figure \ref{fig:Bragg3}(a)-\ref{fig:Bragg3}(b). The results about the other pair can be obtained in a straightforward manner, simply by changing the sign of $Fr$ from positive to negative and vice-versa. What matters is that whether the sign of mean flow and that of surface/interfacial waves are in the same direction or the opposite. The positive shear in this particular case will imply that the velocity at the surface/pycnocline is in the direction of the propagation of the interfacial wave $\mathcal{IG^+}$. Therefore, increasing  $Fr$ will lead to an increase in the frequency of $\mathcal{IG^+}$ but a non-monotonic change in the frequency of the $\mathcal{SG^-}$ mode, as shown in the figure \ref{fig:Bragg3}(b). Even a small value of shear, the effect on the speed of $\mathcal{IG^+}$ is significant but the $\mathcal{SG^-}$ is relatively less affected. However, for a large value of $Fr$, there may exist multiple values of $k_r$ for a given $k_i$ as can be seen from the figure \ref{fig:Bragg3}(a), $Fr=0.3,0.4,0.5,0.6$. The reason is simply a non-monotonic behaviour of frequency of $\mathcal{SG^-}$ with respect to the wavenumber as can be seen from figure \ref{fig:Bragg3}(b). For a higher value of $Fr$, the frequency of $\mathcal{SG^-}$ becomes positive and the triads formed by the positive part of $\mathcal{SG^-}$ are shown in dashed lines in figure \ref{fig:Bragg3}(a). For these triads, the bottom's wavenumber is $k_r-k_i$ whereas for triads marked by solid line, the bottom's wavenumber is $k_r+k_i$.

\begin{figure}
        \centering
        \subfloat[]{\includegraphics[width=6cm]{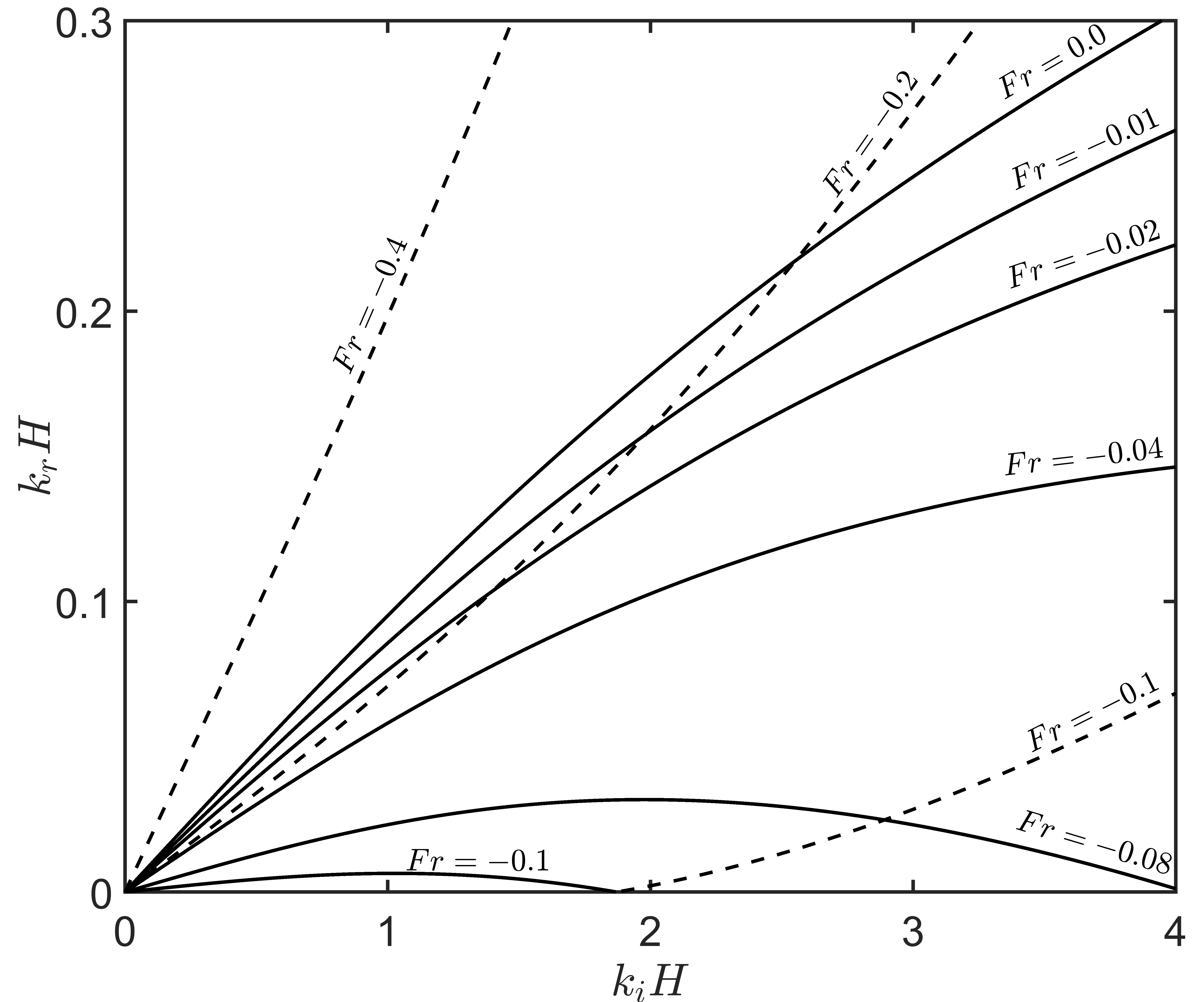}
}
\subfloat[]{\includegraphics[width=6cm]{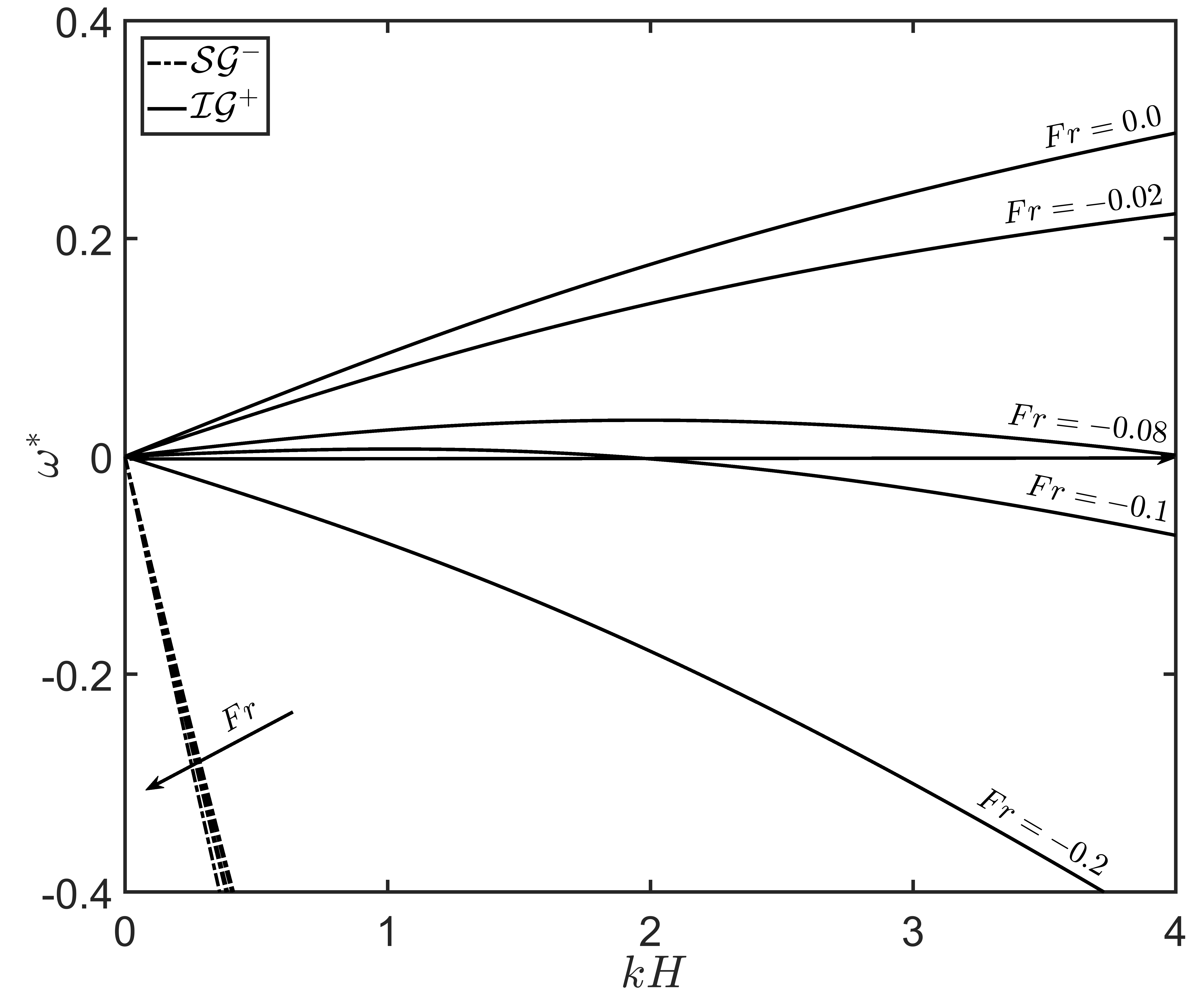}
}
        \caption{(a) Different combinations of $k_r$ on $\mathcal{SG^-}$ such that $k_i$ is on $\mathcal{IG^+}$, for various negative $Fr$ values when  shear is in the lower layer. Here $R=0.95$ and $h_u/h_l=1/3$. For solid lines, $k_b=k_i+k_r$ but for dashed lines, $k_b=|k_i-k_r|$. (b) Dispersion relations for the same case for different negative $Fr$. Direction of arrows imply increasingly negative $Fr$. Here solid lines represent $\mathcal{IG^+}$ modes and dashed lines represent $\mathcal{SG^-}$. }
        \label{fig:Bragg4}
\end{figure}

If the Froude number is negative (see figure \ref{fig:Bragg4}(a)-\ref{fig:Bragg4}(b)),  the frequency of $\mathcal{SG^-}$ increases monotonically but that of $\mathcal{IG^+}$ may become non-monotonic; shown in figure \ref{fig:Bragg4}(b) for the case $Fr=-0.08$. Because the frequency of $\mathcal{SG^-}$ plotted in \ref{fig:Bragg4}(b) is restricted, not much difference in the dispersion curves is obtained. For a higher $Fr$, the frequency changes sign within the chosen limit of $kH=4$ and becomes negative ($Fr=-0.1$). For a further increase in the velocity, the frequency becomes completely negative for the $\mathcal{IG^-}$ ($Fr=-0.2$). This change in the frequency is reflected in the change in the resonance condition and the change can be visualised in the figure \ref{fig:Bragg4}(a). For $Fr=-0.01,-0.02,-0.04$, for a single $k_r$, only one $k_i<4$ exists but for higher $Fr$, a single $k_r$ maybe resonant by multiple $k_i$ ($Fr=-0.08,-0.1$). For $Fr=-0.1$, the bottom's wavenumber is $k_i+k_r$ for the solid line part in figure \ref{fig:Bragg4}(a) and $k_i-k_r$ for the dashed line part i.e. the explosive triads.

\begin{figure}
        \centering
        \subfloat[]{\includegraphics[width=6cm]{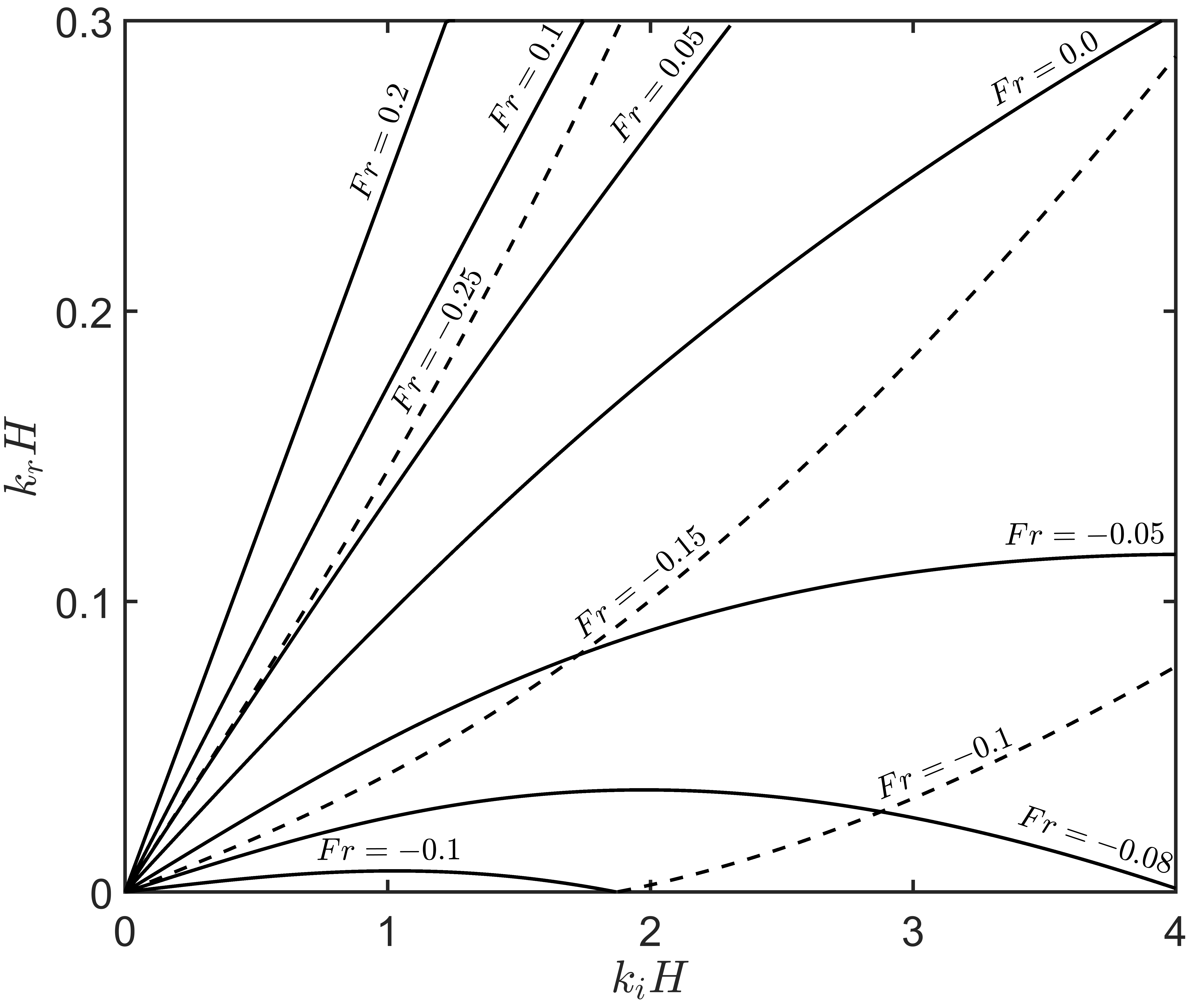}
}
\subfloat[]{\includegraphics[width=6cm]{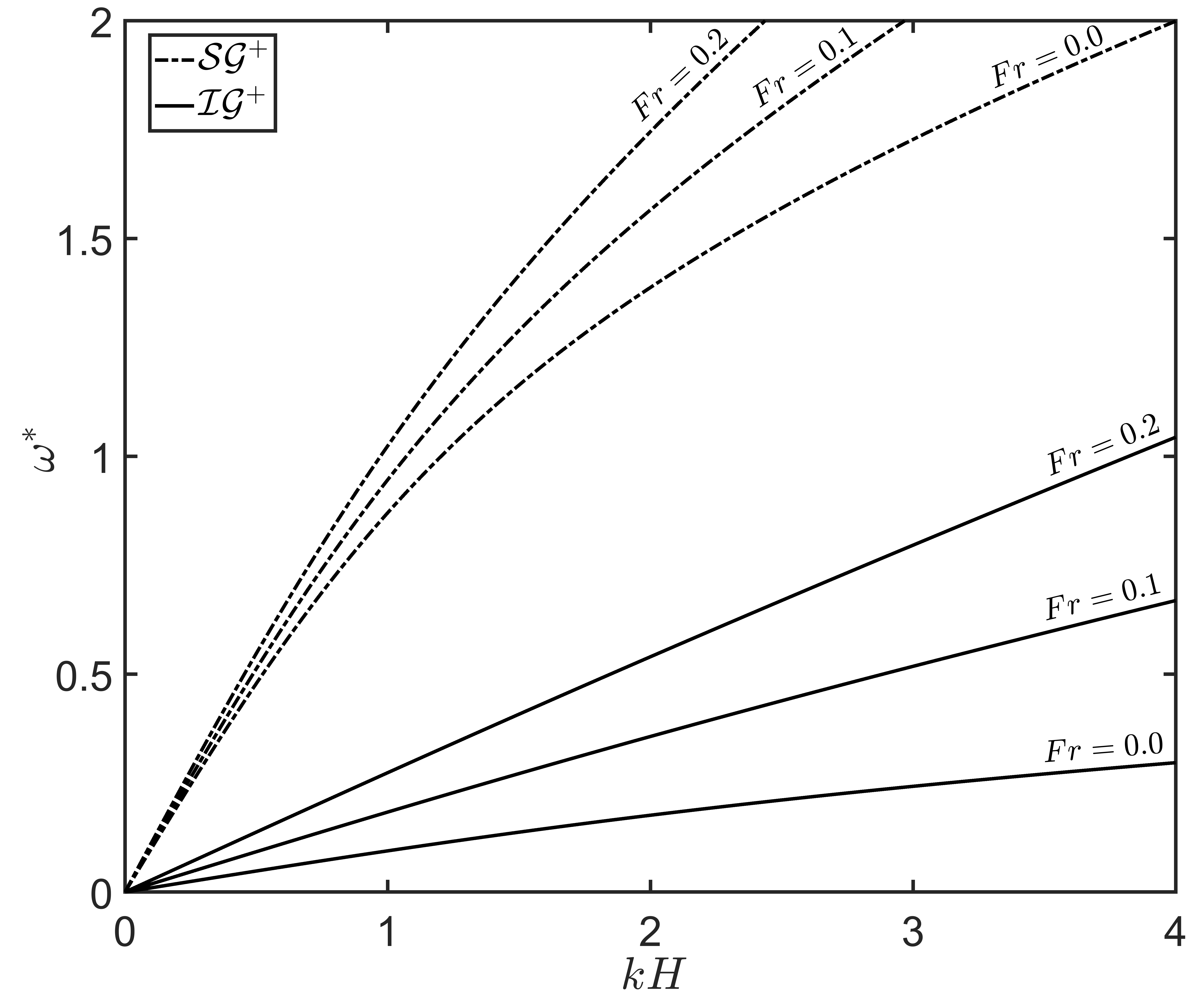}
}
        \caption{(a) Different combinations of $k_r$ on $\mathcal{SG^+}$ such that $k_i$ is on $\mathcal{IG^+}$ for various values of positive $Fr$ and for low values of negative $Fr$  for the case of shear in lower layer. $R=0.95$, $h_u/h_l=1/3$. For solid lines, $k_b=|k_i-k_r|$ but for dashed lines, $k_b=k_i+k_r$. (b) Dispersion relation for positive $Fr$. Here solid lines represent $\mathcal{IG^+}$ modes and dashed lines represent $\mathcal{SG^+}$. }
        \label{fig:Bragg5}
\end{figure}

Finally, we deal with the case when the incident/resonant modes are in the same direction i.e. $\mathcal{IG^+}/\mathcal{SG^+}$ or $\mathcal{IG^-}/\mathcal{SG^+}$. Again, without a loss of generality, we study only the resonance between $\mathcal{IG^+}/\mathcal{SG^+}$ modes and in this case, positive $Fr$ will imply a flow in the same direction of the waves; see figure \ref{fig:Bragg5}(a)-\ref{fig:Bragg5}(b). Because the all the involved waves and the flow are in the same direction, there is no question of sign changing of the frequency of any wave. The frequencies of all the waves increases progressively with increasing $Fr$ (figure \ref{fig:Bragg5}(b)). In figure \ref{fig:Bragg5}(a)-\ref{fig:Bragg5}(b), however, we have plotted both the positive $Fr$ and the negative $Fr$ having low magnitudes. Increasing the $Fr$ will mean that for a given $k_i$, a higher $k_r$ will be needed for resonance which can be seen from the figure \ref{fig:Bragg5}(a)-\ref{fig:Bragg5}(b) (positive $Fr$).

\begin{figure}
        \centering
        \subfloat[]{\includegraphics[width=6cm]{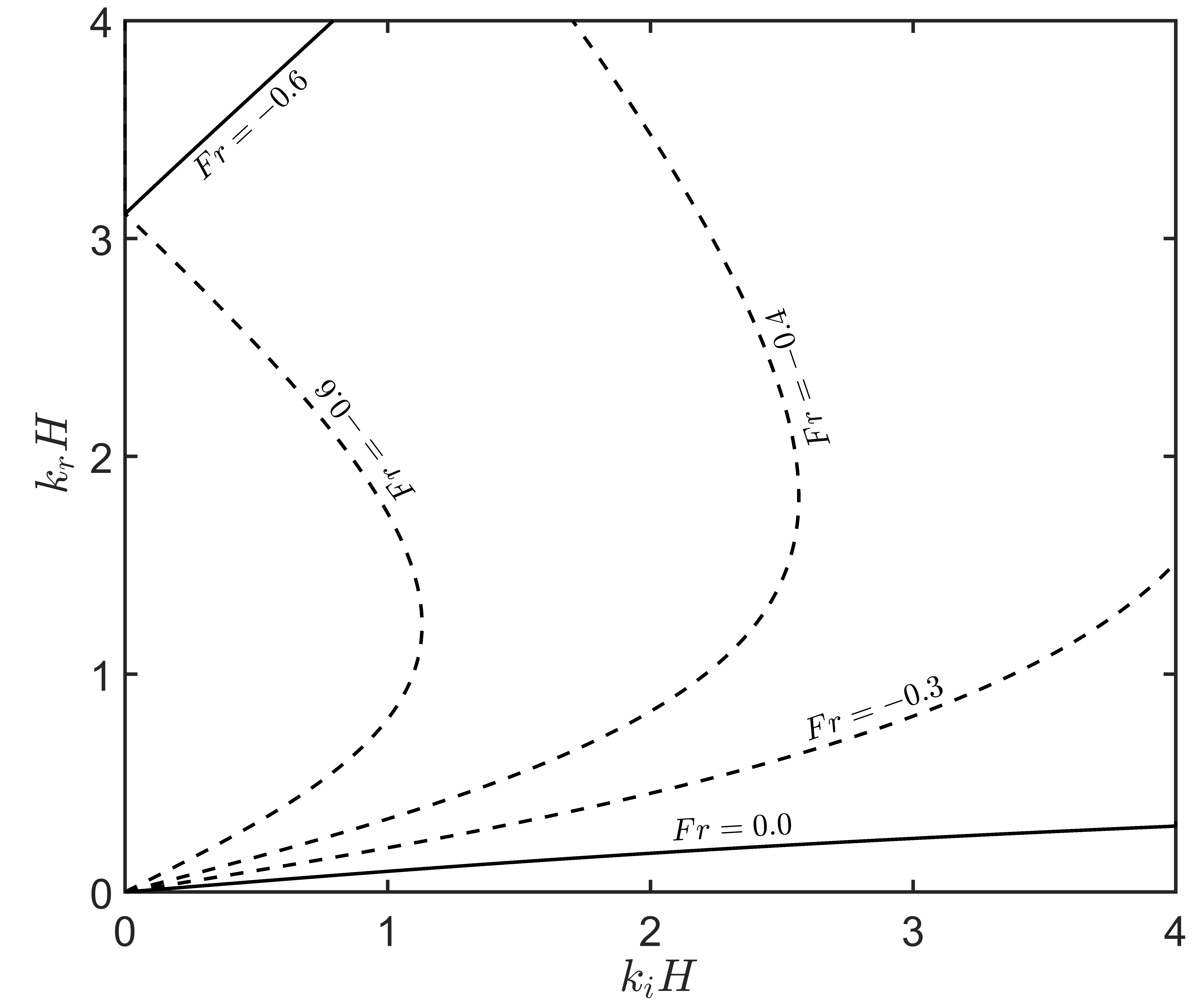}
}
\subfloat[]{\includegraphics[width=6cm]{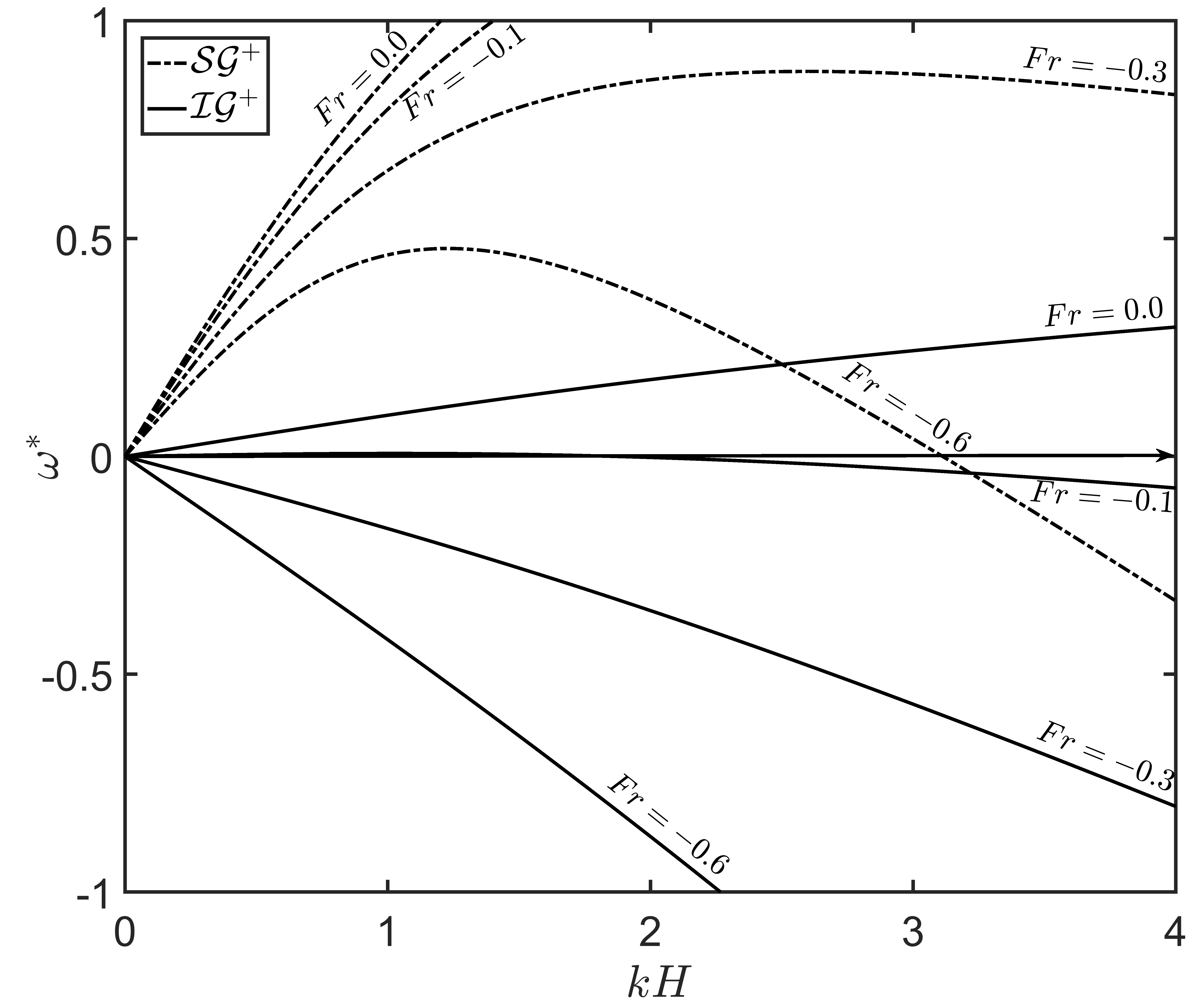}
}
        \caption{(a) Different combinations of $k_r$ on $\mathcal{SG^+}$ such that $k_i$ is on $\mathcal{IG^+}$ for high values of negative $Fr$  for the case of shear in the lower layer. Here $R=0.95$ and $h_u/h_l=1/3$. For solid lines, $k_b=|k_i-k_r|$ but for dashed lines, $k_b=k_i+k_r$. (b) Dispersion relation for negative $Fr$. Here solid lines represent $\mathcal{IG^+}$ modes and dashed lines represent $\mathcal{SG^+}$. }
        \label{fig:Bragg6}
\end{figure}

For a negative $Fr$ (see figure \ref{fig:Bragg6}(a)-(b)), the dispersion relation is plotted in the figure \ref{fig:Bragg6}(b). For $Fr=-0.1$, $\mathcal{SG^+}$ is positive throughout but a part of $\mathcal{IG^+}$ becomes negative. For a higher negative $Fr$ (say $-0.3$), the $\mathcal{SG^+}$ still remains positive, but $\mathcal{IG^+}$ becomes completely negative. For a further negative $Fr$, the $\mathcal{IG^+}$, remains negative and a part of $\mathcal{SG^+}$ also becomes negative. The effect on the resonance conditions for the case of small negative $Fr$ (upto $-0.25$) has been plotted in the figure \ref{fig:Bragg5}(a) and for higher negative $Fr$ have been plotted in the figure \ref{fig:Bragg6}(a).

\subsection{Shear in the upper layer}

\begin{figure}
        \centering
        \subfloat[]{\includegraphics[width=6cm]{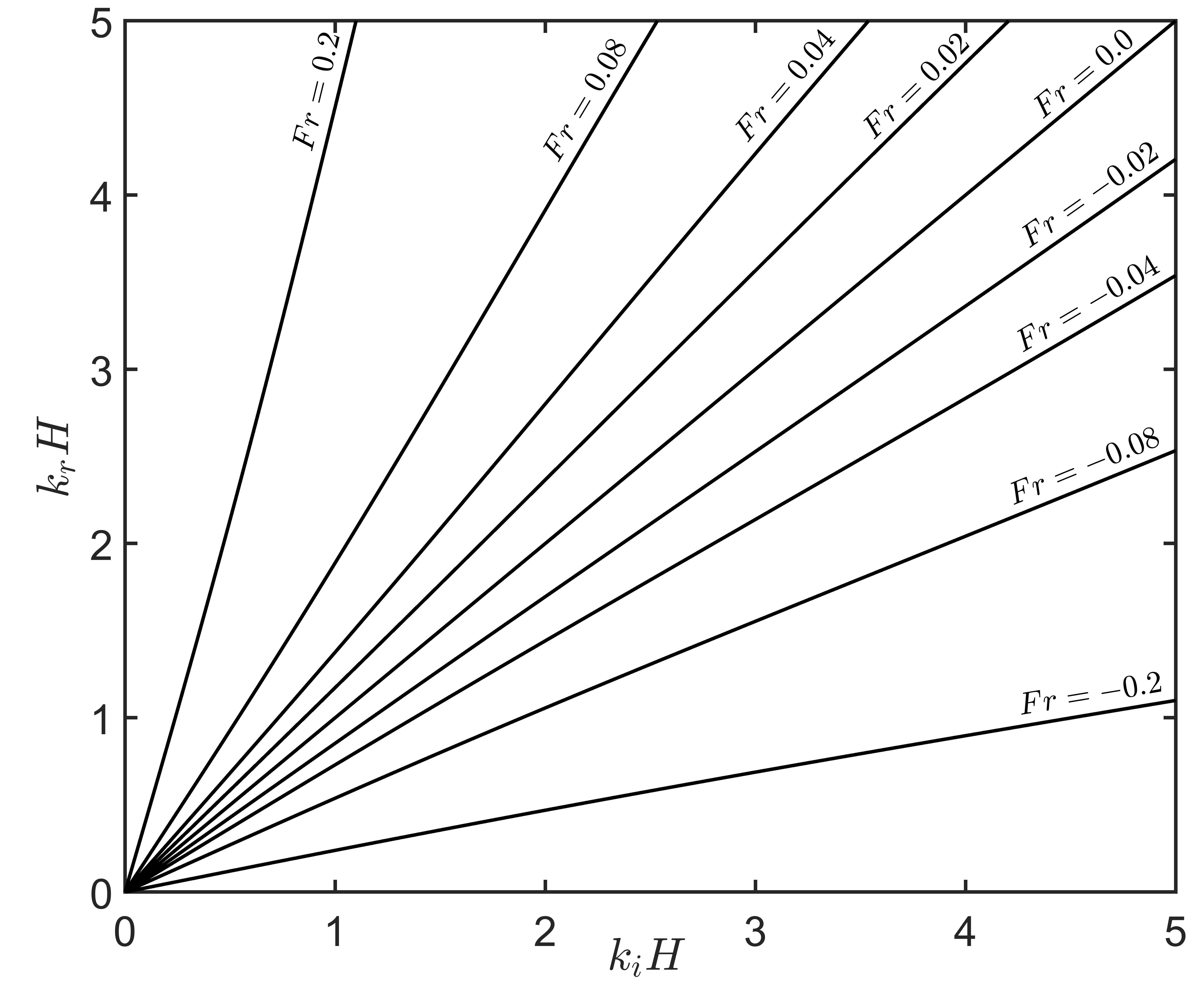}
}
\subfloat[]{\includegraphics[width=6cm]{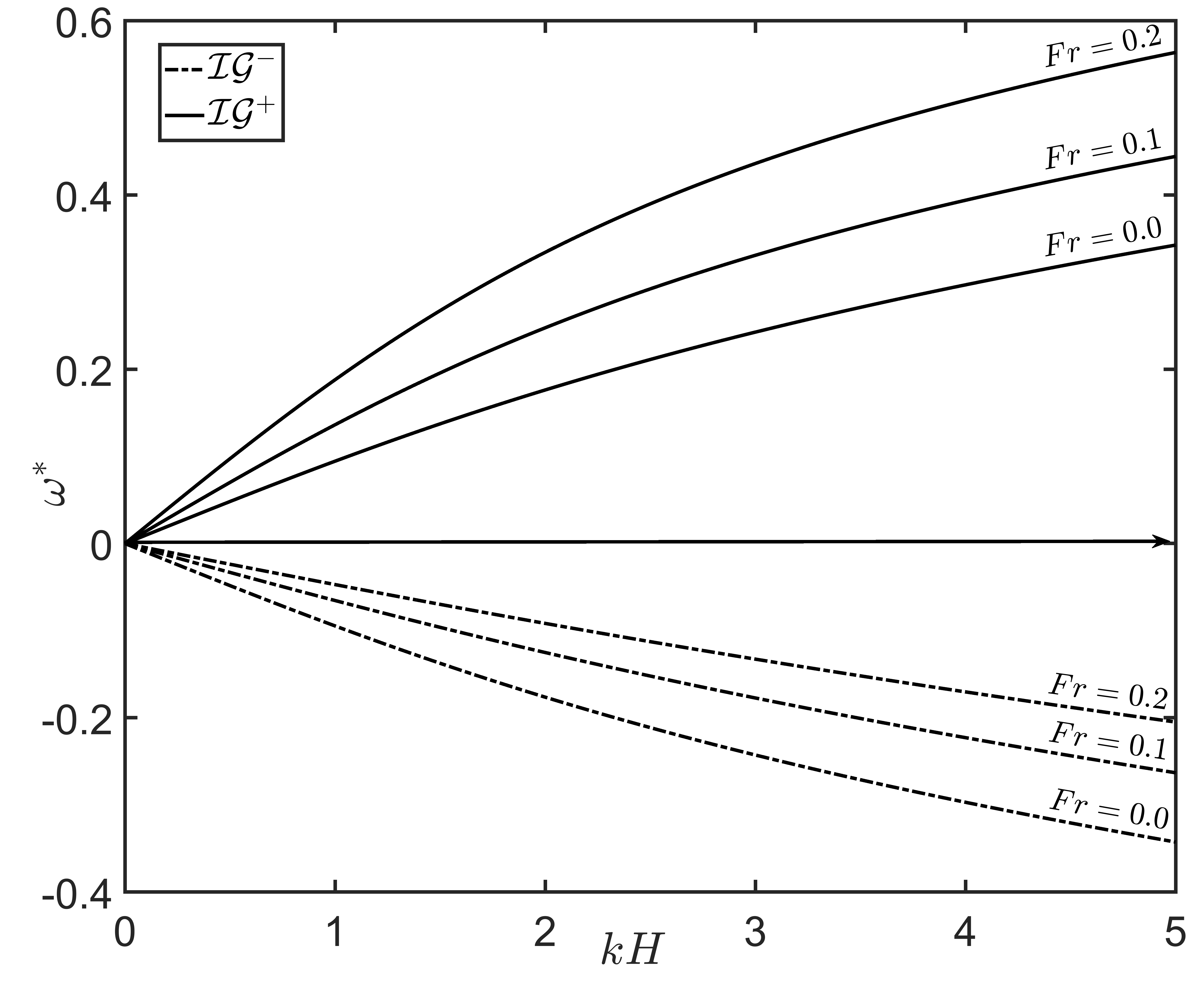}
}
        \caption{(a)  Different combinations of $k_r$ on $\mathcal{IG^-}$ such that $k_i$ is on $\mathcal{IG^+}$, for various values of $Fr\equiv U_u/\sqrt{gH}$ when shear is in the upper layer. Here $U_l=U_b=0$, $R=0.95$, and $h_u/h_l=1/3$. (b) Dispersion relations for the same case for three values of $Fr$.  }
        \label{fig:Case2_Internal}
\end{figure}

As we have mentioned earlier,  shear in the upper layer causes a relative Doppler shift between the surface and the pycnocline, as well as the bottom ripple. Further, the change in the intrinsic frequencies of the surface modes will be minimal compared to the change in the intrinsic frequencies of the interfacial modes. In \S \ref{sec:shear_lower}, we have  performed a detailed study on how the Doppler shift changes the resonance conditions. Here we focus on the case when the intrinsic frequency of the waves get changed, and because the intrinsic frequencies of the interfacial waves are more prone to change, the  case of substantial interest is the resonant interaction between $\mathcal{IG^+}$ and $\mathcal{IG^-}$ modes. Since  shear is only in the upper layer and the pycnocline has no local base velocity, there is no role of  Doppler shift. However, when  shear in the upper layer is positive ($Fr>0$), the $\mathcal{IG^+}$ is sped up but the $\mathcal{IG^-}$ mode is slowed down (we note that the wave at the interface is a vorticity-gravity wave). Although for the $Fr=0$ case, the resonant wave is $k_r=k_i$, the conditions change when $Fr\neq 0$. For $Fr>0$, we have $k_r<k_i$, and for $Fr<0$, we get $k_r>k_i$. The change in the resonance condition is shown in  figure \ref{fig:Case2_Internal}(a) and the dispersion relation for $Fr>0$ has been plotted in  figure \ref{fig:Case2_Internal}(b).



\section{Wave Triad in the presence of a base velocity field}
\label{sec:4}
\subsection{Uniform flow and consequences of shear}
\label{sec:UFCS}
Wave triad interaction is energy exchange between waves on the surface and the pycnocline and there is no direct involvement of bottom topography. Therefore, a velocity field with a uniform flow (figure \ref{fig:1} case 1) will Doppler shift the waves on the surface and the pycnocline by the same velocity $U$ and there won't be any consequences on the resonance condition. To illustrate this, we take three waves having wavenumbers $(k_1,k_2,k_3)$ and corresponding frequencies  $(\omega_1, \omega_2, \omega_3)$  such that $k_1+ k_2-k_3=0$ but $\omega_1+ \omega_2-\omega_3 \neq 0$. In the presence of a constant base velocity $U$, every frequency $\omega_i$ ($i=1,2,3$) would be Doppler shifted by an amount $Uk_i$. In such a case, however, the intrinsic frequencies of the waves will undergo no change. Therefore, the modified frequency condition would be
\begin{align*}
&(\omega_1+Uk_1)+(\omega_2+Uk_2)-(\omega_3+Uk_3)\\
=&\,\omega_1+\omega_2-\omega_3+U(k_1+k_2-k_3)\\
=&\,\omega_1+\omega_2-\omega_3\\ 
\neq & 0.
\end{align*}
Thus, a mere Doppler shift by the same velocity $U$ wouldn't change the resonance condition for the wave triad interaction. However, if the surface and the interface were to be Doppler shifted by  different amounts,  there can be changes in the resonance conditions and naturally the three waves satisfying the resonance condition in  the absence of shear might not do so in the presence of a shear. Alternatively, three waves not satisfying a resonance condition might do so in the presence of velocity shear. This can be elucidated using a simple example: let $k_1+ k_2-k_3=0$ but $\omega_1+ \omega_2-\omega_3 \neq 0$, i.e., the waves do not satisfy the resonant condition in the absence of a base velocity shear. Let us assume that the waves $1$ and $2$ are at the surface, which now has a base velocity $U_u$, while wave $3$ is at the interface, which travels with a velocity $U_l$ (different from $U_u$ because of shear). Then, the frequency condition reads
\begin{align*}
&(\omega'_1+U_uk_1)+(\omega'_2+U_uk_2)-(\omega'_3+U_lk_3)\\
=&\,\omega'_1+\omega'_2-\omega'_3+U_u(k_1+k_2)-U_lk_3\\
=&\,0 \,\,\,\,\,\mathrm{iff}\,\,\,\,\, \omega'_1+\omega'_2 -\omega'_3=k_3(U_l-U_u).
\end{align*}
The primes in the frequencies denote that  the frequencies will get modified due to  shear.
\subsection{Shear in the lower layer}

\begin{figure}
        \centering
        \subfloat[]{\includegraphics[width=6cm]{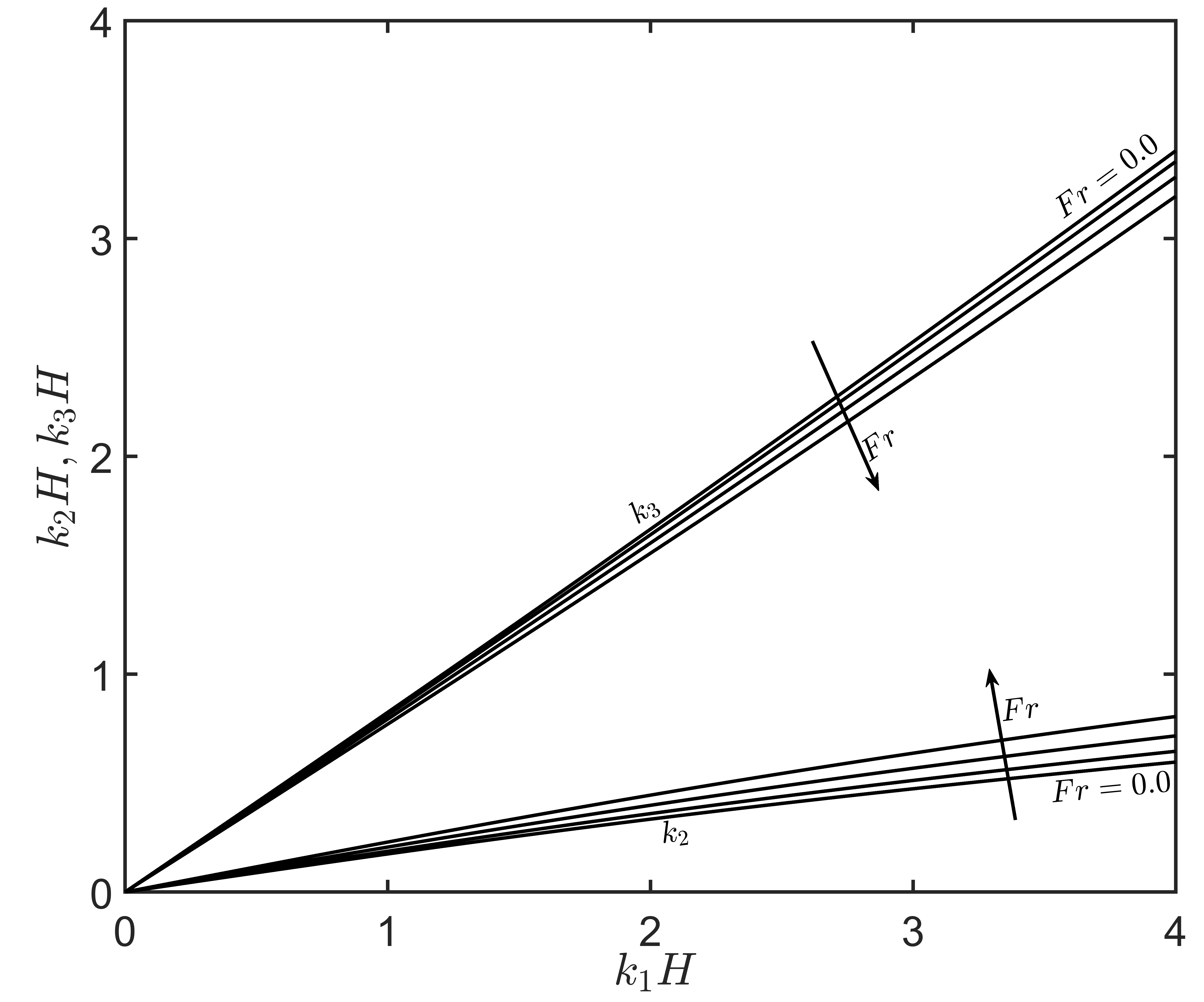}
}
\subfloat[]{\includegraphics[width=6cm]{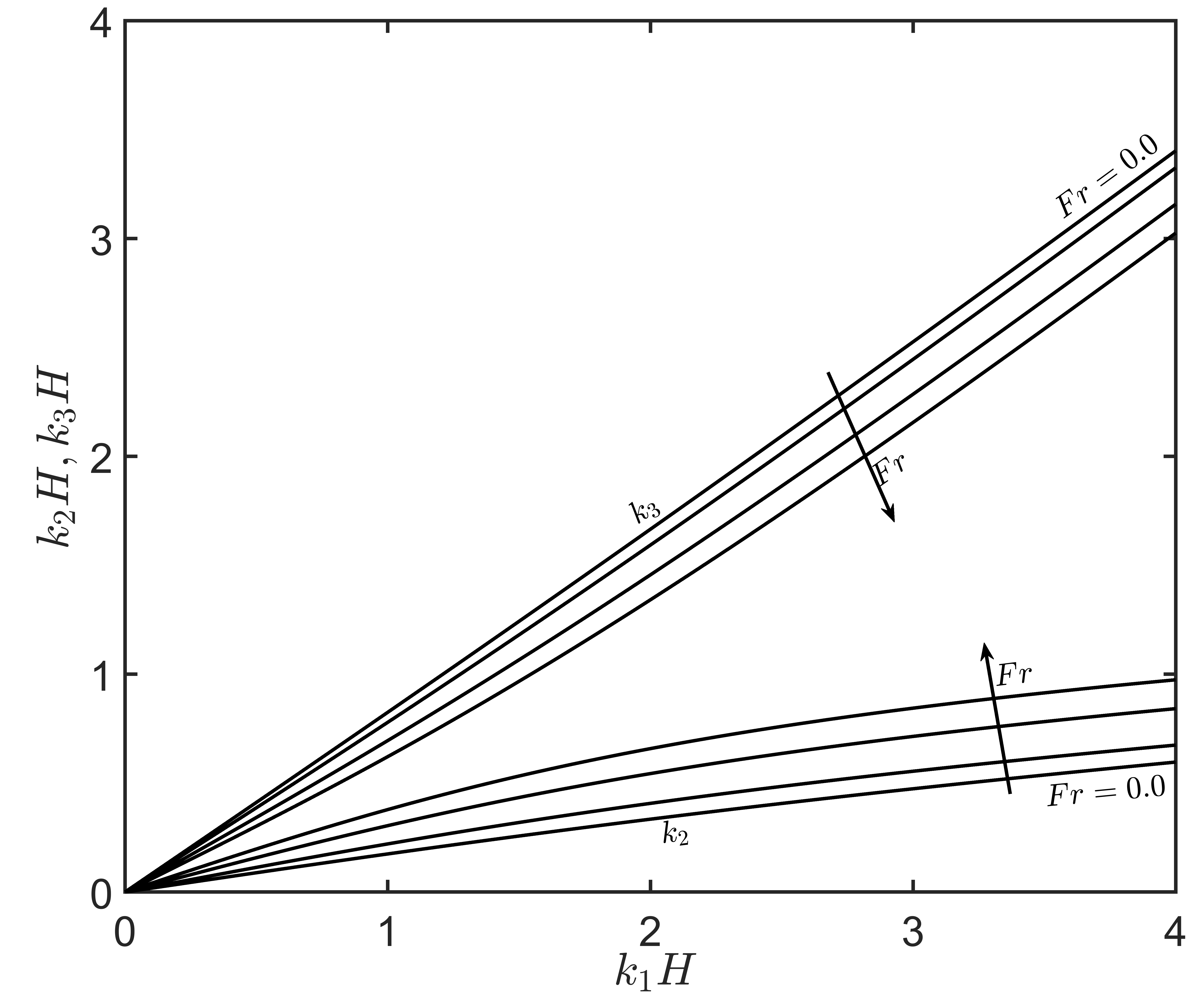}
}
        \caption{Different combinations of $k_1$, $k_2$ and $k_3$ on $\mathcal{IG^-}$, $\mathcal{SG^-}$, $\mathcal{IG^+}$ respectively forming a resonance triad for (a) shear in bottom layer only.  $U_u^*=U_l^*=(0.0,0.2,0.4,0.6)$, $U^*_b=0$, $R=0.95$ and $h_u/h_l=1/3$. (b) Shear in top layer only. $U_u^*=(0.0,0.2,0.4,0.6)$,  $U_l^*=U_b^*=0$, $R=0.95$, and $h_u/h_l=1/3$.    }
        \label{fig:WaveTriad}
\end{figure}

When shear is present only in the lower layer,  there is no Doppler shift between the two waves and the only effect of the shear is felt in modifying the intrinsic frequencies of the waves. Although the shear jump is only at the pycnocline, the effect of it at lower wavenumbers would be felt in the surface mode as well. In figure \ref{fig:WaveTriad}(a), we have shown the change in the resonance condition for three interacting waves having $k_1, k_2$ and $k_3$ on $\mathcal{IG^-}$, $\mathcal{SG^-}$ and $\mathcal{IG^+}$ respectively. The Froude numbers are as follows: $Fr=(0.0,0.2,0.4,0.6)$. As the shear is increased in the positive direction, for a given $k_1$ on $\mathcal{IG^-}$, $k_2$ on $\mathcal{SG^-}$ decreases but $k_3$ on $\mathcal{IG^+}$ increases. The figure reveals that the change is not as significant as that in the Bragg resonance case even at high values of shear. 

\subsection{Shear in the upper layer}

The presence of a shear in the upper layer modifies the flow in two ways. Firstly, there now exists a jump in base shear both at the surface and at the interface, which changes the intrinsic frequencies of all four modes. Secondly, the presence of  shear automatically means that the local mean velocity at the surface and at the pycnocline are different from each other, which implies a relative Doppler shift between the two. Although such a situation may give rise to shear instabilities, such shear instabilities tend to occur at  higher wavenumbers which have  very low growth rates. In figure \ref{fig:WaveTriad}(b), we have shown the resonance condition for three interacting waves having $k_1, k_2$ and $k_3$ respectively on $\mathcal{IG^-}$, $\mathcal{SG^-}$ and $\mathcal{IG^+}$. Yet again, the behaviour is similar to that of previous case but the change in the resonance condition is more prominent here due to the Doppler shifting of the surface and the interface.


\section{Numerical method}\label{sec:5} 
Higher Order Spectral (HOS) method is a highly accurate and efficient numerical method  developed by \cite{Dommermuth} for studying wave propagation and wave-topography interaction in a single layered fluid. Among other things, they studied the collision of two wave packets. The method was further expanded to a two-layered density stratified fluid by \cite{Alam2} to study various cases of Bragg resonance. Although we have derived the evolution equations analytically assuming the resonance conditions are exactly satisfied, the HOS code allows to simulate the near-resonance conditions as well. Furthermore, the study of multiple resonances, which would be a  tedious analytical exercise, becomes simpler on using the HOS method. Here our objective is to  extend the versatile HOS method to incorporate a piecewise linear velocity field. The base velocity field thus introduced will be continuous but its $z$-derivative might be discontinuous at the interfaces, thus giving rise to vorticity gravity waves. In our formulation, we specify the values of the base velocities at $h=\{0,-h_u,-h_u-h_l\}$ as $U=\{U_u,U_l,U_b\}$, using which we  get various sub-cases. For $U_u=U_l=U_b=0$, our system will reduce to the system studied in \cite{Alam2}, i.e. having four pure gravity waves. Furthermore, setting $U_u=U_l=U_b\neq 0$ would lead to gravity waves whose frequencies are simply Doppler shifted with respect to the bottom.  

In the HOS method, we solve the evolution of the surface and interface elevations and velocity potentials associated with them.  Rest of the variables in the fluid bulk are  solved analytically using the boundary conditions.  Since the major part of the computation is limited to the surface and the interface,  the HOS method is highly computationally efficient. We proceed similar to what has been described in \citealt{Alam2}, using similar notations. The continuity equations read
\begin{subequations}
\begin{align}
    \nabla^2 \phi_u &=0\qquad\textrm{$-h_u+\eta_l<z<\eta_u$},\label{HOS_Lap1}\\
    \nabla^2 \phi_u&=0\qquad  \textrm{$-h_u-h_l+\eta_b<z<-h_u+\eta_l$}.\label{HOS_Lap2}
\end{align}
\end{subequations}
The kinematic boundary conditions are as follows:
\begin{subequations}
\begin{align}
	\eta_{u,t}+(U+\phi_{u,x})\eta_{u,x}&=\phi_{u,z} \qquad \textrm{at $z=\eta_u$},\label{HOS_KB1}\\
    \eta_{l,t}+(U+\phi_{u,x})\eta_{l,x}&=\phi_{u,z} \qquad  \textrm{at $z=-h_u+\eta_l$},\label{HOS_KB2}\\
    \eta_{l,t}+(U+\phi_{l,x})\eta_{l,x}&=\phi_{l,z} \qquad \; \textrm{at $z=-h_u+\eta_l$},\label{HOS_KB3}\\
    (U+\phi_{l,x})\eta_{b,x}&=\phi_{l,z} \qquad \;\textrm{at $z=-h_u-h_l+\eta_b$}. \label{HOS_KB4}
\end{align}
\end{subequations}
Likewise, for the dynamic boundary conditions, we have
\begin{subequations}
\begin{align}
	\phi_{u,t}+\frac{1}{2} \left( \phi_{u,x}^2+\phi_{u,z}^2\right)+U\phi_{u,x}-\Omega_u\psi_u+g\eta_u&=0 \qquad \textrm{at $z=\eta_u$}, \label{HOS_DB1}\\
    \begin{split}
    	 \rho_u \left[ \phi_{u,t}+\frac{1}{2} \left( \phi_{u,x}^2+\phi_{u,z}^2\right)+U\phi_{u,x}-\Omega_u\psi_u+g\eta_l\right]\\
         -\rho_l\left[ \phi_{l,t}+\frac{1}{2} \left( \phi_{l,x}^2+\phi_{l,z}^2\right)+U\phi_{l,x}-\Omega_l\psi_l+g\eta_l\right]&=0 \qquad \textrm{at $z=-h_u+\eta_l$}. \label{HOS_DB2}
    \end{split}
\end{align}
\end{subequations}
The governing equation for the potential are simply the Laplace equations, which can't accommodate time evolution in itself. However, there is a time evolution equation for the potentials at the surface and the interface, which are given by the dynamic boundary conditions. We define a surface potential and an interface potential, whose evolution can be tracked using the two dynamic boundary conditions:
\begin{subequations}
\begin{align}
\phi^S(x,t)&\equiv\phi_u(x,\eta_u(x,t),t),\\
\phi_u^I(x,t)&\equiv\phi_u(x,-h_u+\eta_l(x,t),t),\\
\phi_l^I(x,t)&\equiv\phi_l(x,-h_u+\eta_l(x,t),t).
\end{align}
\end{subequations}
Further, we define a new potential at the interface using the above defined potentials:
\begin{align}
\phi^I(x,t)&\equiv\phi_l^I(x,t)-R\phi_u^I(x,t).
\end{align}
Additionally, we  define surface and interface streamfunctions
\begin{subequations}
\begin{align}
\psi^S(x,t)&\equiv\psi_u(x,\eta_u(x,t),t),\\
\psi^I(x,t)&\equiv\psi_u(x,-h_u+\eta_l(x,t),t)\\
\psi^I(x,t)&\equiv\psi_l(x,-h_u+\eta_l(x,t),t).
\end{align}
\end{subequations}
Using the kinematic and dynamic boundary conditions, we obtain the evolution equations for the surface potential, $\phi^S$, the interface potential $\phi^I$, the surface elevation, $\eta_u$ and the interface elevation, $\eta_l$: 
\begin{subequations}
\begin{align}
	\eta_{u,t}&=-\eta_{u,x}[\phi^S_{u,x}+U_u+\Omega_u\eta_u]+(1+\eta^2_{u,x})\phi_{u,z} \qquad \textrm{at $z=\eta_u$},\label{HOS_EV1}\\
    \eta_{l,t}&=-\eta_{l,x}[\phi^I_{l,x}+U_l+\Omega_l\eta_l]+(1+\eta^2_{l,x})\phi_{l,z} \qquad \textrm{at $z=-h_u+\eta_l$},\label{HOS_EV2}\\
    \phi^S_{,t}&=-g\eta_u-\frac{1}{2}(\phi^S_{u,x})^2+\frac{1}{2}(1+\eta^2_{u,x})\phi^2_{u,z}-(U_u+\Omega_u\eta_u)\phi^S_{u,x}+\Omega_u\psi^S\qquad \textrm{at $z=\eta_u$},\label{HOS_EV3}\\
    \phi^I_{,t}&=\frac{1}{2}( R(\phi^I_{u,x})^2-(\phi^I_{l,x})^2)+ \frac{1}{2}(1+\eta^2_{l,x})(\phi_{l,z}^2-R\phi_{u,z}^2)  -g\eta_{l}(1-R)+U_l( R\phi^I_{u,x}-\phi^I_{l,x})\nonumber \\
    & +R\Omega_u\psi_u^I-\Omega_l\psi_l^I +\eta_l(R\Omega_u\phi^I_{u,x}-\Omega_l\phi^I_{l,x})  \qquad \textrm{at $z=-h_u+\eta_l$}.\label{HOS_EV4}
\end{align}
\end{subequations}
In the above equations, we have substituted the Taylor expansion for $U$ (see \eqref{eq:Tay1} and \eqref{eq:Tay2}) The velocity potential and the streamfunctions are expanded in a perturbation series:
\begin{equation}
\phi_{u/l}(x,z,t)=\sum_{m=1}^M\phi^{(m)}_{u/l}(x,z,t)\qquad;\qquad \psi_{u/l}(x,z,t)=\sum_{m=1}^M\psi^{(m)}_{u/l}(x,z,t).
\end{equation}

At every order $m$, we further write the velocity potentials as a sum of basis functions (Fourier basis function in this case). Assuming solutions to be periodic in the $x$-direction, we express the solutions as a discrete Fourier series\footnote{It is necessary to filter out the high wavenumbers by applying a low pass filter, so that the amplification of round off errors at higher wavenumbers does not happen; see \S 3.2.2 of \cite{Dommermuth}. }. Furthermore, we use the Laplace equations to find out the function form of the solutions, and we finally get
\begin{align}
    \phi_u^{(m)}&= \sum_{n=-N}^{N-1}\[A_n^{(m)}(t)\frac{\cosh{k_n(z+h_u)}}{\cosh{(k_nh_u)}} + B_n^{(m)}(t)\frac{\sinh{(k_nz)}}{\cosh{(k_nh_u)}} \] \eee^{\ii k_nx},\label{Four1}\\
    \phi_l^{(m)}&= \sum_{n=-N}^{N-1}\[C_n^{(m)}(t)\frac{\cosh{k_n(z+h_u+h_l)}}{\cosh{(k_nh_l)}} + D_n^{(m)}(t)\frac{\sinh{k_n(z+h_u+h_l)}}{\cosh{(k_nh_l)}} \] \eee^{\ii k_nx}.  \label{Four2}
\end{align}
However, it would not be convenient to directly substitute \eqref{Four1} and \eqref{Four2} in the boundary conditions to obtain the unknown coefficients because at the surface and the interface, $z$ will have a dependence on $x$. Hence, we would expand the surface and interface potentials as a Taylor Series about the respective mean level, so as to eliminate the implicit $x$-dependence of the eigenfunctions:
\begin{equation}
    \phi^S(x,t)=\sum_{m=1}^{M} \phi_u^{(m)}(x,\eta_u,t)=\sum_{m=1}^{M}\sum_{k=0}^{M-m} \frac{\eta_u^k}{k!}\frac{\partial^k}{\partial z^k}\phi_u^{(m)}(x,z,t)\bigg\rvert_{z=0}.\label{HOS_DR1}
\end{equation}
The above equation   can be written as a sequence of Dirichlet boundary conditions at each order $m$. Here, the boundary conditions at each order depends on product of the  terms which have  already been found out at the leading orders, therefore making the problem effectively linear at every order $m$. Further details on the derivation of the boundary conditions can be found in appendix \ref{app:B}. We have
   \begin{equation}
    	\phi_u^{(m)}(x,0,t)=f_u^{(m)},\label{HOS_BC1}
    \end{equation}
    where
    \begin{align}
    	f_u^{(1)}&=\phi^S,\\
        f_u^{(m)}&=-\sum_{k=1}^{m-1}\frac{\eta_u^k}{k!}\frac{\partial^k}{\partial z^k}\phi_u^{(m-k)}(x,z,t)\bigg\rvert_{z=0}.
    \end{align}
        Similarly, for the interface we have a similar sequence of Dirichlet boundary conditions:
    \begin{equation}
    	\Phi^{(m)}(x,-h_u,t)=f_{l1}^{(m)},\label{HOS_BC2}
    \end{equation}
    where
    \begin{align}
    f_{l1}^{(1)}&=\phi^I,\\
        f_{l1}^{(m)}&=-\sum_{k=1}^{m-1}\frac{\eta_l^k}{k!}\frac{\partial^k}{\partial z^k}\Phi^{(m-k)}(x,z,t)\bigg\rvert_{z=-h_u}.
    \end{align}
Here we have defined
$\Phi(x,z,t)\equiv\phi_l(x,z,t)-R\phi_u(x,z,t)$. As for the third boundary condition, we write
\begin{equation}
   	\varphi_{,z}(x,z,t)=\eta_{l,x}\varphi_{,x}(x,z,t)+\eta_l\eta_{l,x}(\Omega_u-\Omega_l)\quad \mathrm{at}\;\;z=-h_u+\eta_l,
\end{equation}
with $\varphi(x,z,t)\equiv\phi_{u}(x,z,t)-\phi_{l}(x,z,t)$. Using the Taylor expansion of $\varphi(x,z,t)$ about the mean interface level along with the Laplace equation, we finally get a sequence of Neumann boundary conditions:
\begin{equation}
    \varphi_{,z}^{(m)}(x,-h_l,t)=f_{l2}^{(m)},\label{HOS_BC3}
\end{equation}
    where
   \begin{align}
    f_{l2}^{(1)}&=0,\\
    f_{l2}^{(2)}&=\frac{\partial}{\partial x}\left[\eta_l\varphi_{,x}^{(1)}(x,z,t)\bigg\rvert_{z=-h_u}\right]+\eta_l\eta_{l,x}(\Omega_u-\Omega_l)\\
        f_{l2}^{(m)}&=\sum_{k=1}^{m-1}\frac{\partial}{\partial x}\left[\frac{\eta_l^k}{k!}\frac{\partial^{k-1}}{\partial z^{k-1}}\varphi_{,x}^{(m-k)}(x,z,t)\bigg\rvert_{z=-h_u}\right].
    \end{align}
Finally, for the bottom boundary condition, we have a similar impenetrability boundary condition:
    \begin{equation}
    \phi_{l,z}^{(m)}(x,-h_u-h_l,t)=f_{b}^{(m)},\label{HOS_BC4}
\end{equation}
    where
   \begin{align}
    f_{b}^{(1)}&=U_b\eta_{b,x},\\
    f_{b}^{(2)}&=\frac{\partial}{\partial x}\left[\eta_b\phi_{l,x}^{(1)}(x,z,t)\bigg\rvert_{z=-h_u-h_l}\right]+\eta_b\eta_{b,x}\Omega_l\\\\
        f_{b}^{(m)}&=\sum_{k=1}^{m-1}\frac{\partial}{\partial x}\left[\frac{\eta_b^k}{k!}\frac{\partial^{k-1}}{\partial z^{k-1}}\phi_{l,x}^{(m-k)}(x,z,t)\bigg\rvert_{z=-h_u-h_l}\right].
    \end{align}
Using the four boundary conditions, we obtain the value of unknown coefficients $A_n$, $B_n$, $C_n$ and $D_n$ at every order $m$. Now we have a full solution of $\phi_u^{(m)}$ and $\phi_l^{(m)}$  at the order $m$. At the next order $m+1$, the functions $f_u$, $f_{l1}$, $f_{l2}$, $f_{b}$ can be evaluated by using the velocity potentials and their derivatives, which were already found out at the previous order $m$. Again the boundary value problem at the order $m+1$ can be solved, and in this way we can proceed further to obtain  $\phi_u^{(m)}$ and $\phi_l^{(m)}$ at each order. It is interesting to note here that the all the above four boundary conditions -- \eqref{HOS_BC1}, \eqref{HOS_BC2}, \eqref{HOS_BC3}, \eqref{HOS_BC4} are the same as that in \cite{Alam2}, i.e.\ without background velocity. Therefore, the function form of the coefficients $A_n$, $B_n$, $C_n$ and $D_n$ in terms of the functions $f_u$ to $f_{b}$ remain the same.  Using these coefficients, we can find any derivative of the velocity potentials at any location. After solving the boundary value problem, we march forward in time using 4-th order Runge-Kutta method. The domain size is chosen to be $2\pi$ and the number of points in real space equals $2N+1$ such that variables are periodic in $x$.
    
\subsection{Validation}

\begin{figure}
\centering\includegraphics[width=70mm]{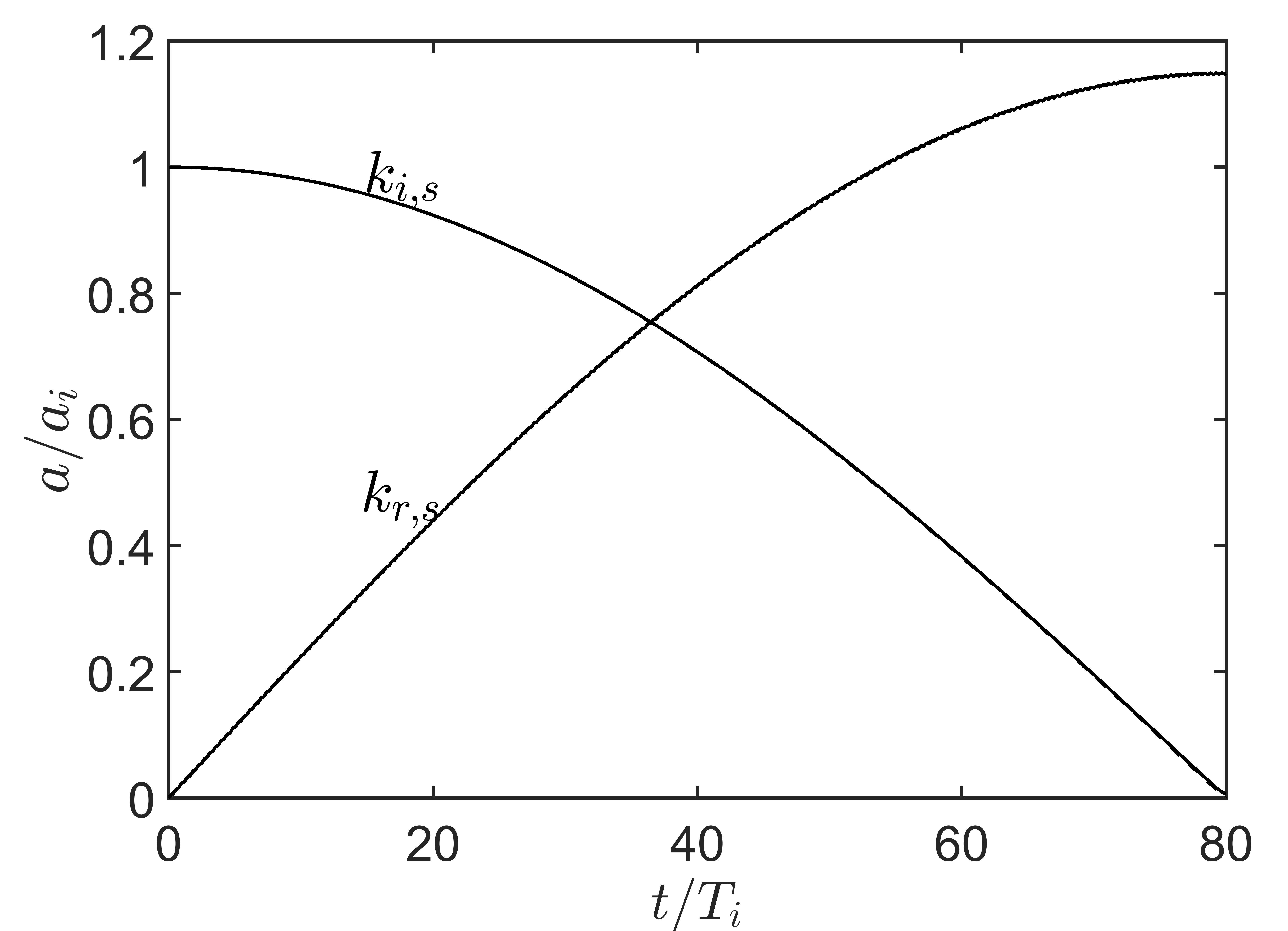} \caption{Code Validation for $R=0.98$, $k_iH=0.086$, $k_rH=0.1140$, $k_bH=0.2$, $\omega^*_i=0.0982$, $\omega^*_r=-0.0982$, $U_u^*=0.1864$, $U_l^*=0.0083$, $M=3$, $N=2048$, $T_i/\Delta T=512$. The analytical and numerical solutions are indistinguishable. }
\label{fig:Dispersion}
\end{figure}

A comprehensive benchmarking of the HOS method for two layers without a velocity field has been performed in \cite{Alam2}. In this paper, we have extended the method to incorporate the velocity field by adding requisite terms. For validation of the code, we have simulated a case of Bragg resonance in which the surface mode interacts with the bottom to generate another surface mode having an intrinsic frequency of the opposite sign. We have compared the solution of the HOS code to the analytically obtained solution. The parameters used are mentioned in the caption of  figure \ref{fig:Dispersion}. It can be seen that the analytical solution and the numerical solution are graphically indistinguishable. 

\subsection{Numerical results}

\begin{figure}

        \centering
        \subfloat[]{\includegraphics[width=6cm]{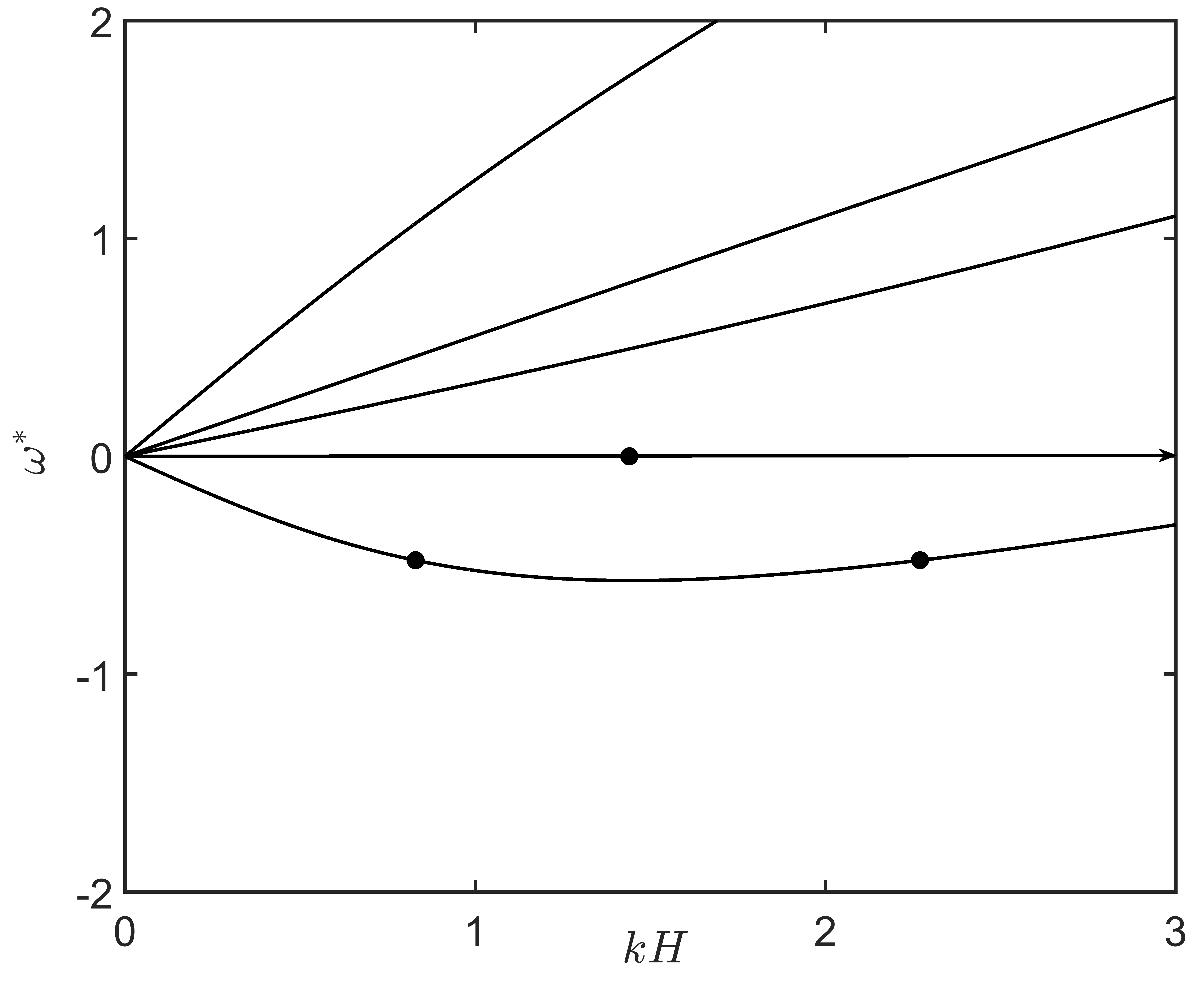}
}
\subfloat[]{\includegraphics[width=6cm]{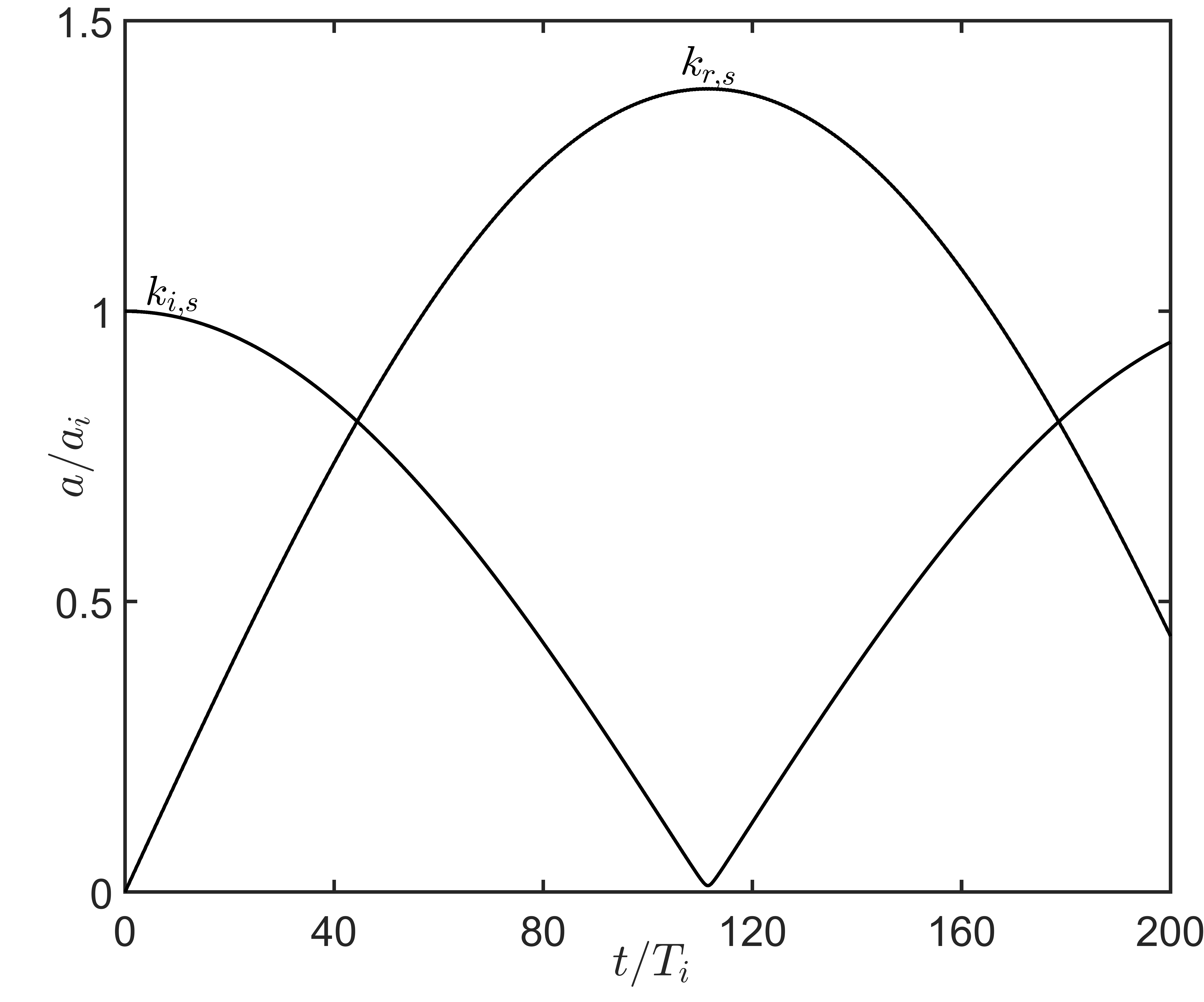}
}

        \caption{(a) Dispersion relation showing the location of resonant triad. Both the incident and the resonant wave lie on the $\mathcal{SG^-}$ curve. (b) Numerical simulation using the HOS code: $a_i=0.00005H$, $a_b=0.02H$, $U_u^*=0.5016$, $U_l^*=U_b^*=0$, $\omega_i^*=-0.4770$, $R=0.95$, $k_iH=0.83$, $k_rH=2.27$, $k_bH=1.44$, $h_u/h_l=1/3$, $M=3$, $N=1024$, $T_i/\Delta T=2048$. $T_i$ is the time period of the incident wave.   }
        \label{fig:NumSim2}

\end{figure}

We have simulated a resonance between the waves on the same branch ($\mathcal{SG^-}$) of the dispersion curve for the case 2, i.e.\ shear only in the lower layer.  The incident wave has the wavenumber $k_iH=0.83$, while the resonant wave has the wavenumber $k_rH=2.27$. These two waves have the same direction of propagation and have the same frequency of $\omega^*=-0.4770$. Because the direction of propagation of the waves is the same, the wavenumber of the bottom is the difference of the wavenumber of  the incident and the resonant waves, i.e. $k_bH=1.44$. The velocities are $U_u^*=0.5016, U_l^*=0, U_b^*=0$. Other relevant physical parameters are $h_u/h_l=1/3$ and $ R=0.95$. The dispersion relation is plotted in figure \ref{fig:NumSim2}(a) and the corresponding HOS simulation is shown in figure \ref{fig:NumSim2}(b).

\begin{figure}

        \centering
        \subfloat[]{\includegraphics[width=6cm]{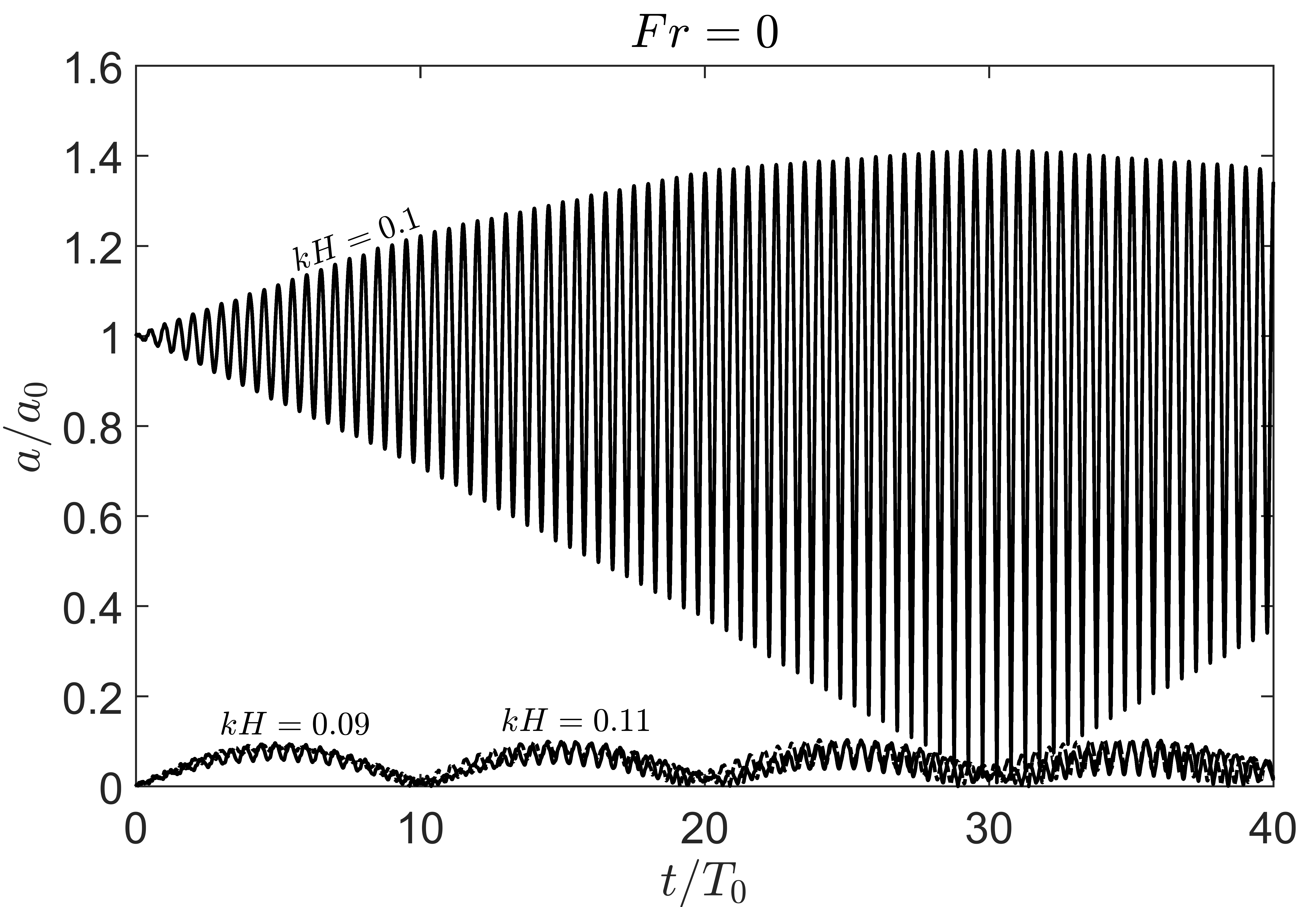}
}
\subfloat[]{\includegraphics[width=6cm]{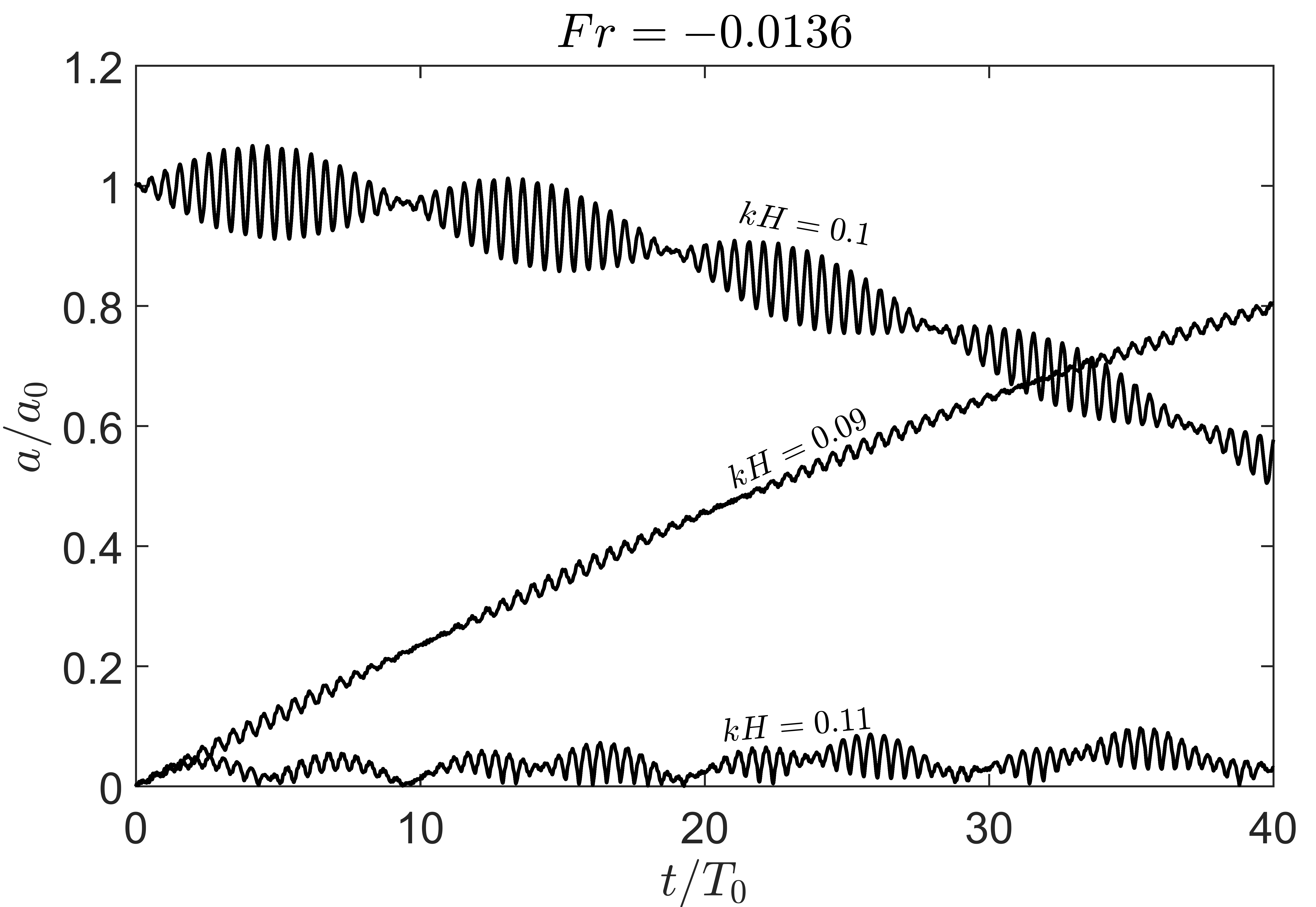}
}

\subfloat[]{\includegraphics[width=6cm]{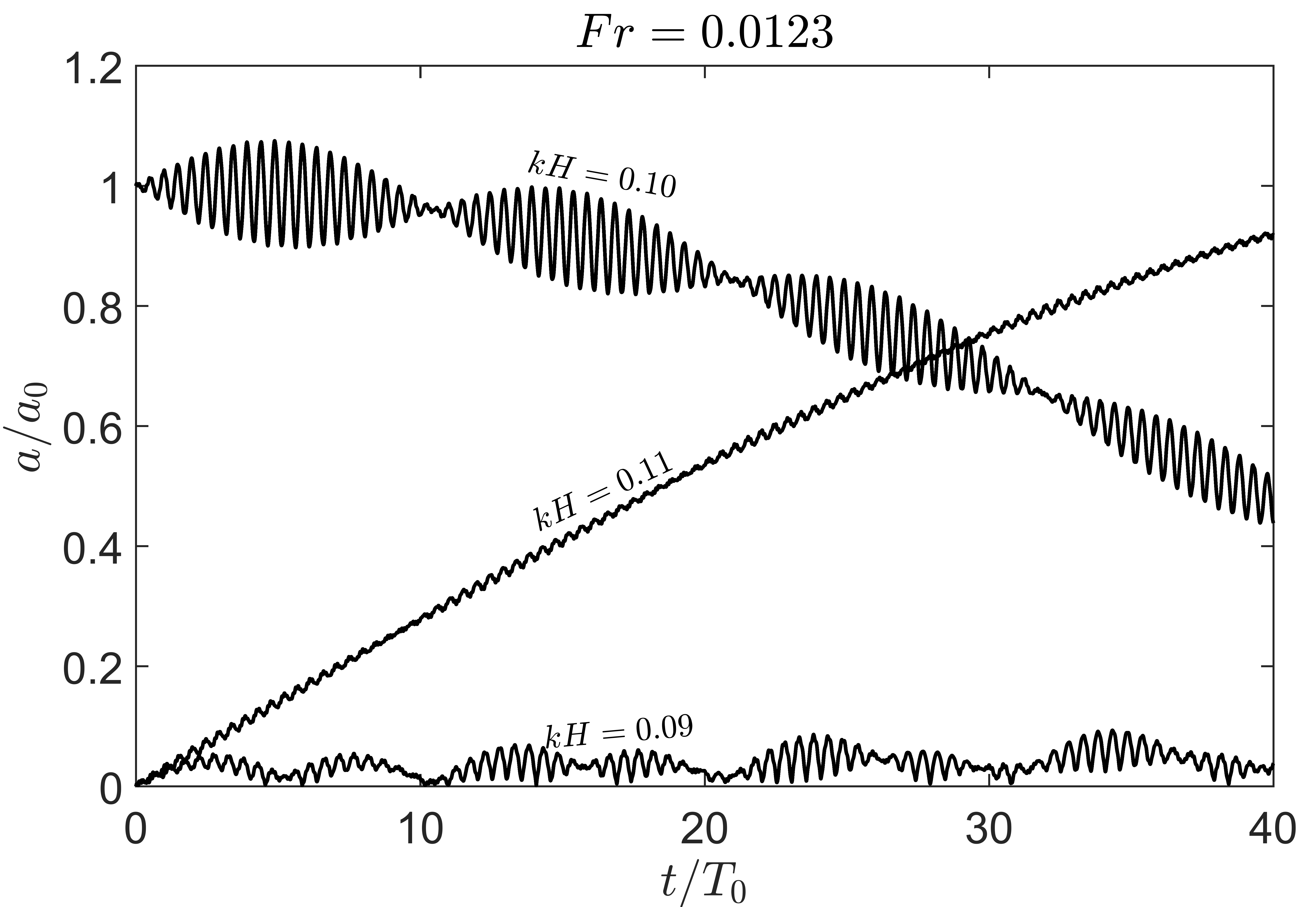}
}
        \caption{Amplitude vs time plot for different wavenumbers on the interface for $R=0.95, h_u/h_l=1/3$, $a_i=0.00005H_0$, $a_b=0.05H_0$, and $U_l^*=U_b^*=0$. (a) $U_u^*=0$, (b) $U_u^*=-0.0136$, and (c) $U^*_u$=0.0123. Parameters for the  simulation are $M=3, N=512$, $T_0/\Delta T=512$. Here, $T_0$ is the time period of the wave $k_i$ in absence of any background velocity. }
        \label{fig:NumSim1}

\end{figure}

Next, we have studied the effect of shear in upper layer on the Bragg resonance between two oppositely travelling internal modes, i.e. $\mathcal{IG}^+$ and $\mathcal{IG}^-$. Because the shear is in the upper layer only, there is no Doppler shift of the concerned waves (both on the pycnocline), and the changes are only in the  the intrinsic frequencies. In the absence of a base flow, it is known that the resonant wavenumber will be the same as the incident wavenumber. However,  we have shown analytically in  \S \ref{sec:UFCS} that shear changes the resonance condition. To illustrate this here, we take our incident wave on $\mathcal{IG}^+$ having wavenumber $k_iH=0.10$ and frequency $\omega_i^*=0.0097$. The bottom ripple consists of three different wavenumbers: $k_{b1}H=0.19, k_{b2}H=0.20$ and $k_{b3}H=0.21$. Thus, the incident wave will interact with the bottom and may generate three different waves having wavenumbers $k_{r1}H=0.09,\,k_{r2}H=0.10,\,k_{r3}H=0.11$ and frequencies $\omega_{r1}^*=-0.0088,\,\omega_{r2}^*=0.0097,\,\omega_{r3}^*=-0.0107$ respectively. Only one of these wavenumbers may satisfy a resonance condition for a given velocity field and the other wavenumbers will be generated in a `near-resonant' way \citep{Craik}. We plot the time evolution of amplitude of all these three wavenumbers in the absence of  shear; see figure \ref{fig:NumSim1}(a). As expected, the maximum growth is only in the wavenumber $k_{r2}H=0.10$. The amplitude plotted is simply the spatial Fourier transform of the interface, and a rapidly changing amplitude corresponding to $kH=0.10$ signifies an oppositely travelling wave increasing in amplitude. At about $t/T_0\approx 30$, both the positively and the negatively travelling waves have the same amplitude. We also observe a small growth in the wavenumbers $kH=0.09$ and $kH=0.11$. These two wavenumbers do not satisfy the exact resonant condition and hence are generated only near resonantly.

Next,  we make the shear negative in the upper layer to yield $U_u^*=-0.0136$, while keeping $U_l^*=U_b^*=0$. Due to the velocity field, the incident wave's frequency gets modified to $\omega_i^*=0.0092$. The frequencies of the three possible resonant waves respectively become $\omega_{r1}^*=-0.0092,\omega_{r2}^*=-0.0103$ and $\omega_{r3}^*=-0.0103$. In this case, we observe that the incident wave frequency is equal to $\omega_{r1}$, therefore the dominant resonating wavenumber is $k_{r1}H=0.09$. The amplitude evolution has been plotted in see figure \ref{fig:NumSim1}(b).  Wiggles in the plot indicate near-resonant generation of oppositely travelling waves.

Likewise, we make the shear positive in the upper layer to yield $U_u^*=0.0123$, while keeping $U_l^*=U_b^*=0$. The incident wave's frequency changes to $\omega_i^*=0.0102$. The frequency of the three possible resonant waves become $\omega_{r1}^*=-0.0083,\,\omega_{r2}^*=-0.0093$ and $\omega_{r3}^*=-0.0102$. We observe that $\omega_i=\omega_{r3}$ and hence, the dominant wavenumber generated is  $k_{r3}H=0.11$. This also corroborates with the  figure \ref{fig:Case2_Internal}, where it can be seen that for a $k_i$ on $\mathcal{IG^+}$ mode, increasing the shear results in increase in $k_r$ on $\mathcal{IG^-}$ mode. The amplitude evolution has been plotted in see figure \ref{fig:NumSim1}(c). Again, wiggles indicate near-resonant generation of oppositely travelling waves. {Thus, we we see that the exclusion of shear present may substantially change the condition for resonant triads. Hence, practical applications, such as broadband cloaking \citep{alam2012broadband}, in which bottom corrugations are designed in a particular way to `cloak' the offshore structures, may need to account for oceanic shear for an optimum design.}

\section{Summary and conclusion}\label{sec:6}
 Four wave modes, two at the surface and two at the pycnocline, exist in  two-layered density stratified flows. A set of three modes can form a triad and undergo weakly nonlinear interactions when a certain resonance condition is met. A rippled bottom topography, if present, can act as a stationary wave and mediate weakly nonlinear interactions -- a process known as  `Bragg resonance'. The conventional approach towards deriving the standard resonance conditions for weakly nonlinear wave triads, as well as Bragg scattering, fails to incorporate the effect of background velocity, especially of background shear. This is because these approaches are based on the potential flow theory, which dramatically simplifies the problem and allows one to  solve for the interfaces only. Since atmospheric and oceanic flows  always have background velocity, it is imperative to account for the background flow  in studying triads and Bragg resonances. 
We have taken a step forward in this direction by including piecewise linear velocity profile, while still using the potential flow approximation. Although piecewise linear velocity means piecewise constant shear, and apparently cannot be dealt using  potential flow theory, we show that the perturbed flow remains  potential, even though the base flow has shear.

On incorporating background velocity, the resonance conditions for wave triads and Bragg scattering 
get strongly modified. Background velocity  influences the resonance conditions in two ways: (i) by causing unequal Doppler shifts between the surface,  pycnocline, and the bottom (at least two of them), and (ii) by changing the intrinsic frequencies of the waves. We have explored various kinds of velocity fields - uniform, constant shear in the lower layer, constant shear in the upper layer, and constant shear in both layers, to form a broad understanding of the effect of background velocity on triads and Bragg resonances. 
For Bragg resonance, even a uniform velocity field  changes the resonance condition. In the absence of background shear, Bragg resonance only occurs when the two wave modes (the third `wave' is the bottom ripple) lie on two distinct branches of the dispersion curve. However with shear (in the lower layer), we show that resonant triads appear even when the two wave modes lie on the same branch of the dispersion curve. In this regard  interfacial modes are more susceptible than surface modes; modest Froude numbers are required for  causing surface modes on the same branch to resonate; however, small Froude numbers are sufficient to do the same for the interfacial modes.    

Using multiple scale analysis along with the Fredholm alternative, we have analytically obtained the equations governing the (slow) time evolution of the amplitudes of the waves forming both classical and Bragg triads up to $\mathcal{O}(\epsilon^2)$. The formalism that we have developed has also been added  to the Higher Order Spectral (HOS) method, a highly efficient and accurate numerical technique that can incorporate several triads up to any prescribed order of nonlinearity, which traditionally does not include background velocity. Using the `modified' HOS we have numerically studied two problems on Bragg resonance: (i) the case when shear is present in the lower layer and leads to resonance between two wave modes lying on the same branch of the dispersion curve, and (ii) shear in the upper layer, which strongly affects the intrinsic frequencies. In the second case, we consider a bottom ripple consisting of three wavenumbers (chosen close to each other); a given incident wave resonantly generates only one wave, however two additional waves are generated via near-resonant interactions. Imposing the velocity field leads to change in the standard resonance condition and the wave generated in a near-resonant way may become resonant. This mechanism of near-resonant generation and effect of velocity field on resonance condition has been captured using the modified HOS method.

\section*{Acknowledgments}
This work has been partially supported by the following Grant Nos.: IITK/ME/2014338, STC/ME/2016176 and ECR/2016/001493.

\appendix 
\section{Derivation of dynamic boundary condition in the presence of a piecewise linear background shear}\label{app:A}

The inviscid Navier-Stokes equation within the bulk of a fluid of constant density $\rho$ is
\begin{equation}\label{eq:A1}
	\rho \left[  \mathbf{u}_{,t} +\frac{1}{2} \nabla(\mathbf{u}\cdot\mathbf{u})- \mathbf{u}\times(\nabla\times \mathbf{u})\right]=-\nabla p -\nabla (\rho gz).
\end{equation}
Using the fact that there is no base vorticity generation in the bulk, we have $\nabla \times \mathbf{u}=\nabla \times \mathbf{\bar{u}}$=$\Omega\hat{j}$, where $\Omega$ is  constant for each layer. Besides we use
\begin{equation}\label{eq:A2}
\mathbf{u}\times (\Omega\hat{j})=\Omega\nabla \psi=\nabla (\Omega\psi).
\end{equation}
Substituting \eqref{eq:A2} in \eqref{eq:A1} and removing the mean flow part, we are left with
\begin{equation}
	\rho \left[ \ \mathbf{u'}_{,t}+\frac{1}{2} \nabla(\mathbf{u}'\cdot\mathbf{u}')+\nabla(\mathbf{\bar{u}}\cdot\mathbf{u}')- \nabla (\Omega\psi')\right]=-\nabla p' -\nabla (\rho g\eta).
\end{equation}
Since the perturbed flow is irrotational, we introduce $\mathbf{u}'=\nabla \phi'$. Moreover, since density is constant within each layer, we obtain
\begin{equation}
	\nabla\left[\rho \left( \phi'_{,t}+\frac{1}{2} \nabla \phi'\cdot\nabla\phi'+\mathbf{\bar{u}}\cdot\nabla\phi'- \Omega\psi'+g\eta\right)+p'\right]=0.
\end{equation}
Since this is true for any arbitrary curve inside the domain, we have on integration
\begin{equation}
\rho \left(\phi'_{,t}+\frac{1}{2} \nabla \phi'\cdot\nabla\phi'+\mathbf{\bar{u}}\cdot\nabla\phi'- \Omega\psi'+g\eta\right)+p'=c,
\end{equation}
where $c$ is an arbitrary function of time, which turns out to be zero in order to satisfy the unperturbed far-field condition. Thus, equating the pressure just above and just below the interface $z=\eta(x,t)$, at which the base flow velocity is $\mathbf{\bar{u}}=U\hat{i}$, we  obtain
\begin{multline}
\rho_1 \left( \phi'_{1,t}+\frac{1}{2} \nabla \phi'_1\cdot\nabla\phi_1'+U\hat{i}\cdot\nabla\phi_1'- \Omega_1\psi'+g\eta\right)\\=\rho_2 \left( \phi'_{2,t}+\frac{1}{2} \nabla \phi_2'\cdot\nabla\phi_2'+U\hat{i}\cdot\nabla\phi_2'- \Omega_2\psi'+g\eta\right).
\end{multline}
Dropping the primes, we get
\begin{multline}
\rho_1 \left[ \phi_{1,t}+\frac{1}{2} \left( \phi_{1,x}^2+\phi_{1,z}^2\right)+U\phi_{1,x}-\Omega_1\psi_1+g\eta\right]
         =\\ \rho_2\left[\phi_{2,t}+\frac{1}{2} \left( \phi_{2,x}^2+\phi_{2,z}^2\right)+U\phi_{2,x}-\Omega_2\psi_2+g\eta\right].
\end{multline}

\section{Relevant Coefficients}
\label{app:B}
The coefficients  $\mathbf{n}_j$ are same for the case of wave triad interaction and the case of Bragg resonance. They are simply the  null vector of the transpose of the matrix $\dunderline{\mathfrak{D}}(k_j,\omega_j)$. The coefficients of time derivatives of $\mathcal{O}(\epsilon)$ terms, i.e. the vector $\mathbf{r}_j$ also remains the same both for wave triad interaction and Bragg resonance. 

The components of the vector $\mathbf{r}_j$ are
\begin{align*}
r_j(1)&=1,\\
r_j(2)&=T_j,\\
r_j(3)&=T_j,\\
r_j(4)&=-\ii Q_j,\\
r_j(5)&=-\frac{\ii}{k_j}\[T_j(\omega_j-k_jU_l)\(\tanh{k_jh_u}+\frac{1}{\tanh{k_jh_l}}\)+\frac{Q_j}{\cosh{k_jh_u}}\],\\
r_j(6)&=0.
\end{align*}

\begin{align*}
T_j&=\frac{\cosh{(k_jh_u)}[(\omega_j-U_uk_j)^2+(\omega_j-U_uk_j)\Omega_u\tanh{(k_jh_u)}-gk_j\tanh{(k_jh_u)}]}{(\omega_j-U_lk_j)(\omega_j-U_uk_j)},\\
Q_j&=\frac{\Omega_u}{k_j}+\frac{g}{U_uk_j-\omega_j}.
\end{align*}

For the case of wave triad interaction the coefficients of $\mathbf{v}_1$ are as follows:

\begin{align*}
v_1(1)&=-\ii gk_1\left(\frac{\Omega_u}{g}+\frac{k_2}{U_uk_2-\omega_2}+\frac{k_3}{U_uk_3-\omega_3}\right),\\
v_1(2)&=-\ii k_1\left[\frac{k_2T_3Q_2}{\cosh{k_2h_u}}+\frac{k_3T_2Q_3}{\cosh{k_3h_u}}+T_2T_3\left(\frac{\omega_2-U_lk_2}{\tanh{k_2h_u}}+\frac{\omega_3-U_lk_3}{\tanh{k_3h_u}}-\Omega_u\right)\right],\\
v_1(3)&=-\ii T_2T_3k_1\left(\frac{k_2U_l-\omega_2}{\tanh{k_2h_l}}+\frac{k_3U_l-\omega_3}{\tanh{k_3h_l}}-\Omega_l\right),\\
v_1(4)&=\frac{T_2(U_lk_2-\omega_2)}{\cosh{k_2h_l}}\left(k_2U_u-\omega_2+k_3Q_3\tanh{k_3h_l}\right)\\
&+\frac{T_3(U_lk_3-\omega_3)}{\cosh{k_3h_l}}\left(k_3U_u-\omega_3+k_2Q_3\tanh{k_3h_l}\right)\\
&+(U_uk_3-\omega_3)k_3Q_3\tanh{k_2h_u}+(U_uk_2-\omega_2)k_2Q_2\tanh{k_2h_u}\\
&-Q_2Q_3k_2k_3(1-\tanh{k_2h_u}\tanh{k_3h_u})+\frac{T_2T_3(\omega_2-U_lk_2)(\omega_3-U_lk_3)}{\cosh{k_2h_u}\cosh{k_3h_l}},\\
v_1(5)&=T_2T_3(R-1)\left[(k_3U_l-\omega_3)^2+(k_2U_l-\omega_2)^2+(k_2U_l-\omega_2)(k_3U_l-\omega_3)\right]\\
&-\frac{Rk_2k_3Q_2Q_3}{\cosh{k_2h_u}\cosh{k_3h_u}}+Rk_2T_3Q_2\frac{(k_3U_l-\omega_3)\tanh{k_3h_u}}{\cosh{k_2h_u}}+ Rk_3T_2Q_3\frac{(k_2U_l-\omega_2)\tanh{k_2h_u}}{\cosh{k_3h_u}}\\
&-T_2T_3(k_2U_l-\omega_2)(k_3U_l-\omega_3)\left(R\tanh{k_2h_u}\tanh{k_3h_u}-\frac{1}{\tanh{k_2h_l\tanh{k_3h_u}}}\right),\\
v_1(6)&=0.
\end{align*}

Similarly, the terms of the vectors $\mathbf{v}_2$ and $\mathbf{v}_3$ can be obtained by changing the indices in a cyclic order, i.e.\ the substitution $\{1\rightarrow2,2\rightarrow3,3\rightarrow1\}$ in above equations.\\

For the case of Bragg resonance, where $k_3\equiv k_b=k_1+k_2$ and $\omega_3=\omega_b=0$ we have two cases,

Case 1: $U_b=0$:
\begin{align*}
v_1(1)&=v_1(2)=v_1(3)=v_1(4)=v_1(5)=0,\\
v_1(6)&=\ii\frac{k_1(\omega_2-U_lk_2)T_2}{\sinh{k_2h_u}};\\
v_2(1)&=v_2(2)=v_2(3)=v_2(4)=v_2(5)=0,\\
v_2(6)&=\ii\frac{k_2(\omega_1-U_lk_1)T_1}{\sinh{k_1h_u}}.
\end{align*}
The vector $\mathbf{r}_j$ remains the same as before.

Case 2: $U_b\neq 0$:

In this case, the bottom boundary condition would be inhomogeneous. This will mean that there will exist a time independent particular solution of the system at $\mathcal{O}(\epsilon)$ having

\begin{align*}
    \hat{\eta}_u&=-\frac{U_uU_bU_l^2k_b^6}{\mathfrak{D}(0,k_b)\cosh{k_bh_u}\cosh^2{k_bh_l}}\hat{\eta}_b\equiv X_{b1}\hat{\eta}_b,\\
    \hat{\eta}_l&=-\frac{U_bU_lk_b^5(U_u^2k_b\cosh{k_bh_u}-(\Omega_uU_u+g)\sinh{k_bh_u})}{\mathfrak{D}(0,k_b)\cosh{k_bh_u}\cosh^2{k_bh_l}}\hat{\eta}_b \equiv X_{b2}\hat{\eta}_b,\\ 
    \hat{A}&=-\ii\frac{U_bU_l^2k_b^5(\Omega_uU_u+g)}{\mathfrak{D}(0,k_b)\cosh{k_bh_u}\cosh^2{k_bh_l}}\hat{\eta}_b\equiv \ii X_{b3}\hat{\eta}_b,\\
    \hat{B}&=-\ii\frac{U_bU_l^2k_b^5(\cosh{k_bh_u}U_u^2k_b-\sinh{k_bh_u}(\Omega_uU_u+g))}{\mathfrak{D}(0,k_b)\cosh{k_bh_u}\cosh^2{k_bh_l}}\hat{\eta}_b\equiv \ii X_{b4}\hat{\eta}_b,\\ 
    \hat{C}&=-\ii \frac{U_b(k_b^6U_u^2U_l^2+\mathfrak{D}(0,k_b)\cosh^3{k_bh_l}-k_b^5U_l^2(\Omega_uU_u+g)\tanh{k_bh_u})}{\mathfrak{D}(0,k_b)\cosh{k_bh_l}\sinh{k_bh_l}}\hat{\eta}_b\equiv \ii X_{b5}\hat{\eta}_b,\\
    \hat{D}&=\ii U_b\cosh{k_bh_l}\hat{\eta}_b\equiv \ii X_{b6}\hat{\eta}_b.
\end{align*}

The coefficients $v_2(1),v_2(2),v_2(3),v_2(4), v_2(5)$ may not be zero if $U_b\neq0$ and will be given as:


\begin{align*}
    v_2(1)&=-\ii k_2(k_bQ_1X_{b1}+k_bX_{b3}+\Omega_uX_{b1}),\\
    v_2(2)&=\ii k_2\left(\frac{X_{b2}(T_1(k_1U_l-\omega_1)\sinh{k_1h_u}-k_1Q_1)}{\cosh{k_1h_u}}+\frac{T_1(-X_{b4}\sinh{k_3h_u}+X_{b3})k_b}{\cosh{k_bh_u}}+\Omega_uT_1X_{b_2}\right),\\
    v_2(3)&=-\ii k_2T_1\left(\frac{X_{b2}(k_1U_l-\omega_1)}{\tanh{k_1h_l}}-X_{b6}k_1\tanh{k_1h_l}+k_bX_{b5}-\Omega_lX_{b2}\right),\\
    v_2(4)&=T_1X_{b1}\frac{(k_1U_l-\omega_1)}{k_1U_u-\omega_1}{\cosh{k_1h_u}}+k_1Q_1 X_{b1}(k_1 U_u-\omega_1)\tanh{k_1h_u}\\
        &-k_b\left(X_{b3}\tanh{k_bh_u}+\frac{X_{b4}}{\cosh{k_bh_u}}\right)\left(\frac{T_1(k_1U_l-\omega_1)}{\cosh{k_1h_u}}-k_bU_u\right)\\
        &-Q_1k_1X_{b3}k_b(\tanh{k_1h_u}\tanh{k_bh_u}+1)-\frac{Q_1k_1X_{b4}k_b\tanh{k_1h_u}}{\cosh{k_bh_u}},\\
    v_2(5)&=T_1\left(\frac{\Omega_lX_{b6}k_b}{\coth{k_bh_l}}-\frac{U_lX_{b5}k_b^2}{\coth{k_bh_l}}+RU_lX_{b4}k_b^2-X_{b6}k_1U_lK_b\right)\\
    &+T_1(k_1U_l-\omega_1)\left[-\frac{Rk_bX_{b4}}{\coth{k_1h_u}\coth{k_bh_u}}+(k_1U_l-\omega_1)X_{b2}(R-1)\right.\\
    &\left.+X_{b6}(k_1+k_b)-RX_{b4}k_b+\frac{k_b\tanh{k_1h_u}RX_{b3}}{\cosh{k_bh_u}}+\frac{k_bX_{b5}}{\tanh{k_1h_l}}+\frac{k_bX_{b5}}{\coth{k_bh_l}}\right]\\
    &+Rk_1k_bQ_1\frac{X_{b4}\sinh{k_bh_u}-X_{b3}}{\cosh{k_1h_u}\cosh{k_bh_u}}+\frac{\Omega_lk_bT_1}{\sinh{k_1h_l}}\left(\frac{1}{\cosh{k_1h_l}}-\cosh{k_1h_l}\right),\\
    v_2(6)&=\ii\frac{k_2(\omega_1-U_lk_1)T_1}{\sinh{k_1h_u}}.
    \end{align*}
    
    The coefficients $v_1(1),v_1(2),v_1(3),v_1(4), v_1(5)$ will be given by swaping $k_2$ aand $k_1$ in the above equations.

\section{Boundary conditions}

\subsection{Dirichlet Boundary conditions}
We expand the velocity potential as a perturbation series up to an order `$(m)$'. So, from (\ref{HOS_DR1}) we have,
   $$\phi^S(x,t)=\sum_{m=1}^{M}\sum_{k=0}^{M-m} \frac{\eta_u^k}{k!}\frac{\partial^k}{\partial z^k}\phi_u^{(m)}(x,z,t)\bigg\rvert_{z=0}.$$
   Collecting the terms of leading order that is terms of $\mathcal{O}(\epsilon$), we have,
   $$\phi_u(x,0,t)=\phi^S(x,t).$$
   Further, collecting the terms of $\mathcal{O}(\epsilon^m$) from (\ref{HOS_DR1}), we have,
   \begin{align*}
   &\sum_{k=0}^{m-1} \frac{\eta_u^k}{k!}\frac{\partial^k}{\partial z^k}\phi_u^{(m-k)}(x,z,t)\bigg\rvert_{z=0}&=&\qquad 0\\
   \Rightarrow & \sum_{k=1}^{m-1} \frac{\eta_u^k}{k!}\frac{\partial^k}{\partial z^k}\phi_u^{(m-k)}(x,z,t)\bigg\rvert_{z=0}+\phi_u^{(m)}(x,z,t)\big\rvert_{z=0}&=&\qquad 0,\\
   \Rightarrow & \qquad\qquad\qquad\qquad\phi_u^{(m)}(x,0,t)&=& -\sum_{k=1}^{m-1} \frac{\eta_u^k}{k!}\frac{\partial^k}{\partial z^k}\phi_u^{(m-k)}(x,z,t)\bigg\rvert_{z=0}.
   \end{align*}
   So, combining the boundary conditions at every order, we can write,
    \begin{equation}
    	\phi_u^{(m)}(x,0,t)=f_u^{(m)},
    \end{equation}
    where
    \begin{align}
    	f_u^{(1)}&=\phi^S,\\
        f_u^{(m)}&=-\sum_{k=1}^{m-1}\frac{\eta_u^k}{k!}\frac{\partial^k}{\partial z^k}\phi_u^{(m-k)}(x,z,t)\bigg\rvert_{z=0}.
    \end{align}
    In a very similar way, we can derive the Dirichlet boundary condition at the pycnocline.
    
    \subsection{Neumann Boundary Condition}
   Firstly, we subtract the two kinematic boundary conditions at the pycnocline to get rid of the time derivative, and obtain
   \begin{equation}
   	\varphi_{,z}(x,z,t)=\eta_{l,x}\varphi_{,x}(x,z,t)+\eta_l\eta_{l,x}(\Omega_u-\Omega_l)\qquad z=-h_u+\eta_l. \label{B4}
\end{equation}
   Note, that we have expanded the base velocity $U$ in a Taylor series about the mean surface. Here, $\varphi(x,z,t)\equiv\phi_{u}(x,z,t)-\phi_{l}(x,z,t)$. We expand $\varphi(x,z,t)$ in a Taylor expansion about the mean height of the interface to get
 $$\varphi(x,-h_u+\eta_l,t)=\sum_{m=1}^{M}\sum_{k=0}^{M-m} \frac{\eta_l^k}{k!}\frac{\partial^k}{\partial z^k}\varphi^{(m)}(x,z,t)\bigg\rvert_{z=-h_u}.$$
 Substituting this in the (\ref{B4}), while ignoring the term $\eta_l\eta_{l,x}(\Omega_u-\Omega_l)$, for now, we have,
 $$\sum_{m=1}^{M}\sum_{k=0}^{M-m} \frac{\eta_l^k}{k!}\frac{\partial^k}{\partial z^k}\varphi_{,z}^{(m)}(x,z,t)\bigg\rvert_{z=-h_u} = \frac{\partial \eta_l}{\partial x}\sum_{m=1}^{M}\sum_{k=0}^{M-m} \frac{\eta_l^k}{k!}\frac{\partial^k}{\partial z^k}\varphi_{,x}^{(m)}(x,z,t)\bigg\rvert_{z=-h_u}.$$
 Again, collecting the terms of $\mathcal{O}(\epsilon^m$) from the above equation we have
 \begin{align}
 &\sum_{k=0}^{m-1} \frac{\eta_l^k}{k!}\frac{\partial^k}{\partial z^k}\varphi_{,z}^{(m-k)}\bigg\rvert_{z=-h_u} &= 
 \frac{\partial \eta_l}{\partial x}\sum_{k=1}^{m-1} \frac{\eta_l^{k-1}}{(k-1)!}\frac{\partial^{k-1}}{\partial z^{k-1}}\varphi_{,x}^{(m-k)}\bigg\rvert_{z=-h_u}\nonumber \\
 \Rightarrow & \sum_{k=1}^{m-1} \frac{\eta_l^k}{k!}\frac{\partial^k}{\partial z^k}\varphi_{,z}^{(m-k)}\bigg\rvert_{z=-h_u}+\varphi_{,z}^{(m)}\big\rvert_{z=-h_u}&=\sum_{k=1}^{m-1} \frac{\partial \eta_l}{\partial x}\frac{\eta_l^{k-1}}{(k-1)!}\frac{\partial^{k-1}}{\partial z^{k-1}}\varphi_{,x}^{(m-k)}\bigg\rvert_{z=-h_u}. \label{B5}
 \end{align}
 Now, using the continuity equation we have,
\begin{align*}
&\frac{\partial^2 \phi_u}{\partial z^2}=-\frac{\partial^2 \phi_u}{\partial x^2}\\
\Rightarrow &\frac{\partial^{k }\phi_{u,z}}{\partial z^k}=-\frac{\partial^{k-1}}{\partial z^{k-1}}\phi_{u,xx} \quad \textrm{for} \quad k>0.
\end{align*}
Similarly, we have,
\begin{align*}
\frac{\partial^{k }\phi_{l,z}}{\partial z^k}=-\frac{\partial^{k-1}}{\partial z^{k-1}}\phi_{l,xx}
\end{align*}
Subtracting the above two equations, we obtain,
\begin{align*}
\frac{\partial^{k }\varphi_{,z}}{\partial z^k}=-\frac{\partial^{k-1}}{\partial z^{k-1}}\varphi_{,xx}
\end{align*}
Now using the above result in (\ref{B5}), we have,
\begin{align}
-\sum_{k=1}^{m-1} \frac{\eta_l^k}{k!}\frac{\partial^{k-1}}{\partial z^{k-1}}\varphi_{,xx}^{(m-k)}\bigg\rvert_{z=-h_u}+\varphi_{,z}^{(m)}\big\rvert_{z=-h_u}=\quad\sum_{k=1}^{m-1} \frac{\partial \eta_l}{\partial x}\frac{\eta_l^{k-1}}{(k-1)!}\frac{\partial^{k-1}}{\partial z^{k-1}}\varphi_{,x}^{(m-k)}\bigg\rvert_{z=-h_u}.
\end{align}
Rearranging the terms, we have,
\begin{align}
 &\varphi_{,z}^{(m)}\big\rvert_{z=-h_u} &=&\quad \sum_{k=1}^{m-1} \frac{\partial^{k-1}}{\partial z^{k-1}}\left[\frac{\eta_l^{k-1}}{(k-1)!}\frac{\partial \eta_l}{\partial x}\varphi_{,x}^{(m-k)}+\frac{\eta_l^k}{k!}\varphi^{(m-k)}_{,xx}\right]_{z=-h_u}\\
\Rightarrow & \varphi_{,z}^{(m)}\big\rvert_{z=-h_u} &=&\quad \sum_{k=1}^{m-1} \frac{\partial^{k-1}}{\partial z^{k-1}}\frac{\partial}{\partial x}\left[\frac{\eta_l^k}{k!}\varphi_x^{(m-k)}\right]_{z=-h_u}\\
\Rightarrow &\varphi_{,z}^{(m)}\big\rvert_{z=-h_u} &=&\quad \sum_{k=1}^{m-1} \frac{\partial}{\partial x}\left[\frac{\eta_l^k}{k!}\frac{\partial^{k-1} }{\partial z^{k-1}}\varphi_x^{(m-k)}\right]_{z=-h_u}.
\end{align}

Now, we include the term $\eta_l\eta_{l,x}(\Omega_u-\Omega_l)$, the effect of which will be only in the $\mathcal{O}(\epsilon^2)$ terms. So, finally we have
   \begin{equation}
    \varphi_{,z}^{(m)}(x,-h_u,t)=f_{l2}^{(m)},
\end{equation}
    where
 \begin{align}
    f_{l2}^{(1)}&=0,\\
    f_{l2}^{(2)}&=\frac{\partial}{\partial x}\left[\eta_l\varphi_{,x}^{(1)}(x,z,t)\bigg\rvert_{z=-h_u}\right]+\eta_l\eta_{l,x}(\Omega_u-\Omega_l)\\
        f_{l2}^{(m)}&=\sum_{k=1}^{m-1}\frac{\partial}{\partial x}\left[\frac{\eta_l^k}{k!}\frac{\partial^{k-1}}{\partial z^{k-1}}\varphi_{,x}^{(m-k)}(x,z,t)\bigg\rvert_{z=-h_u}\right].
    \end{align}
    At $\mathcal{O}(\epsilon)$
In a similar way we can derive the bottom boundary condition, which also is a Neumann boundary condition.

\bibliographystyle{jfm}
\bibliography{references}

\end{document}